%% file: champI.5flat.tex
\documentclass[usenatbib,usegraphicx]{mn2e}
\pdfoutput=1

\usepackage{color}

\newcommand{\solar}{$_{\odot}$}

\newcommand{\ceto}{C$^{18}$O}
\newcommand{\hcop}{HCO$^+$}

\newcommand{\nnh}{N$_2$H$^+$}

\newcommand{\vlsr}{$V_{\rm{LSR}}$}
\newcommand{\joz}{$J=1\rightarrow0$}

\newcommand{\fto}{$F_1=2\rightarrow1$}
\newcommand{\kms}{\,km\,s$^{-1}$}
\newcommand{\hii}{H\textsc{ii}}

\newcommand{\brg}{Br$\gamma$}
\newcommand{\htwo}{H$_2$}
\newcommand{\voz}{$v$=1$\rightarrow$0}
\newcommand{\vto}{$v$=2$\rightarrow$1}
\def\lapp{\ifmmode\stackrel{<}{_{\sim}}\else$\stackrel{<}{_{\sim}}$\fi}
\def\gapp{\ifmmode\stackrel{>}{_{\sim}}\else$\stackrel{>}{_{\sim}}$\fi}

\title[Signposts of Molecular Clump Evolution \& Star Formation]{Millimetre-Wave and Near-Infrared Signposts of Massive \\ Molecular Clump Evolution and Star Cluster Formation}
\author[Barnes et al.]{Peter J. Barnes$^{1}$\thanks{E-mail: pjb@astro.ufl.edu}, Stuart D. Ryder$^{2}$, Stefan N. O'Dougherty$^{1}$, Luis E. Alvarez$^{1}$, 
 \newauthor 
Adriana S. Delgado-Navarro$^{1}$, Andrew M. Hopkins$^{2}$, and Jonathan C. Tan$^{1}$\vspace{2mm}\\
$^{1}$Astronomy Department, University of Florida, Gainesville, FL 32611, USA\\
$^{2}$Australian Astronomical Observatory, PO Box 915, North Ryde, NSW 1670, Australia
}

\begin{document}

\setcounter{page}{1}
\date{Accepted 2013 April 3. Received 2013 March 27; in original form 2012 February 22}

\pagerange{\pageref{firstpage}--\pageref{lastpage}} \pubyear{2013}

\maketitle

\label{firstpage}

\begin{abstract}
We report new near-infrared and mm-wave observational data on a selection of massive Galactic molecular clumps (part of the CHaMP sample) and their associated young star clusters.  The clumps show, for the first time in a ``dense gas tracer'', a significant correlation between \hcop\ line emission from cold molecular gas and \brg\ line emission of associated nebulae.  This correlation arises in the \hcop\ line's brightness, not its linewidth.  In contrast, the correlation between the \nnh\ line emission and \brg\ is weak or absent.  The \hcop/\nnh\ line ratio also varies widely from clump to clump: bright \hcop\ emission tends to be more closely associated with \brg\ nebulosity, while bright \nnh\ emission tends to avoid areas that are bright in \brg.  Both molecular species show correlations of weak significance with infrared \htwo\ \voz\ and \vto\ line emission, in or near the clumps.   The \htwo\ emission line ratio is consistent with fluorescent excitation in most of the clumps, although thermal excitation is seen in a few clumps.  We interpret these trends as evidence for evolution in the gas conditions due to the effects of ongoing star formation in the clumps, in particular, the importance of UV radiation from massive YSOs as the driving agent that heats the molecular gas and alters its chemistry.  This suggests that some traditional dense gas tracers of molecular clouds do not sample a homogeneous population of clumps, i.e., that the \hcop\ brightness in particular is directly related to the heating and disruption of cold gas by massive young stars, whereas the \nnh\ better samples gas not yet affected by this process.  We therefore suggest that the \hcop--\nnh--\brg\ relationship is a useful diagnostic of a molecular clump's progress in forming massive stars.
\end{abstract}

\begin{keywords}
astrochemistry --- infrared lines: ISM --- ISM: clouds --- ISM: molecules --- radio lines: ISM --- stars: formation.
\end{keywords}

\section{Introduction\label{intro}}

Our understanding of the most important processes in massive star and cluster formation is still very inadequate, despite much effort in the past decade \citep{bcm07}.  While a number of important surveys have been completed, detailed multi-wavelength studies often suffer from small sample size \cite[e.g.,][]{pzc07,h09}, while larger surveys often suffer from various selection effects and limited wavelength coverage \cite[e.g.,][]{L07,w10}.  Thus, there is little agreement on the essential physical, chemical, and dynamical features of massive molecular clumps as they give rise to star clusters, or likewise on the evolution of molecular cores that make massive stars.  There are also enduring mysteries, such as the source of turbulence in massive molecular clouds \citep{NL05,NL11,m07}.

In large part, progress in our understanding of massive molecular clump evolution and cluster formation has been hindered by the triple obstacles of (i) large distances due to their relative rarity, (ii) their rapid evolution, and (iii) the rich variety of phenomena (masers, outflows, \hii\ regions, a high density of objects) that luminous sources exhibit, and which a paradigm of massive star formation must account for.  These obstacles make conducting large-scale, high-resolution and -sensitivity surveys difficult and time-consuming.  Recently, however, a number of star formation surveys in the Milky Way have been undertaken, such as GLIMPSE \citep{b03}, BGPS \citep{agd11}, Hi-GAL \citep{m10}, and several others, that are at last providing the resolution, sensitivity, and statistical basis for a more thorough understanding of this problem.  While these mainly broadband surveys are indeed transformative, rarer molecular-spectroscopic data from surveys such as CHaMP (described next) can provide even keener insights into massive molecular clump evolution.

\vspace{-3mm}
\section{Observations\label{obs}}

\subsection{The CHaMP Sample\label{champ}}
The results reported here were obtained as part of the Galactic {\em Census of High- and Medium-mass Protostars} \cite[CHaMP;][]{YAK05,BYR10,BYF11}.  This is a complete, unbiased, multi-wavelength survey of massive, dense molecular clumps and associated young star clusters within a 20$^{\circ}\times$6$^{\circ}$ sector of the southern Milky Way.  CHaMP's main objectives are to systematically characterise the physical processes and timescales in massive star and cluster formation.  The first complete catalogues and results of this project were presented by \citet[][hereafter Paper I]{BYF11}.  Here, 209 parsec-scale ``Nanten clumps'' were identified from 4$'$-resolution maps in the \joz\ transitions of \ceto\ and \hcop\ at $\lambda$ = 3 mm wavelength: these comprise the Nanten Master Catalogue (hereafter NMC), denoted by their catalogue numbers (e.g., BYF 73).  The brightest 121 of these were mapped at 40$''$ resolution using the Mopra telescope (see \S\ref{mopra} for more details) to resolve the molecular emission into sub-parsec-scale ``Mopra clumps,'' denoted by subcomponents of the NMC (e.g., BYF 77a).  The powerful Mopra backend maps many molecular species simultaneously: over the initial observed range of 85--93\,GHz, the brightest of these is the \hcop\ line, and analysis of the Mopra \hcop\ maps yielded the 303 entries in the catalogue of massive, dense clumps in Paper I (hereafter MHC, for Mopra \hcop\ Catalogue).  Other species mapped are as listed in Table 3 of Paper I, and the analysis of these will appear in future papers.  Among these, the species \nnh\ and HCN are also fairly bright, but much less widespread than \hcop.  Physically, \nnh\ and HCN should trace similar conditions to \hcop, but their abundances are thought to depend on astrochemical factors \cite[e.g.,][]{CWZ02}.

The main results of 
Paper I include: (i) the first recognition that the Milky Way's dense molecular 
ISM is dominated by a vast, subthermally-excited population of massive clumps, which significantly outnumber the brighter and traditionally better-studied star forming regions; (ii) if the clumps evolve by slow contraction, up to 95\% of this population may represent a long-lived stage of pressure-confined, gravitationally stable massive clump evolution, and the CHaMP clump population may not engage in vigorous massive star formation until the last 5\% of their lifetimes; (iii) the brighter sources are denser, more massive, more highly pressurised, and closer to gravitational instability than the less bright sources.  The MHC clumps' properties are: integrated line intensity 1--30\,K\,\kms, peak line brightness 1--7\,K, linewidth 1--10\,\kms, integrated line luminosity 0.5--200\,K\,\kms\,pc$^2$, size 0.2--2.5\,pc, mean projected axial ratio 2, optical depth 0.08--2, surface density 30--3000\,M\solar\,pc$^{-2}$, number density 0.2--30$\times10^9$\,m$^{-3}$, mass 15--8000\,M\solar, virial parameter 1--55, and gas pressure 0.3--700\,pPa.

Paper I has therefore already provided a significant and well-advanced component of CHaMP, devoted to molecular spectroscopic mapping of these clumps at 3 mm wavelength, but the planned coverage of these clouds at other wavelengths is still in its early stages \cite[e.g.,][]{BYR10}.  Nevertheless, from among this sample of star-forming clouds, we can already discern new insights from near-infrared (NIR) imaging of a subset of these clumps, and from a comparison of these images with selected mm-wave molecular maps.  In order to statistically characterise the major stages of massive star and cluster formation, starting from the cold, prestellar gas stages and proceeding to the newly-emerged pre-MS stars, we need both a suitable sample of objects and a judicious choice of probes.  Here we compare, for a subset of 60 members of the NMC, the \hcop\ maps from Paper I together with companion \nnh\ maps presented here for the first time, plus new near-IR images which we describe next.  As explained below, this subset of the CHaMP sample provides a good representation of the range of clump parameters contained in the MHC.

\begin{table}
\centering
\begin{minipage}{140mm}
\caption{AAT Observing Log.\label{aatlog}}
\begin{tabular}{@{}cll@{}}
\hline
UT Date & Field$^a$ & Filters \\
\hline
  2006 May 13 & BYF\,40 & all narrow \\ 
  2006 May 13 & BYF\,54 & all narrow \\ 
  2006 May 13 & BYF\,73$^b$ & all narrow \\ 
  2006 May 13 & BYF\,77 & all narrow \\ 
  2006 May 13 & BYF\,128 & all narrow \\ 
  2007 May 26 & BYF\,66 & all narrow \\ 
  \\
  2011 Feb 18 & BYF\,5 & \brg, $K$-cont \\
  2011 Feb 18 & BYF\,60 & all narrow+broad \\
  2011 Feb 18 & BYF\,61 & all narrow+broad \\
  2011 Feb 18 & BYF\,62 & all narrow+broad \\
  2011 Feb 18 & BYF\,63 & all narrow+broad \\
  2011 Feb 18 & BYF\,64 & all narrow+broad \\
  2011 Feb 18 & BYF\,83 & all narrow+broad \\
  2011 Feb 18 & BYF\,85 & all narrow+broad \\
  2011 Feb 18 & BYF\,118 & all narrow+broad \\
  2011 Feb 18 & BYF\,123 & all narrow+broad \\
  2011 Feb 18 & BYF\,208 & all narrow+broad \\
  \\
  2011 Mar 21 & BYF\,66 & all narrow+broad \\
  2011 Mar 23 & BYF\,67 & all narrow+broad \\
  2011 Mar 23 & BYF\,68 & all narrow+broad \\
  2011 Mar 23 & BYF\,69 & all narrow+broad \\
  2011 Mar 23 & BYF\,71 & all narrow+broad \\
  2011 Mar 24 & BYF\,72 & all narrow+broad \\
  2011 Mar 24 & BYF\,73 & all narrow+broad \\
  2011 Mar 24 & BYF\,76 & all narrow+broad \\
  2011 Mar 24 & BYF\,77 & \brg, \htwo\ \voz, $K$-cont \\
\hline
\end{tabular}\\
$^a$ From the Nanten Master Catalogue in Paper I.\\
$^b$ Analysed by \cite{BYR10}.
\end{minipage}
\end{table}

\subsection{Near-IR Imaging\label{aat}}
Being a hydrogen recombination line, Brackett $\gamma$ naturally traces \hii\ regions, i.e. ionised gas at temperatures $\sim$7--9,000\,K.  However, being in the NIR, \brg\ is far less affected by dust extinction than the traditional optical tracer of ionised hydrogen, H$\alpha$.  Radio recombination lines (RRLs) are completely unaffected by dust extinction, however traditional surveys of cm-wave RRLs are still not sensitive to the youngest Compact and Ultracompact \hii\ regions, since the optical depth to free-free emission in such objects may be quite high.  Thus, \brg\ can conveniently probe the earliest stages of \hii\ region formation around very young, deeply embedded massive stars that H$\alpha$ and cm-wave RRL surveys (for example) may simply be unable to see.  Similarly, NIR \htwo\ lines sample regions where the molecular hydrogen has been heated (whether by shocks or fluorescence) to temperatures $\sim$1--4,000\,K.  Thus, these lines can potentially probe gas which has been warmed, but not yet ionised, by the radiation from very young stars.  Such line imaging has been used to study a number of Galactic star-forming regions \cite[e.g.,][]{CGN06}, and recently a \htwo\ \voz\ survey of more than half of the Milky Way's first quadrant has been completed \citep{f11}.  But apart from a few such studies, remarkably little systematic, large-scale, near-IR survey work of the kind described in this paper has been performed.

We obtained near-IR images of 26 fields, of which 22 are non-redundant, covering 60 CHaMP clumps (about 20\% of the MHC) using the IRIS2 camera \citep{TRE04} on the Anglo-Australian Telescope (Table \ref{aatlog}).  We began with a pilot study using AAT service time in 2006 and 2007, obtaining images of 6 clumps using only the narrowband filters (fractional bandwidths $\sim$1.4\%) for the emission lines \brg\ ($\lambda=2.17\mu$m), \htwo\ \voz\ $S$(1) (2.12 $\mu$m) and \vto\ $S$(1) (2.25 $\mu$m), plus a fourth ``$K$-continuum'' narrowband filter in a line-free part of the band (2.25--2.28 $\mu$m, although this filter may still contain some residual emission from the \vto\ line).  As these filters all lie in the $K$-band, the differential reddening between them should be minimal.\footnote{The labels for the two \htwo\ filters were inadvertently switched in Figure 7a of \cite{BYR10}, but the error does not affect the conclusions of that work.  This labelling error has been corrected here.}
The observation and data reduction procedures were as described by \cite{BYR10}.  Briefly, for each filter 9 $\times$ 60s images (dithered by 1$'$) were obtained of the instrument's 7\farcm7 $\times$ 7\farcm7 field under 0\farcs9 seeing, and reduced using the \textsc{orac--dr} pipeline \citep{CHJ03}.  Subsequent image processing was performed with the \textsc{iraf}\footnote{\textsc{iraf} is distributed by the National Optical Astronomy Observatories, which are operated by the Association of Universities for Research in Astronomy, Inc., under cooperative agreement with the U.S. National Science Foundation.} 
and \textsc{miriad} \citep{stw95} packages.  After astrometric registration using SuperCOSMOS\footnote{This research has made use of data obtained from the SuperCOSMOS Science Archive, prepared and hosted by the Wide Field Astronomy Unit, Institute for Astronomy, University of Edinburgh, which is funded by the UK Science and Technology Facilities Council.} $I$-band images, 
we linearly scaled the spectral-line images to the same relative brightness scale as the $K$-band continuum by matching the integrated fluxes of several stars in each filter, assuming they were of similar colour.  We then subtracted the continuum from the spectral-line images before transforming each image to Galactic coordinates.  Finally, pseudo-colour images (described in \S\ref{general}) were formed from these line images.

In 2011 over 8 nights with variable conditions, we obtained (in the periods when the sky was clear) similar narrowband images of the other fields in Table \ref{aatlog}, plus $JHK$ broadband images.  In this season, however, {\color{black} 3 of the narrowband images were made with 5 $\times$ 60s dithers each, while the \vto\ imaging used 5 $\times$ 120s dithers to compensate for the generally fainter emission seen previously in this line;} the broadband images were each constructed from 5 $\times$ 10s dithers.  {\color{black} Thus, the pilot $K$-cont, \brg, and \voz\ images from 2006--07 are more sensitive than the equivalent 2011 data.}  The \textsc{orac--dr} pipeline for the 2011 observations included astrometric registration using 2MASS\footnote{This publication makes use of data products from the Two Micron All Sky Survey, which is a joint project of the University of Massachusetts and the Infrared Processing and Analysis Center/California Institute of Technology, funded by the National Aeronautics and Space Administration and the National Science Foundation.}
$JHK$ catalogue data, resulting in alignments good to $\sim$1$''$ (no further astrometry with SuperCOSMOS was performed here).  Otherwise the reduction procedures used were as above.  In what follows we discuss only the narrowband images, and leave the analysis of the broadband emission for a later study.

From the above, while it is clear that the fields covered by this IR imaging were not systematically optimised for (e.g.) uniformity, brightness, or other property (due primarily to the vagaries of observing conditions), the resulting dataset nevertheless yields a useful microcosm of the MHC: for example, the 2006--07 data cover many of the brighter clumps, while the 2011 data cover most of ``Region 9'' from Paper I.  We can therefore reliably examine here a range of clumps from the brightest and most massive to the relatively large fraction of fainter and more moderate-mass clouds.

\begin{figure*}
\hspace{-3mm}\includegraphics[angle=-90,scale=0.94]{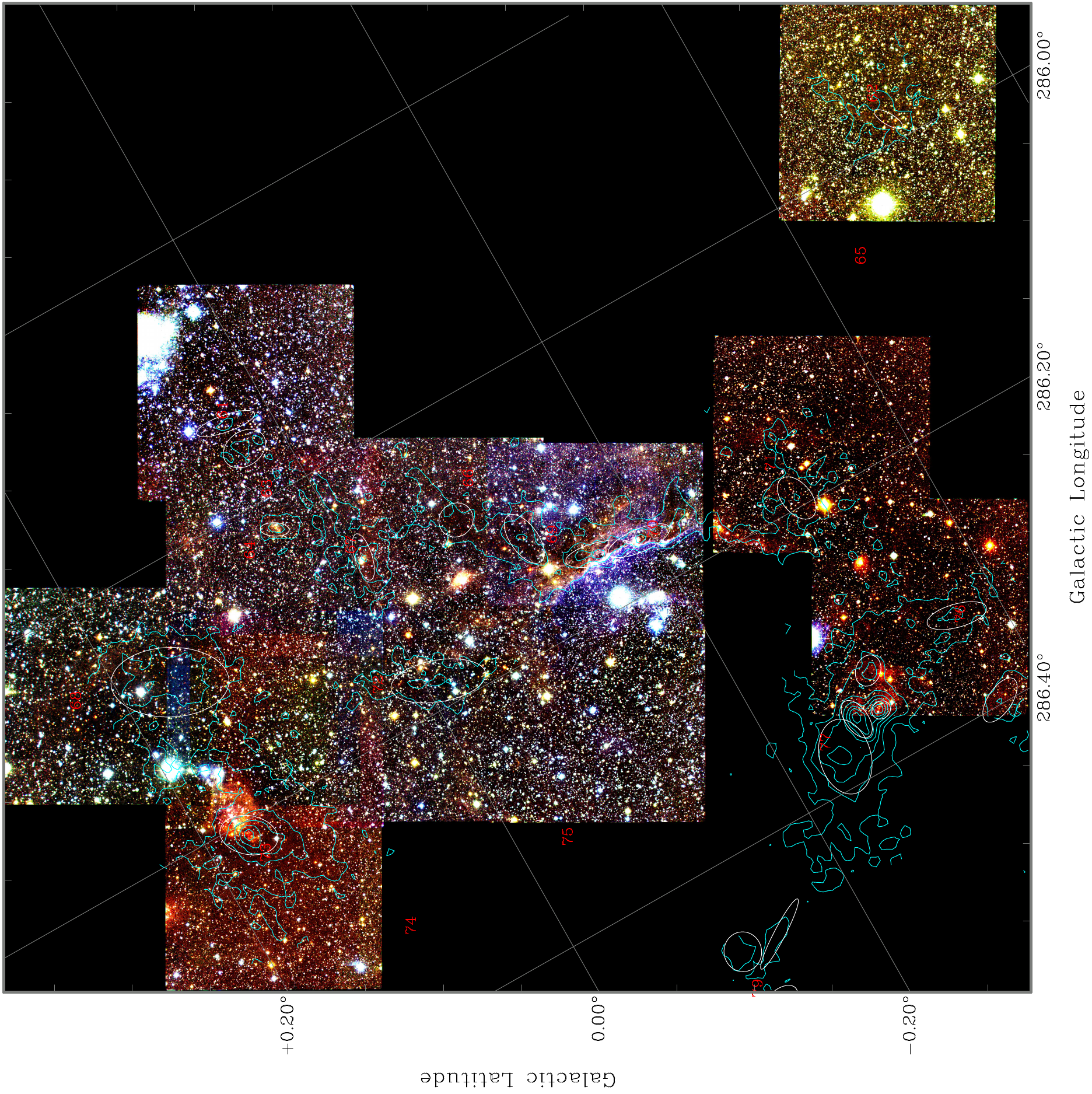}
\caption{Wide-field RGB-pseudo-colour image of broad-band $JHK$ for BYF\,61--78 (most of Region 9 from Paper I) using mosaicked 2011 data, overlaid by Mopra \hcop\ contours (cyan levels at 5, 10, 15, 20, 25, 30, and 35 times the rms level of 0.404\,K\kms).  The white ellipses show gaussian fits to the \hcop\ emission, while the red numbers indicate nominal positions of NMC clumps.  This figure is provided for the convenience of locating the fields of Figs.\,\ref{byf61}--\ref{byf77} in their wider context --- compare this image with Fig.\,34 in Paper I.
At a distance of 2.5\,kpc, the scale is 40$''$ = 0.485\,pc or 1\,pc = 0\fdg0229 = 82\farcs5.
\label{reg9}}
\end{figure*}

\subsection{Mopra Mapping\label{mopra}}
{\color{black} At lower temperatures, \htwo\ is spectroscopically inert: to probe this cold gas, we must turn to trace species.  The CO molecule and its isotopologues are bright and widely-used, yet they typically represent the lower-density envelopes of molecular clouds (number densities $\sim$10$^{8-9}$\,m$^{-3}$ or less, corresponding to mass densities $\sim$60\,M\solar\,pc$^{-3}$).  
In contrast, most star formation is known to occur in only the denser cores and clumps of these clouds, with densities at least 10 times higher \citep{ll03}.  Instead, \hcop\ and \nnh\ are useful higher-density cold gas probes expected to trace similar physical conditions in the molecular gas: both are ions with large dipole moments and similar excitation requirements, sensitive to densities $\sim$10$^{10}$\,m$^{-3}$ or more.}

We obtained maps of the \hcop\ and \nnh \joz\ emission at {\color{black} 89.188\,526 and 93.173\,777} GHz (resp.) using the 22 m diameter Mopra antenna, part of the Australia Telescope National Facility\footnote{The Mopra telescope is part of the Australia Telescope which is funded by the Commonwealth of Australia for operation as a National Facility managed by CSIRO.  The University of New South Wales Digital Filter Bank used for the observations with the Mopra telescope was provided with support from the Australian Research Council.}.
Mopra's MOPS spectrometer collects data on multiple molecular species simultaneously, therefore the relative registration of features in maps of the two molecular species is perfect.  The absolute pointing accuracy was set by $\sim$hourly checks on the SiO maser source R Carinae, and was generally found to be better than 10$''$.  Full details of the hardware, mapping, and data reduction procedures are given in Paper I, and the \hcop\ maps are drawn from that catalogue.  The \nnh\ maps presented here are new, but have been produced in the same way as the \hcop\ maps of Paper I.  Therefore, details of the \nnh\ data such as beam size (HPBW $\sim$ 40$''$), noise characteristics (global average $T_{\rm rms}$ $\sim$ 0.31\,K per 0.11\,\kms\ channel), and the like, are virtually identical to those for \hcop\ in Paper I.


\begin{figure*}
(a)\hspace{-3mm}\includegraphics[angle=-90,scale=0.41]{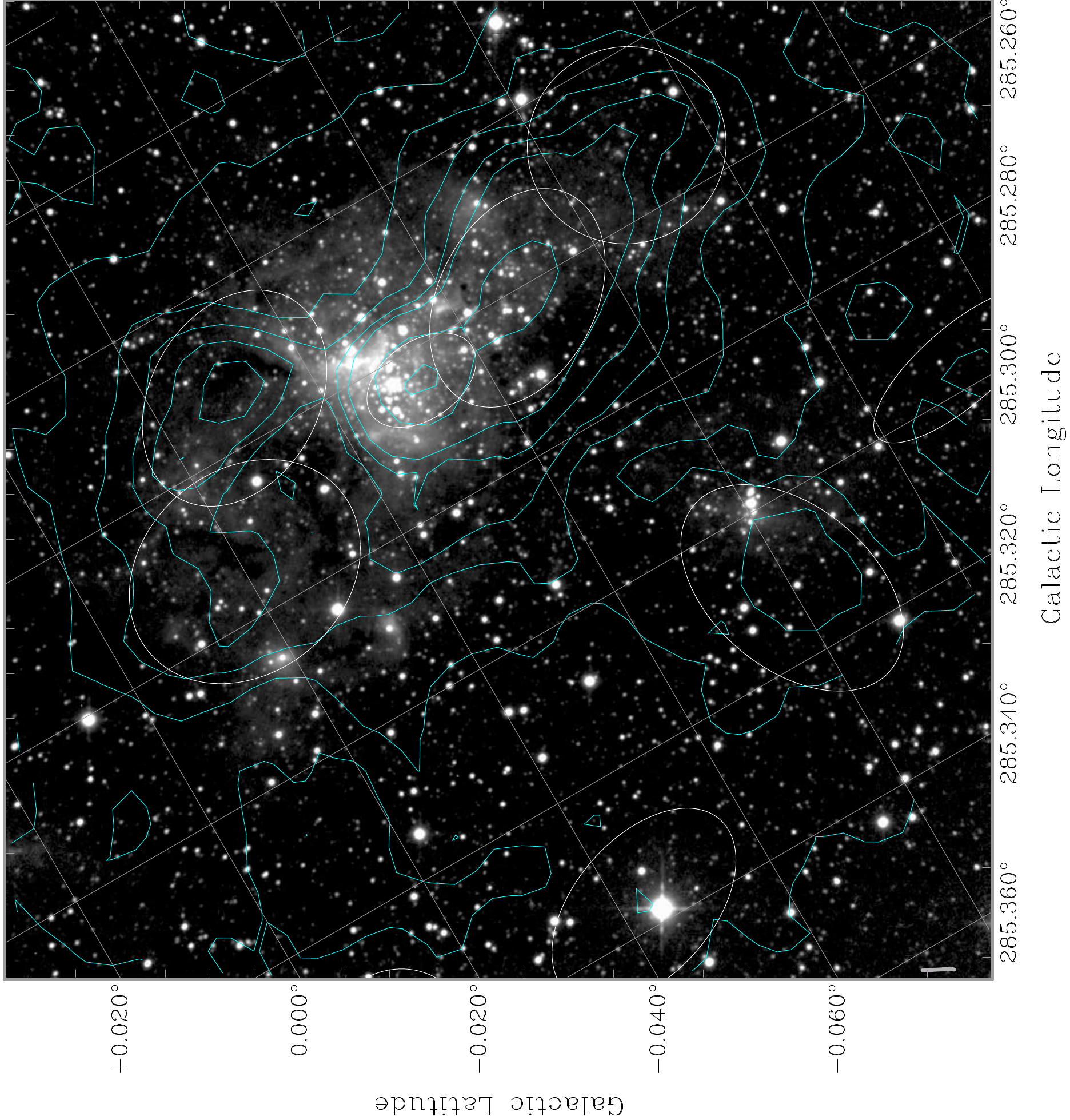}\hspace{3mm}
(b)\hspace{-3mm}\includegraphics[angle=-90,scale=0.41]{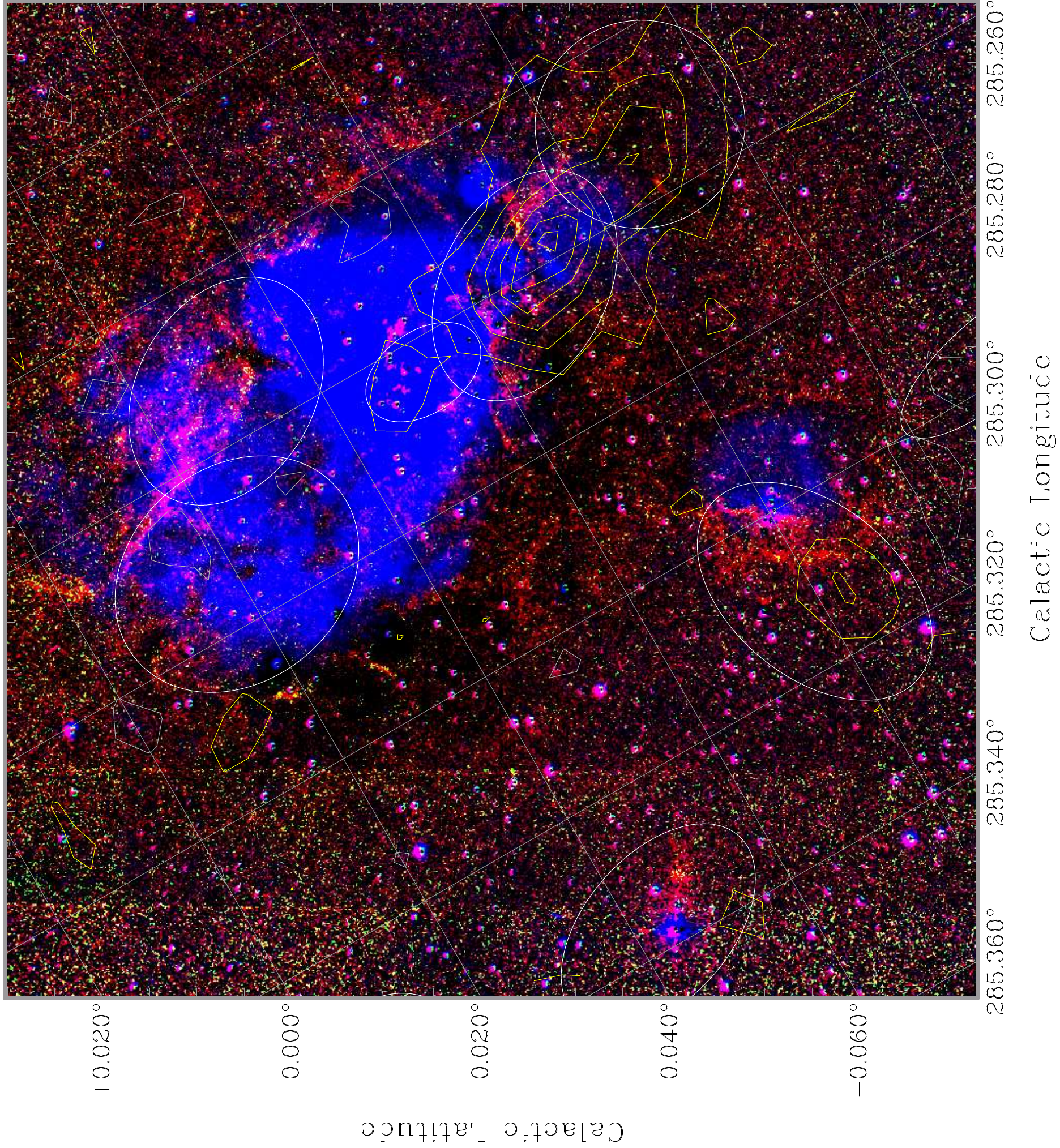} 
\caption{Example of the survey fields presented in the Appendix (available online only).  These images of {\color{black} BYF\,54 (part of Region 8} from Paper I) 
are from the more sensitive 2006 data.  (a) $K$-band line-free continuum image on a linear brightness scale.  Contours are overlaid from Mopra \hcop\ integrated intensity: 
cyan levels at 4, 8, 12, 16, 20, 24, 40, and 56 times the rms at 0.401\,K\kms.  (b) RGB-pseudo-colour image of the continuum-subtracted $K$-band spectral lines on a linear brightness scale.  Here, H$_2$ $S$(1) is shown as red (\voz) \& green (\vto), while \brg\ is shown as blue.  Overlaid here are contours of Mopra \nnh\ integrated intensity (grey levels at --2, and yellow levels at 2, 4, 6, 8, 10, and 12 times the rms of 0.603\,K\kms).  The white ellipses represent gaussian fits to the \hcop\ emission in both panels.  At a distance of 2.5\,kpc, the scale is 40$''$ = 0.485\,pc or 1\,pc = 0\fdg0229 = 82\farcs5.  {\color{black} This field shows perhaps the most obvious example of the \hcop/\nnh\ intensity variations with position, as described in \S\S\ref{mmdata}ff.}
\label{sample}}
\end{figure*}

\section{Images and Analysis\label{results}}

\subsection{General Features\label{general}}
Of the 22 fields imaged and reported here, 12 comprise a large part of Region 9 from Paper I, while the other 10 fields variously cover some of the brighter and fainter clumps from the MHC.  We first present in Figure \ref{reg9} a wide-field, $JHK$ broadband mosaic of the twelve Region 9 fields, to provide a wider context for the individual fields described next.  In Figure \ref{sample} we give a sample narrowband field showing the salient features of our presentation, while all the available narrowband IR images are presented in the Appendix (Figures \ref{byf5}--\ref{byf208}).

In each individual field (see Fig.\,\ref{sample}), we show the line-free $K$-continuum in the left panel, and a pseudo-colour composite image of the continuum-subtracted line emission in the right panel, where (except in Fig.\,\ref{byf5}) the \htwo\ \voz\ and \vto\ are shown as red and green, respectively, while \brg\ is shown as blue.  These colours therefore represent the excitation level in the gas: where only \htwo\ \voz\ exists at gas temperatures $\sim$1,000--2,000\,K, we see red nebulosity, through yellow or green areas where the \vto\ emission picks up or dominates at $\sim$2,000--4,000\,K, to cyan or blue areas where the \brg\ shows ionised gas at $\sim$7,000--9,000\,K.  (The colour stretches in each Figure, however, are not the same, and are chosen to display the most information in each panel.)  In addition, the panels are overlaid by contours of the integrated intensity of the two molecular species \hcop\ (cyan contours, usually in the $K$-cont panel) and \nnh\ (yellow contours, usually in the line emission panel where available).  In some figures, both sets of contours are overlaid on each panel.  We also show in both panels (as white ellipses) gaussian fits to the components of {\em the \hcop\,emission only}, which help to register features between the panels.

While these figures show a wide variety of stellar groupings and emission-line morphologies, there are a few trends in common, which we now explore.  First, we note a somewhat subtle point: IR emission tends to be absent or very weak in the clumps with relatively weaker mm-line emission.  In other words, stronger IR emission tends to occur with stronger mm-line emission, and there may be a distinct threshold of mm emission needed before IR line emission becomes apparent at these levels.  A second trend is more prominent: the IR emission lines tend to be spatially quite distinct, i.e. the \hii\ regions (as traced by the \brg) rarely overlap with the \htwo\ emission, although they are frequently adjacent.  Also, the \hii\ emission is always relatively diffuse, although sometimes a sharp ionisation front is visible (e.g., BYF\,70).  In contrast, the \htwo\ emission is usually (but not always) quite filamentary, with widths often as small as a few arcseconds but lengths up to an arcminute or more (e.g., in or near clumps BYF\,54d, 73, 77a,c; the clump designations are given in Table 4 and Appendix A of Paper I).  Where diffuse (i.e., non-filamentary) \htwo\ emission is occasionally seen, it is almost always centred on one of the stars embedded in the cold gas clump (e.g., BYF\,40a, 77a).  In general, then, there is a high coincidence of \brg\ and \htwo\ emission from within and around the brighter examples of these cold, massive, dense clumps.

\begin{table*}
\centering
\begin{minipage}{175mm}
\caption{Mopra \nnh \joz\ Observed Parameters.\label{NNHsources}}
\begin{tabular}{@{}lccrcrlllccc@{}}
\hline
  BYF & $l$ & $b$ & $\int T_R^* dV$ & V range & $T_p$ & $V_{\rm LSR}$ & $\sigma_V$ & {\color{black} $\sigma_{V,dec}$} & $\theta_{\rm maj}$ & $\theta_{\rm min}$ & PA \\ 
  no. & deg & deg & K\,\kms & \kms & K & \kms & \kms & {\color{black} \kms} & arcsec & arcsec & deg \\ 
\hline
\input{table2a}

\hline
\end{tabular}
\end{minipage}
\end{table*}

\begin{table*}
\centering
\begin{minipage}{175mm}
\contcaption{}
\begin{tabular}{@{}lccrcrlllccc@{}}
\hline
  BYF & $l$ & $b$ & $\int T_R^* dV$ & V range & $T_p$ & $V_{\rm LSR}$ & $\sigma_V$ & {\color{black} $\sigma_{V,dec}$} & $\theta_{\rm maj}$ & $\theta_{\rm min}$ & PA \\ 
  no. & deg & deg & K\,\kms & \kms & K & \kms & \kms & {\color{black} \kms} & arcsec & arcsec & deg \\ 
\hline
\input{table2b}

\hline
\end{tabular} \\
{\textsc{Notes.}  An ``N'' in columns 2 \& 3 and dashes in the other columns indicate clumps for which no \nnh\ emission could be detected, with the rms sensitivity given in column 4.  An ``S'' in columns 7--9 indicates clumps which were detected in \nnh, but for which the S/N was too low to give reliable higher moment measurements.  {\color{black} The values in column 9 are deconvolved from those in column 8, as described in the text.}  Uncertainties in parentheses are in the last digit of the corresponding value.  
Because they were mapped at Mopra in 2005 before the powerful MOPS spectrometer became available, clumps BYF 60a--b and 123a-d were not mapped in \nnh, and so do not appear in this Table.  They do appear below in Table \ref{IRsources} and in Paper I.}
\end{minipage}
\end{table*}

As for the clumps themselves, the \hcop\ and \nnh\ also show interesting relationships.  Overall, the \nnh\ follows the \hcop\ distribution reasonably well, although typically it is $\sim$2--5 times fainter.  In detail, however, the two species differ.  Often the \hcop/\nnh\ brightness ratio (hereafter $\eta$) varies markedly across adjacent clumps.  A very clear example of this is in BYF\,54 (Fig.\,\ref{sample}), where the ratio drops from $\sim$14 to 2.1 to 1.7 going from clumps 54a to 54b to 54c, but the trend is evident in many other locations as well.

When we now intercompare the trends seen in the IR and mm line ratios, we obtain a remarkable result: the \hcop\ emission seems to be more strongly correlated with bright \brg\ emission, while bright \nnh\ seems to avoid such areas of high ionisation.  We further note that the \brg\ or \htwo\ emission is often (but not always) only associated with a single clump, or single interclump location.  In some cases, a single gas clump is associated with both \brg\ and \htwo\ emission (e.g., BYF\,54f), but then the IR lines are adjacent to each other, either on different sides of the clump, or separated by an apparent ionisation front.  We examine these trends in more detail next.

\vspace{-4mm}
\subsection{Millimetre Data\label{mmdata}}
The data analysis procedure for the mm lines has been described in Paper I.  Briefly, each emission clump is characterised by its brightness, angular size, and velocity dispersion or linewidth.  Then, whether one assumes an excitation temperature ($T_{\rm ex}$) for the molecular emission (as was done in Paper I) or derives the $T_{\rm ex}$ from other mm-wave data, one can calculate (if the cloud's distance and molecular abundance are known or assumed) a host of physical parameters, such as the radius, mass, density, etc., for the cold gas so traced.  The \hcop\ data for these clouds were given in Paper I, and the observed data are given for the \nnh\ maps in Table \ref{NNHsources} in the same format as in Paper I, since we found that almost all of the \nnh\ emission {\bf{\em structures}} correlated very closely to the location of the \hcop\ clumps.

Because the \nnh \joz\ transition is split into 7 hyperfine components separated by a few \kms, measuring simple moments of the emission profiles in cold, dense clumps will not give the true \vlsr\ and $\sigma_V$ values of the clumps.  Approximately, when integrating over the central 3 hyperfine components (\fto) as was done here, the \vlsr\ will be measured to be 0.20\,\kms\ too positive.  Likewise for $\sigma_V$, the 3 hyperfine components can be treated as adding a dispersion of 0.58\,\kms\ in quadrature to the intrinsic value of the emission; the measured $\sigma_V$ can then be corrected by inverting this effect.  The reader should therefore note these approximate corrections to the {\em unmodified} moments compiled in Table \ref{NNHsources} (columns 7, 8).  While a full hyperfine treatment of {\em all} \nnh\ maps is deferred to a later paper, the {\em corrected} \nnh\ values (as described here) are given in Table \ref{NNHsources}, column 9 and analysed below (Figs.\,\ref{hsigbrg}, \ref{nsigbrg}, \ref{summary}; \S\S\ref{correl}--\ref{mmdisc}).  These should be sufficiently close to the true values for the purposes of this comparative analysis.

\begin{figure}
\includegraphics[angle=0,scale=0.44]{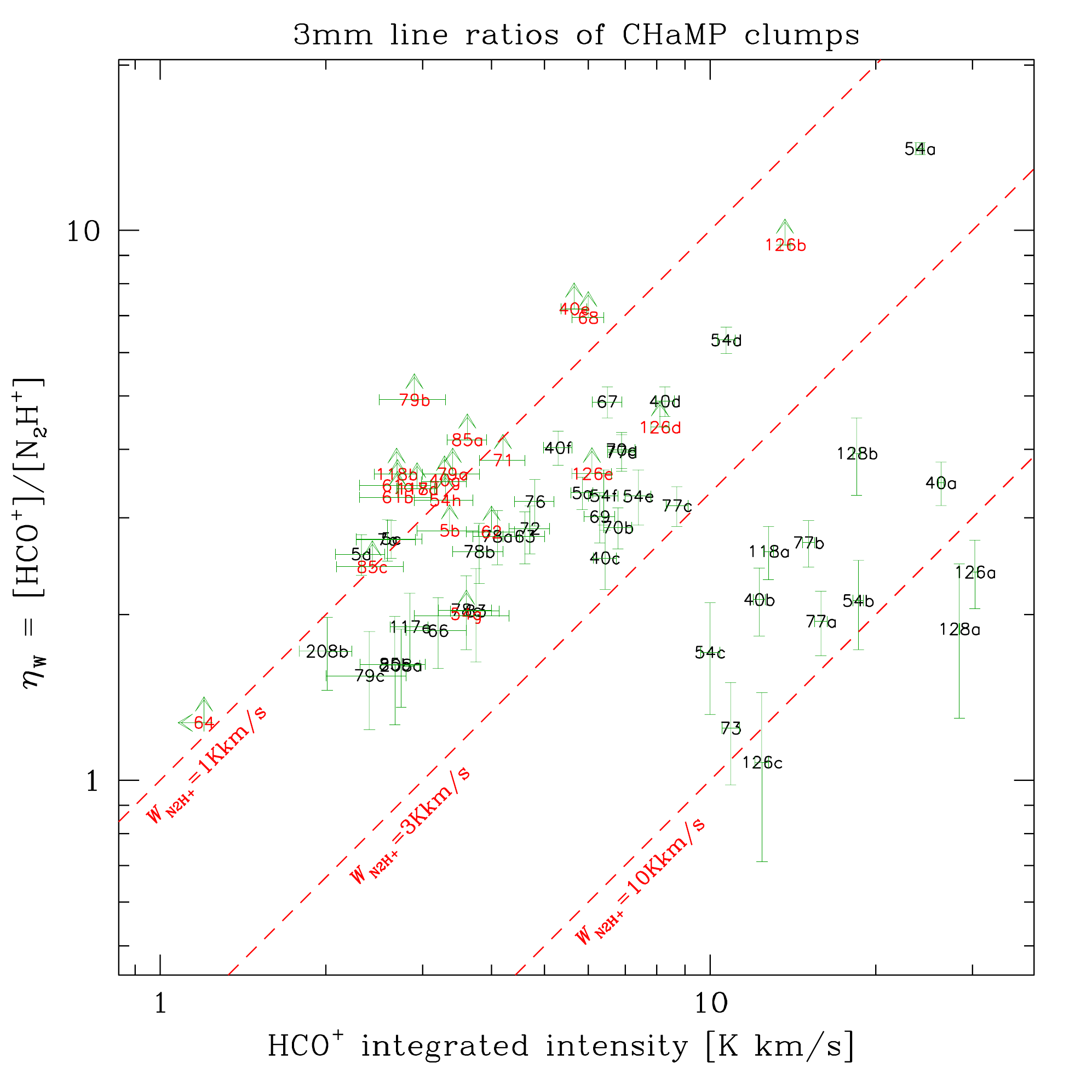}
\caption
{{\color{black} \hcop--\nnh\ line intensity ratio $\eta_W$ as a function of $W$(\hcop), labelled by clump number and overlaid by lines of constant $W$(\nnh).  Where shown, lower limits on $\eta_W$ (shown with red labels) derive from upper limits on $W$(\nnh) at 3 times the rms noise level.}} 
\label{mmlines}
\end{figure}

Likewise, we also defer the calculation of physical parameters from the \nnh\ maps, since these also need a full hyperfine analysis, and moreover, will depend strongly on the assumed molecular abundance $X$(\nnh) of \nnh\ relative to \htwo.  It has already been shown in low-mass cores that $X$(\nnh) is a strong function of CO-depletion processes in the cold gas \cite[e.g.,][]{CWZ02}, and recent work suggests that a similar condition holds in higher-mass clumps \cite[e.g.,][]{ZCP09}.  Therefore we cannot yet meaningfully discuss the physical conditions in the \nnh\-emitting gas.  Instead, we discuss below the relationship of \nnh\ to the other tracers in the context of this understanding of the chemical origin and evolution of \nnh.

\vspace{-3mm}
\subsection{IR Image Analysis\label{irdata}}
We measured the continuum-subtracted IR line emission as follows.  For each field and filter, any IR line emission was first ``assigned'' to a particular clump.  In most cases, clumps are sufficiently separated that this assignment was obvious, but in some areas (e.g., BYF\,54 in Fig.\,\ref{sample}) the IR emission extends broadly across multiple clumps.  In these cases we used the gaussian ellipses fitted to the \hcop\ emission to divide the IR emission into non-overlapping areas associated with each clump.  The basic idea is one of proximity: to which mm clump is the IR emission closest on the sky?

\begin{figure}
\includegraphics[angle=0,scale=0.44]{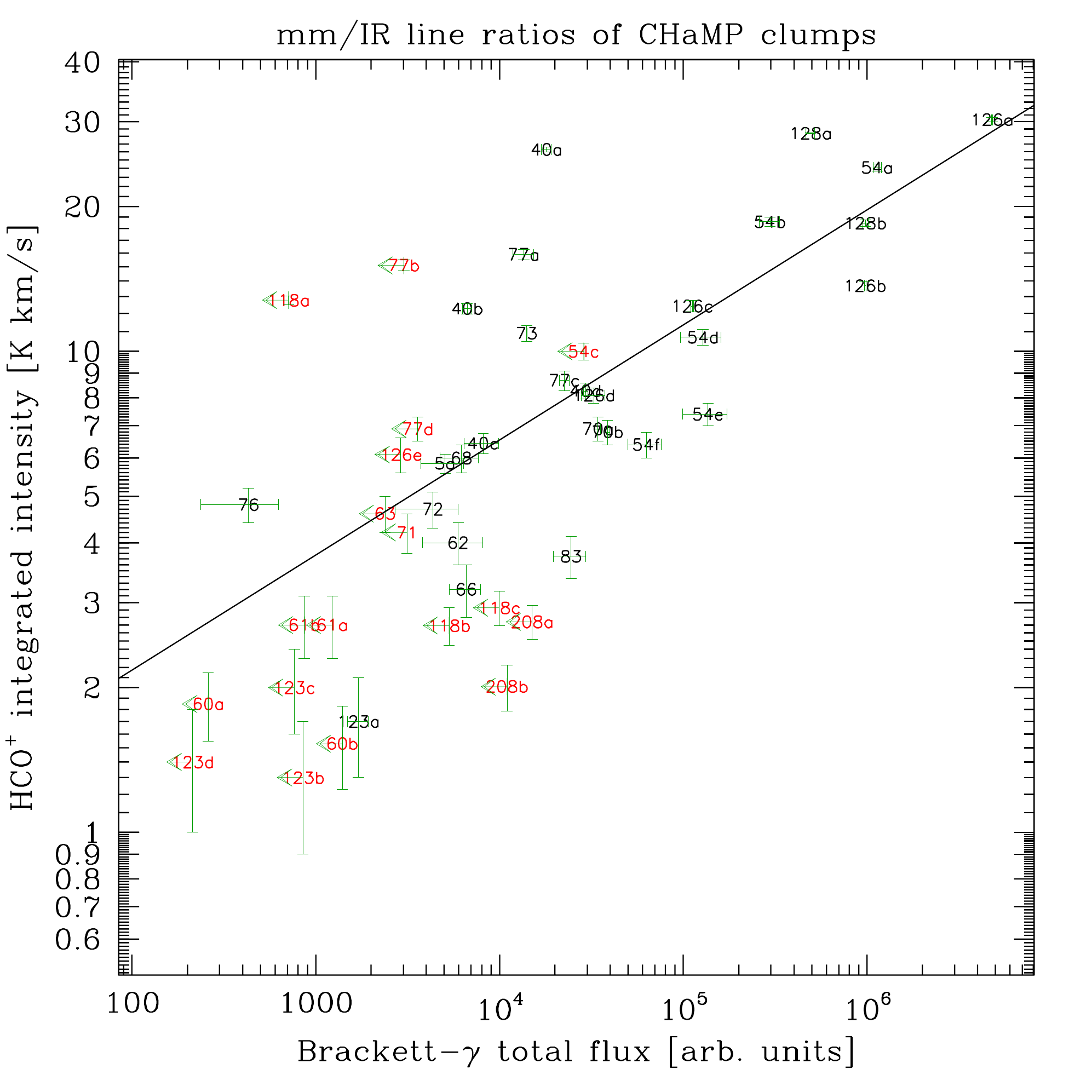}
\caption{\hcop\ line intensity vs. \brg\ integrated flux, labelled by clump number.  Black labels with green error bars denote clumps with reliable measurements in both lines, while red labels with green 3$\sigma$ upper limits show clumps near which no \brg\ was detectable.  Also shown is an unweighted least-squares fit to the well-measured points (black labels), with slope $m=0.24\pm0.04$ {\color{black} and Pearson's correlation coefficient $r^2=0.54$}.  The rms scatter about this trend is a factor of 1.6 in the ordinate (0.20 in log\,$y$).} 
\label{hcopbrg}
\end{figure}

Once these assignments were made, we used the task \textsc{cgcurs} in \textsc{miriad} 
to measure the IR lines' mean brightness over the area assigned to each clump, and to record the area of integration.  Then, one or two neighbouring emission-free areas in the image were also measured to find a suitable mean background brightness level.  This background value was subtracted from the mean emission value, and the result multiplied by the size (in pixels) of the area integrated, to obtain a net line flux for each molecular clump.  The results of these measurements are shown in Table \ref{IRsources}.

\begin{figure}
\includegraphics[angle=0,scale=0.44]{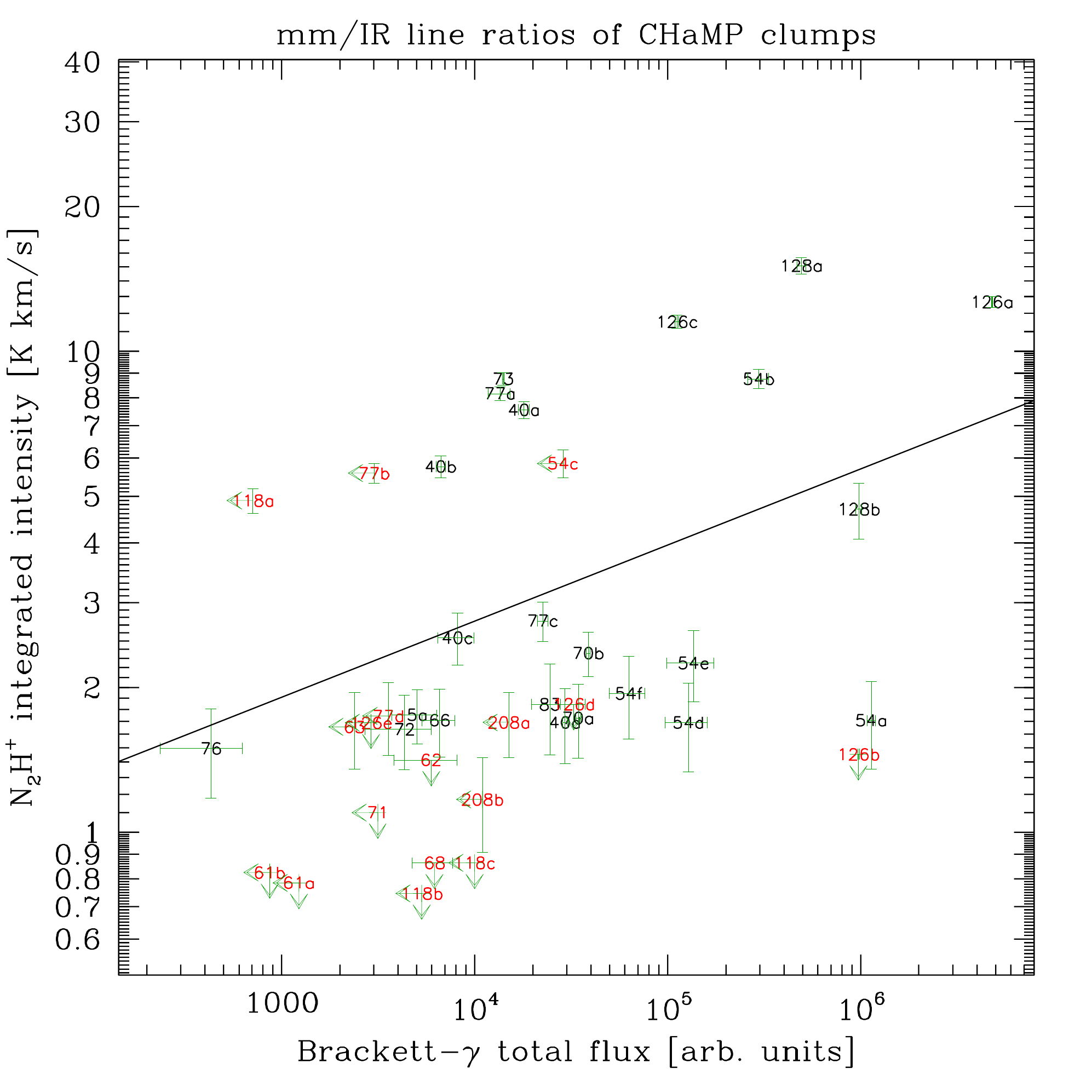}
\caption{\nnh\ line intensity (on the same ordinate scale as Fig.\,\ref{hcopbrg}) vs. \brg\ integrated flux, labelled by clump number.  Other details are as in Fig.\,\ref{hcopbrg}.  Here the unweighted least-squares fit to the well-measured points (black labels) has $m=0.16\pm0.07$, $r^2$ = 0.18, and an rms scatter of 2.2 in the ordinate (0.35 in log\,$y$).} 
\label{n2hpbrg}
\end{figure}

\begin{table*}
\centering
\begin{minipage}{165mm}
\caption{AAT 2$\mu$m Narrowband Filters Observed Parameters.\label{IRsources}}
\begin{tabular}{@{}lrrrrrrrrr@{}}
\hline
  &  & Brackett-$\gamma$ &  &  & \htwo\ \voz &  &  & \htwo\ \vto & \\
  BYF & \multicolumn{3}{c}{--------------------------------------------} & \multicolumn{3}{c}{--------------------------------------------} & \multicolumn{3}{c}{--------------------------------------------} \\
  no. & Area & Net Mean & Net Flux & Area & Net Mean & Net Flux & Area & Net Mean & Net Flux \\
\hline
\input{table3a}
\hline
\end{tabular}
\end{minipage}
\end{table*}

\begin{table*}
\centering
\begin{minipage}{165mm}
\contcaption{}
\begin{tabular}{@{}lrrrrrrrrr@{}}
\hline
  &  & Brackett-$\gamma$ &  &  & \htwo\ \voz &  &  & \htwo\ \vto & \\
  BYF & \multicolumn{3}{c}{--------------------------------------------} & \multicolumn{3}{c}{--------------------------------------------} & \multicolumn{3}{c}{--------------------------------------------} \\
  no. & Area & Net Mean & Net Flux & Area & Net Mean & Net Flux & Area & Net Mean & Net Flux \\
\hline
\input{table3b}
\hline
\end{tabular} \\
{\textsc{Notes.}  The areas of emission are measured in pixels, {\color{black} and the mean values in these areas are in ADUs/pixel (see text)}.  Dashes for certain clumps/filters mean that no emission associated with the clump was discernible in that filter above the immediate surroundings.  Uncertainties in brackets are in the last digit of the corresponding value.  The net flux values and uncertainties {\color{black} (measured in ADUs)} have been divided by 1000 to show significant figures.  Notes on individual sources appear below. \\
7:  At edge of field \\
61b:  \htwo\ \voz\ values uncertain due to proximity to edge \\
70a:  \htwo\ \vto\ emission was split at $l$=286.160, $b$=0.177 \\
117:  At edge of field \\
118a:  Small \htwo\ \voz\ knot found but no evidence of anything else \\
123c:  The only sign of any emission knots was in \htwo\ \voz \\
}
\end{minipage}
\end{table*}

{\color{black} Here we quote the IR line brightnesses on a relative scale.  This scale was set by the analogue-to-digital units (ADU) conversion in the IRIS2 camera and \textsc{orac-dr} pipeline (see \S\ref{aat}).  In good conditions, this scale would normally be stable over the long- and short-term, resulting in well-calibrated images that could easily be converted to an absolute scale by a single numerical factor.  Because of the rather variable sky conditions in the 2011 season, however, our calibration stability was less certain.  To measure this, we examined our NIR data in two ways.

First, we compared the factors we derived for scaling the various narrowband filters to the $K$-cont filter in each observed field (i.e., prior to subtracting them from the continuum image).  The mean $\pm$ standard deviation scaling factors for the \brg\ and \htwo\ \voz\ filters were 0.96 $\pm$ 0.06 and 0.96 $\pm$ 0.11 (resp.), while it was 0.54 $\pm$ 0.08 for \vto\ (as expected for the doubled integration time in this filter).  These factors represent good short-term stability in the calibration, considering the $\sim$40$^m$ timescale for cycling through all filters in a given field.

Second, we compared the unscaled $K$-cont fluxes of several stars in each field to the $Ks$ values from the 2MASS catalogue, in order to measure the longer-term calibration stability over the nights of observation.  Averaged over all fields, we found a mean $\pm$ standard deviation conversion of (3.37\,$\pm$\,1.20)\,$\times$\,10$^5$\,ADUs/Jy.  Perhaps not surprisingly, this dispersion is larger than for the short-term, and means that the calibration between fields is less certain.

Nevertheless, we show below that this calibration uncertainty is well within the scatter of the results presented here.  For example, the scatter in the IR line parameters is typically a factor of 4 or more; thus a calibration dispersion of 35\%, while larger than desirable, cannot affect our results by a meaningful amount.}


\subsection{Detailed Correlations\label{correl}}
In this section we numerically evaluate the trends apparent above (\S\ref{general}).  We first examine the mm line ratios.  For each of the 60 clumps imaged in the IR, we plot in Figure \ref{mmlines} the 
{\color{black} [\hcop]/[\nnh] integrated intensity line ratio $\eta_W$ as a function of $W$(\hcop)}.  
{\color{black} Here we see a wide range of $\eta$ in the clump sample, with measured values $\eta$ $\sim$ 1.5--5 for $W$(\hcop) {\lapp} 9\,K\kms\ (corresponding to $W$(\nnh) $\sim$ 1--3\,K\kms\ in this range), and a possibly wider range $\eta$ $\sim$ 1--14 for $W$(\hcop) {\gapp} 9\,K\kms\ (corresponding to $W$(\nnh) $\sim$ 1.5--15\,K\kms\ in this range), although the range at lower $W$(\hcop) may also be somewhat wider due to undetected \nnh\ in some clumps.  There also appears to be a threshold in $W$(\hcop) around 10\,K\kms, below which bright \nnh\ emission ($W$(\nnh) $>$ 3\,K\kms) is not seen.}
From the general trends noted above, it already seems clear that clumps with high $\eta$ are also bright in \brg, while \nnh-bright (i.e., low-$\eta$) clumps lie further afield from the \brg\ emission.

This disparity between \hcop\ and \nnh\ becomes more striking when we directly compare the individual mm lines to the IR emission.  In Figures \ref{hcopbrg} and \ref{n2hpbrg} we respectively plot the \hcop\ and \nnh\ integrated intensities against the total \brg\ flux integrated around an area appropriate for each clump, as described in \S\ref{irdata}.  We tried both weighted and unweighted least-squares fits to these plots, but found that even a weak inverse-noise weighting tends to produce fits that are strongly dominated by the few points with relatively small errors.  Therefore we only give the results of unweighted least-squares fits in the discussion below.  Fitting in this way only the well-measured points (i.e., not including the upper limits for some clumps), we find from Figure \ref{hcopbrg} that $W$(\hcop) $\propto$ $F$(\brg)$^{0.24\pm0.04}$.  In other words, {\em the brightness of the \hcop\ emission seems directly tied to the \brg\ flux}.  In contrast, Figure \ref{n2hpbrg} shows a much weaker correlation between the \nnh\ emission brightness and the \brg, one that is effectively flat.  In other words, {\em the \nnh\ emission seems to be independent of the \brg\ emission, and may thus better trace gas that is unaffected (or not yet affected) by the presence of massive young stars.}  However, we examine a more subtle effect in this plot below (\S\ref{mmdisc}).

\begin{figure}
\includegraphics[angle=0,scale=0.44]{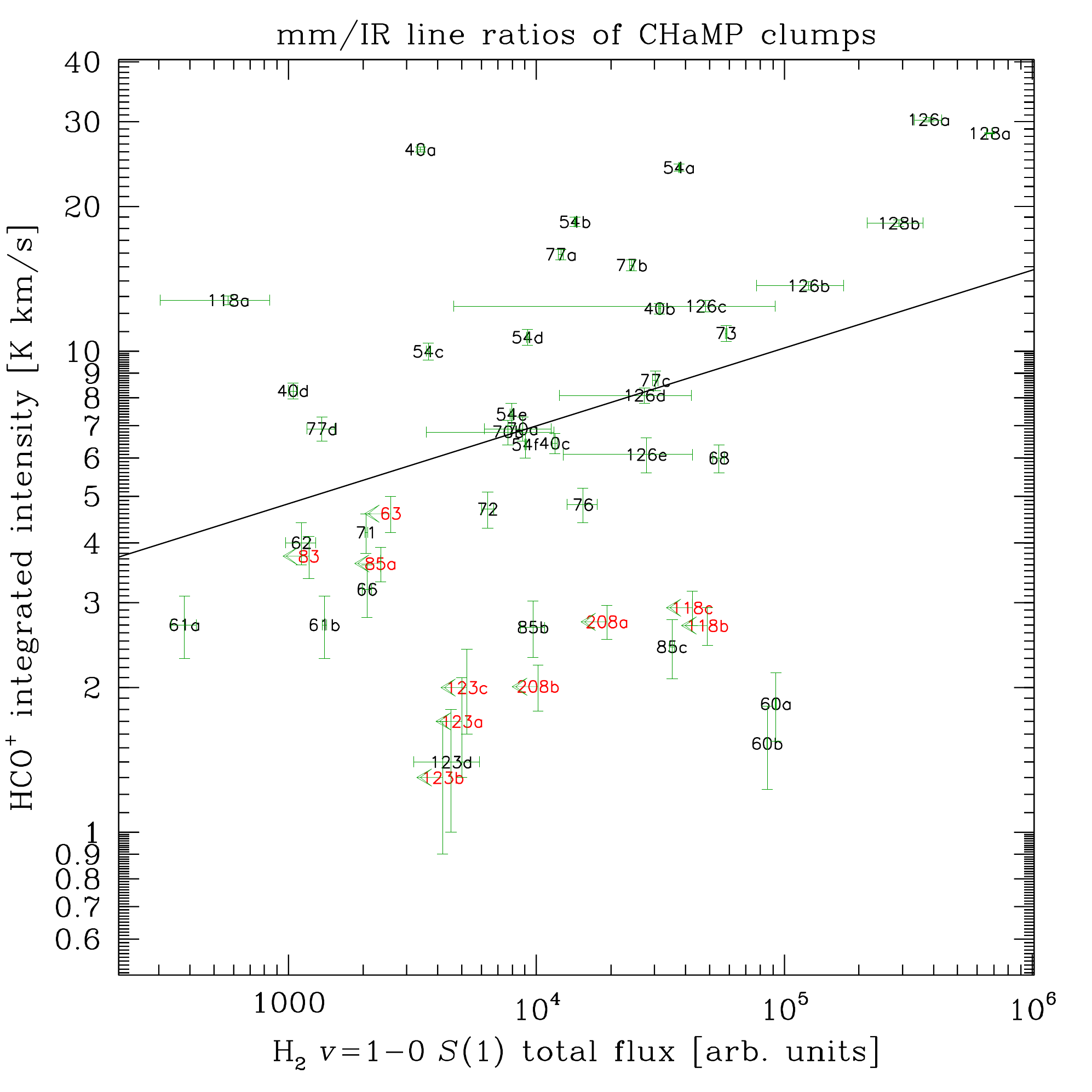}
\caption{\hcop\ line intensity (on the same ordinate scale as Fig.\,\ref{hcopbrg}) vs. \htwo\ \voz\ integrated flux, labelled by clump number.  Other details are as in Fig.\,\ref{hcopbrg}.  Here the unweighted least-squares fit to the well-measured points (black labels) has $m=0.16\pm0.07$, $r^2=0.13$, and an rms scatter about this trend of 2.0 in the ordinate (0.30 in log\,$y$).} 
\label{hcophoz}
\end{figure}

\begin{figure}
\includegraphics[angle=0,scale=0.44]{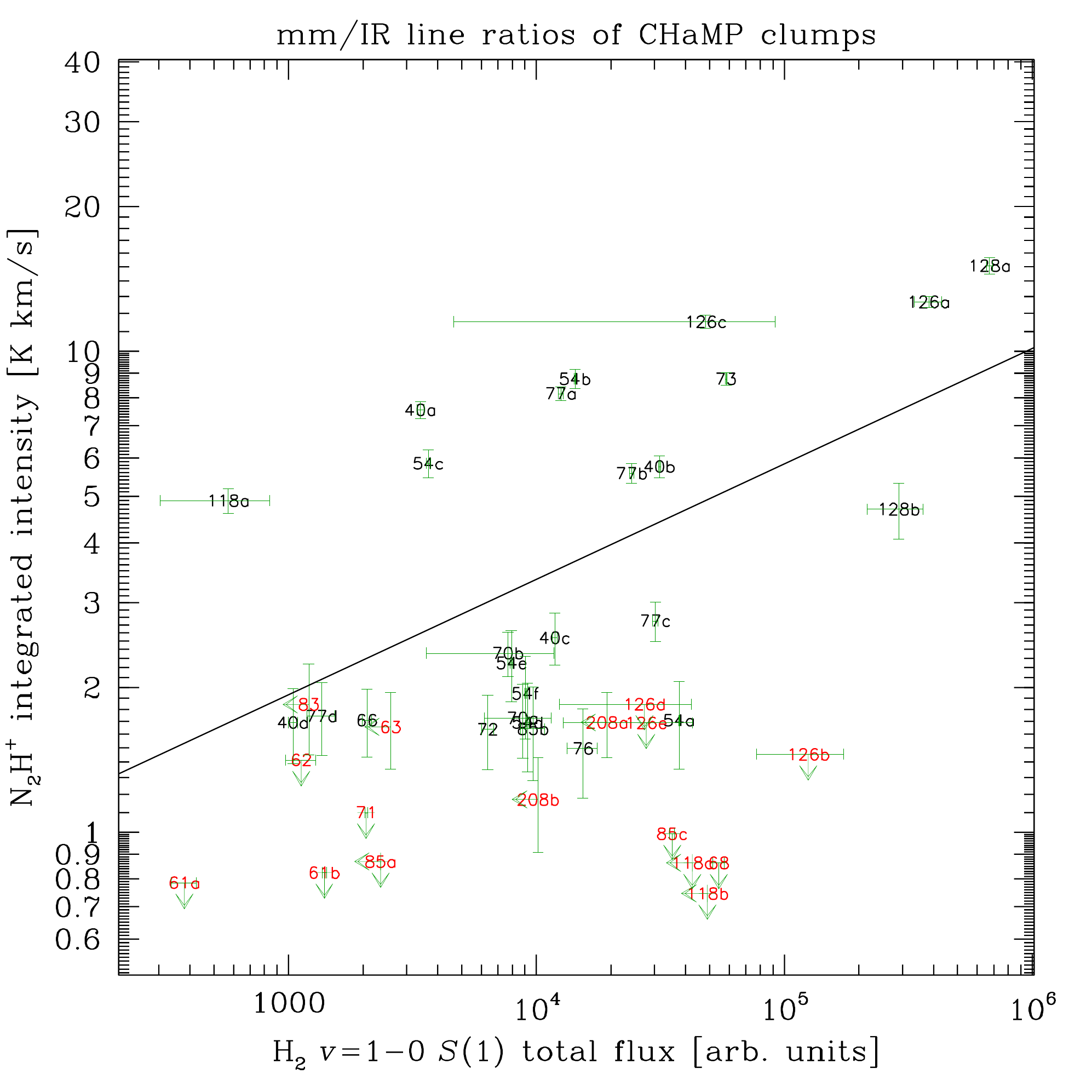}
\caption{\nnh\ line intensity (on the same ordinate scale as Fig.\,\ref{hcopbrg}) vs. \htwo\ \voz\ integrated flux, labelled by clump number.  Other details are as in Fig.\,\ref{hcopbrg}.  Here the unweighted least-squares fit to the well-measured points (black labels) has $m=0.24\pm0.08$, $r^2=0.28$, and an rms scatter about this trend of 2.1 in the ordinate (0.33 in log\,$y$).} 
\label{n2hphoz}
\end{figure}

A similar comparison between the mm lines' emission and the \htwo\ \voz\ nebulosity (Figs.\,\ref{hcophoz}, \ref{n2hphoz}) shows correlations that are between the above two cases.  
Thus, while the slope in the power law from Figure \ref{hcopbrg} is $\sim$5$\sigma$ above zero {\color{black} (Pearson's $r^2$ = 0.54)}, both the [\hcop]-[\htwo\ \voz] and [\nnh]-[\htwo\ \voz] relations (Figs.\,\ref{hcophoz}, \ref{n2hphoz}) seem quite marginal at $\lapp$3$\sigma$ {\color{black} ($r^2$ = 0.13, 0.28 resp.)}, and the [\nnh]-[\brg] relation (Fig.\,\ref{n2hpbrg}) is very weak ($\sim$2$\sigma$, {\color{black} $r^2$ = 0.18}).  Also, while both \hcop\ and \nnh\ seem similarly correlated with \htwo\ \voz, the \hcop\ emission is systematically brighter than the \nnh\ by a factor of $\sim$3.  Finally, both \hcop- and \nnh-bearing clumps exhibit a fairly clear threshold for the detection of \htwo\ \voz: one requires $W$(\hcop) \gapp\ 5\,K\,\kms\ or $W$(\nnh) \gapp\ 2\,K\,\kms\ to reliably detect the \voz\ line.  Below this level the \voz\ is sometimes detected, but usually not.  

For completeness, we also present in Figures \ref{hcophto} and \ref{n2hphto} similar comparisons of the mm line emission with the \htwo\ \vto\ fluxes.  These show similarly marginal trends ($<$3$\sigma$) and wider scatter (factors of $\sim$3 in $y$) than Figures \ref{n2hpbrg}--\ref{n2hphoz}.

A natural question that arises from these plots, especially Figure \ref{hcopbrg}, is whether the excitation traced by either the \brg\ or \htwo\ emission contributes to the internal turbulence of the molecular clumps?  That is, to what extent are the correlations with mm line strengths $W$ determined by the lines' velocity dispersions?  To explore this, we show in Figures \ref{hsigbrg} and \ref{nsigbrg} a comparison between the \hcop\ and \nnh\ (resp.) linewidth with the \brg\ flux.  Clearly if any relationship exists, it shows up only weakly in this sample, with trends that are indistinguishable from scatter plots.  An even weaker result is obtained when comparing the mm linewidths to either of the \htwo\ emission lines (not shown).  It must be concluded that the strong trend in Figure \ref{hcopbrg} is dominated by the brightness of the line emission, rather than due to any contribution from the linewidth.  We see this explicitly in Figure \ref{hteffbrg}, with a similar strong correlation ($\sim$4$\sigma$) to Figure \ref{hcopbrg}, only with larger errors on the individual points.  A plot of \nnh\ peak brightness vs. \brg\ flux (not shown) gives another scatter plot similar to Figure \ref{n2hpbrg}.

These relationships are summarised for convenience in Figure \ref{summary}.

\vspace{-3mm}
\section{Discussion\label{disc}}

\subsection{Comparison of the mm and IR data\label{mmdisc}}
As shown in Paper I, the \hcop\ emission in these massive clumps is almost always optically thin ($\tau$$\ll$1), despite being quite bright in some of them.  This is true whether we assume a fairly low (10\,K) or high (30\,K) value for the excitation temperature ($T_{\rm ex}$) in the \hcop\ line.  So, while the exact values for $\tau$ in each clump will be somewhat degenerate with the assumed $T_{\rm ex}$, the column density calculated from these lines is quite insensitive to the assumed $T_{\rm ex}$ (e.g., see Fig.\,13 in Paper I).  Therefore, the integrated intensity $W$(\hcop) should be a reliable indicator of the \hcop\ column density in a molecular clump.  (Of course, translating this into a gas or mass column density will depend on the \hcop\ abundance $X$(\hcop), which was discussed in Paper I: we do not revisit that issue here.)  This means that {\em the \hcop\ column density is closely correlated with the \brg\ flux from a clump}.  In contrast, the \nnh\ brightness (which to first order, likely also traces its column density) is not obviously correlated to any of the IR emission we have examined here, although there may be a marginally significant relationship with the \htwo\ \vto\ flux, and a more subtle correlation with the \brg\ which we discuss below.  Overall, however, this result seems to support the astrochemical picture of the dependence of \hcop\ and \nnh\ abundances in dense clumps on the gas temperature (via CO freeze-out) and excitation conditions \cite[discussed further below; see][]{CWZ02,ZCP09}.  In contrast, the \htwo\ emission only exhibits weak or marginally significant relationships with either mm line, whereas one might have expected clearer trends than found here.  This may also be consistent with the same astrochemical model, if we suppose that, where \htwo\ emission arises, the destruction of \nnh\ and production and/or heating of \hcop\ has not progressed as far as in areas emitting \brg.  

This hypothesis deserves further examination. The lack of correlation between the mm-lines' velocity dispersions and any of the IR line fluxes suggests that one of the first effects that massive young stars (i.e., those massive enough to ionise significant amounts of gas) have on their natal molecular environment is one of direct excitation through (photonic) heating, rather than one of mechanical energy injection through winds, shocks, and the like (which would tend to increase the velocity dispersions in the gas).  Thus, while we cannot (with these data) confirm the view that turbulent support of these clouds {\em before massive star formation has commenced} is maintained by low-mass outflow activity \citep{m07,NL05,NL11}, we cannot rule it out either, due to an apparent lack of sensitivity to such activity in these kpc-distant clouds.  The lack of correlation of the mm emission (either the lines' brightness or linewidth) with the ``pre-ionised'' warm \htwo\ would simply suggest that gas heated to these intermediate temperatures has not (yet?) spread to significant portions of these massive clumps.

\begin{figure}
\includegraphics[angle=0,scale=0.44]{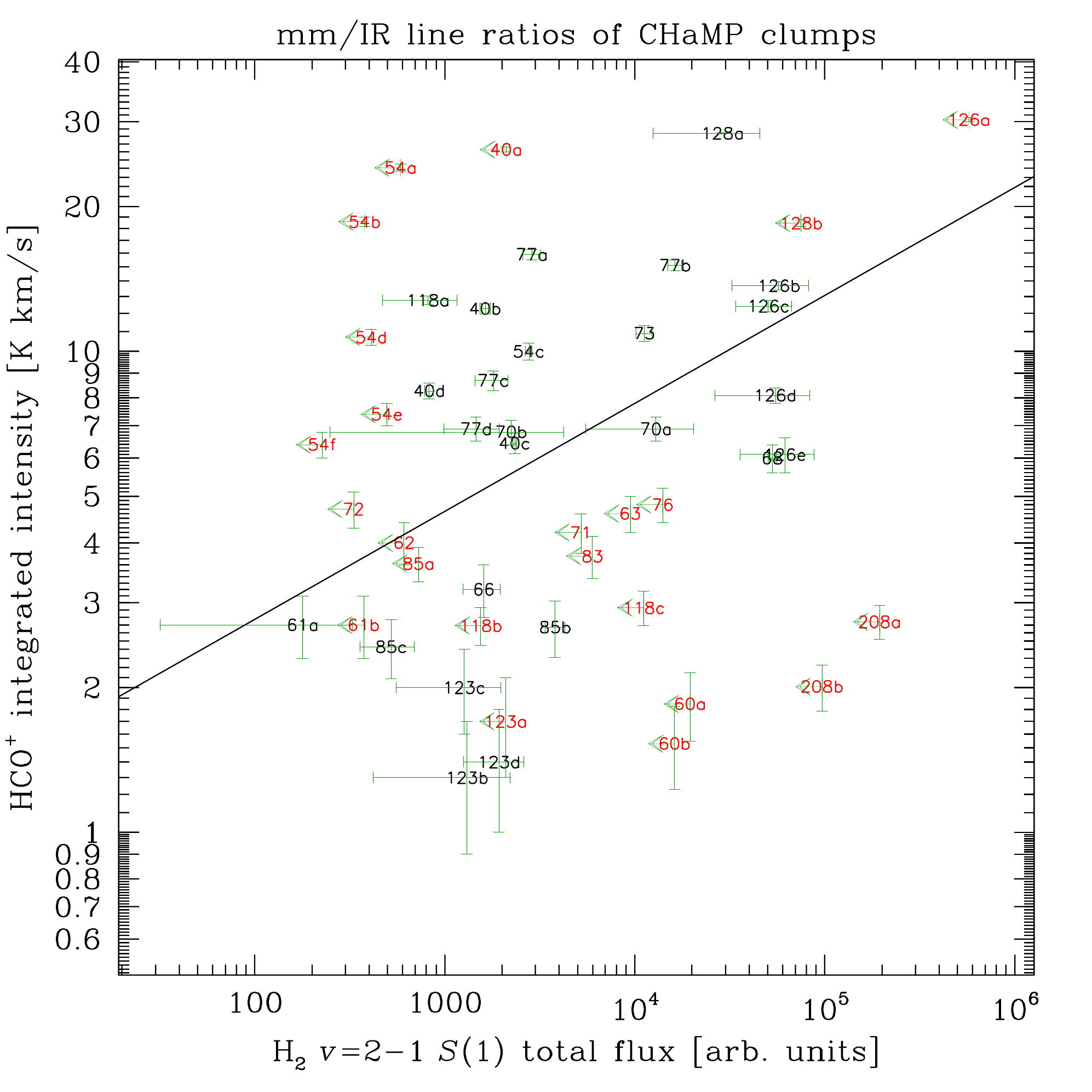}
\caption{\hcop\ line intensity (on the same ordinate scale as Fig.\,\ref{hcopbrg}) vs. \htwo\ \vto\ integrated flux, labelled by clump number.  Other details are as in Fig.\,\ref{hcopbrg}.  Here the unweighted least-squares fit to the well-measured points (black labels) has $m=0.22\pm0.09$, $r^2=0.22$, and an rms scatter about this trend of 2.1 in the ordinate (0.32 in log\,$y$).} 
\label{hcophto}
\end{figure}

\begin{figure}
\includegraphics[angle=0,scale=0.44]{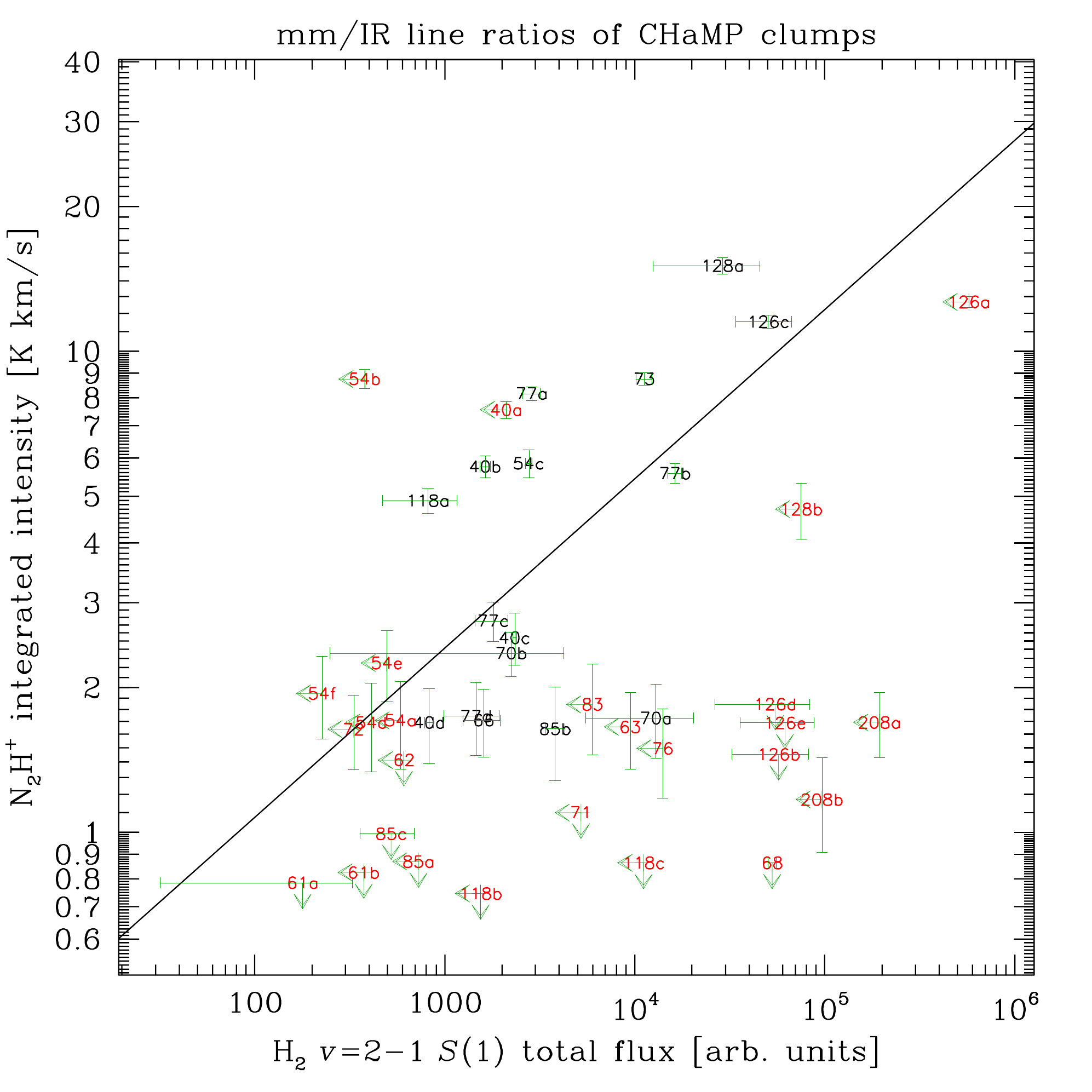}
\caption{\nnh\ line intensity (on the same ordinate scale as Fig.\,\ref{hcopbrg}) vs. \htwo\ \vto\ integrated flux, labelled by clump number.  Other details are as in Fig.\,\ref{hcopbrg}.  Here the unweighted least-squares fit to the well-measured points (black labels) has $m=0.35\pm0.13$, $r^2=0.34$, and an rms scatter about this trend of 3.0 in the ordinate (0.47 in log\,$y$).} 
\label{n2hphto}
\end{figure}

\begin{figure}
\includegraphics[angle=0,scale=0.44]{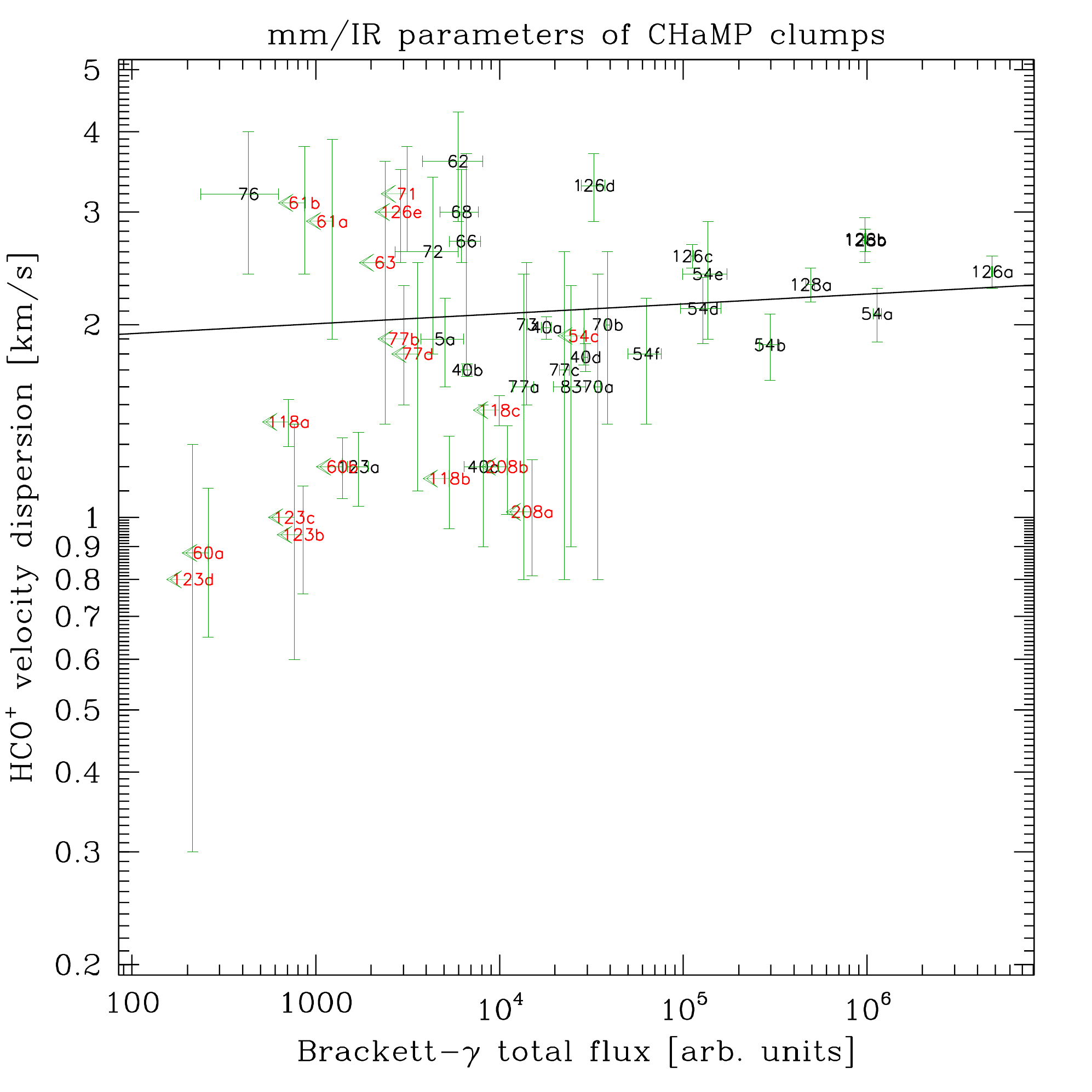}
\caption{\hcop\ velocity dispersion vs. \brg\ integrated flux, labelled by clump number.  Other details are as in Fig.\,\ref{hcopbrg}.  Here the unweighted least-squares fit to the well-measured points (black labels and line) has $m=0.02\pm0.03$, $r^2=0.01$, and an rms scatter about this trend of 1.3 in the ordinate. 
}\label{hsigbrg}
\end{figure}

\begin{figure}
\includegraphics[angle=0,scale=0.44]{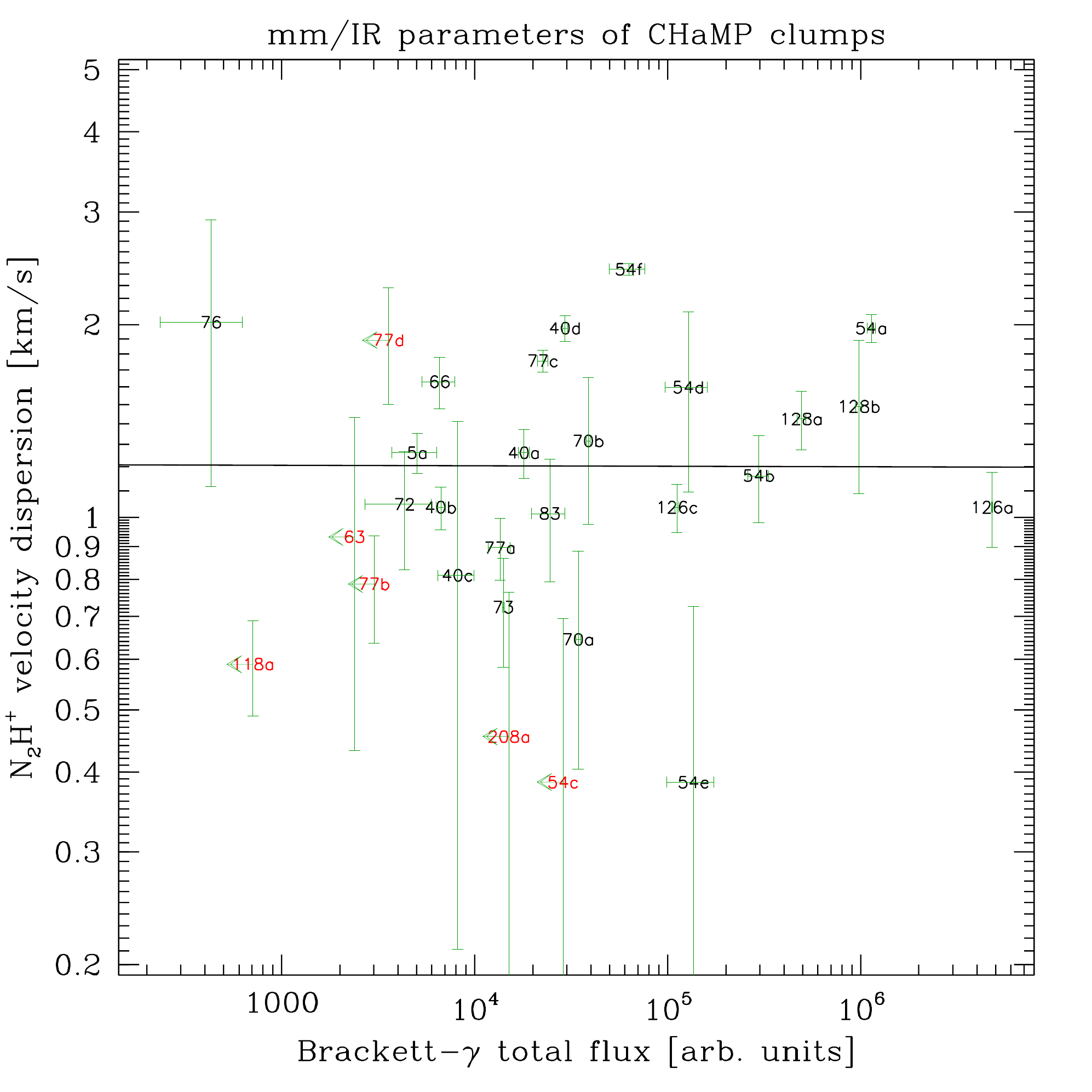}
\caption{\nnh\ velocity dispersion vs. \brg\ integrated flux, labelled by clump number.  Other details are as in Fig.\,\ref{hcopbrg}.  Here the unweighted least-squares fit to the well-measured points (black labels and line) has $m=0.00\pm0.04$, $r^2=0.00$, and an rms scatter about this trend of 1.6 in the ordinate.  
}\label{nsigbrg}
\end{figure}

\begin{figure}
\includegraphics[angle=0,scale=0.44]{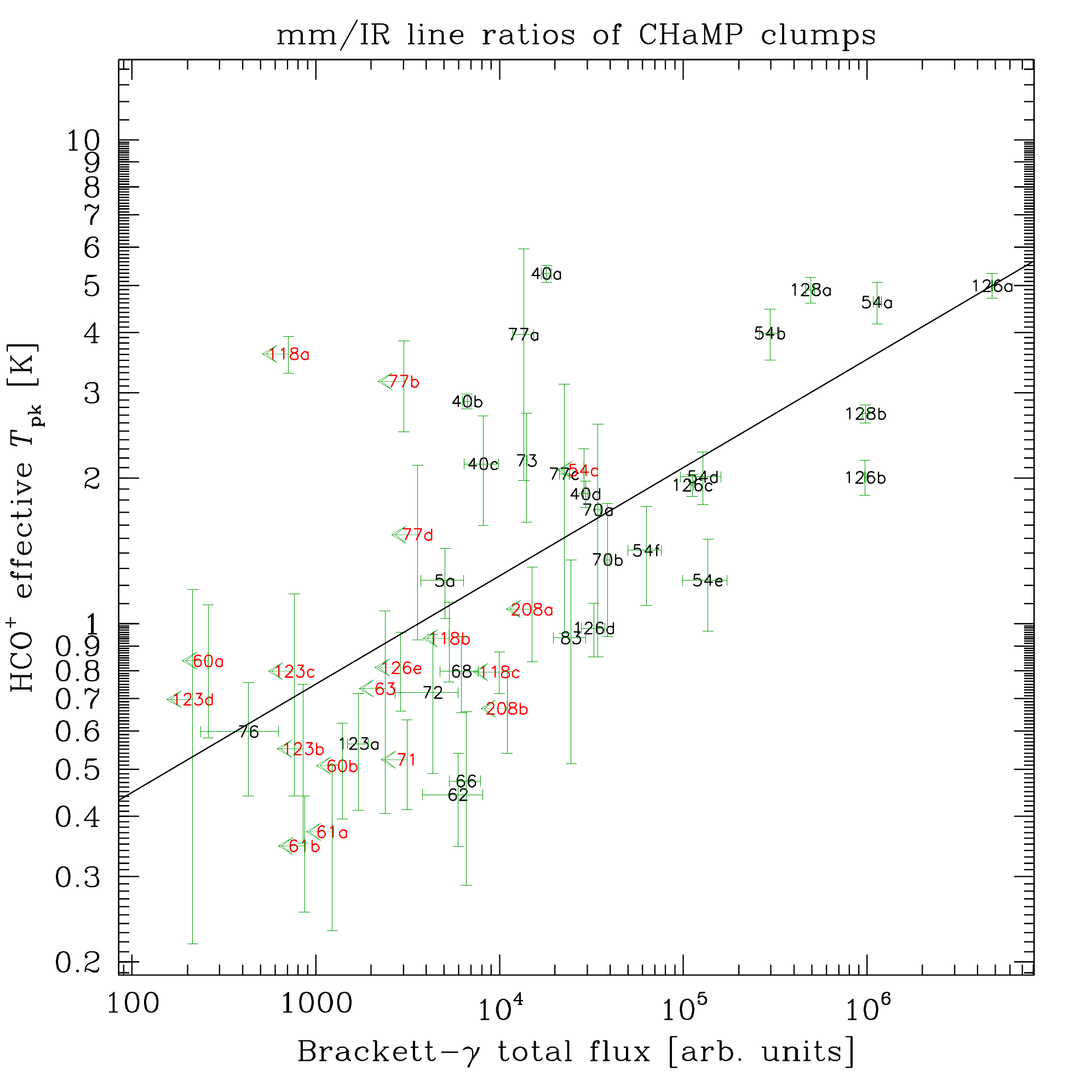}
\caption{\hcop\ effective $T_{\rm pk}$ vs. \brg\ integrated flux, labelled by clump number.  Other details are as in Fig.\,\ref{hcopbrg}.  Here the unweighted least-squares fit to the well-measured points (black labels and line) has $m=0.22\pm0.05$, $r^2=0.43$, and an rms scatter about this trend of a factor of 1.7 in the ordinate.} 
\label{hteffbrg}
\end{figure}

\begin{figure}
(a)\hspace{-4mm}\includegraphics[angle=0,scale=0.44]{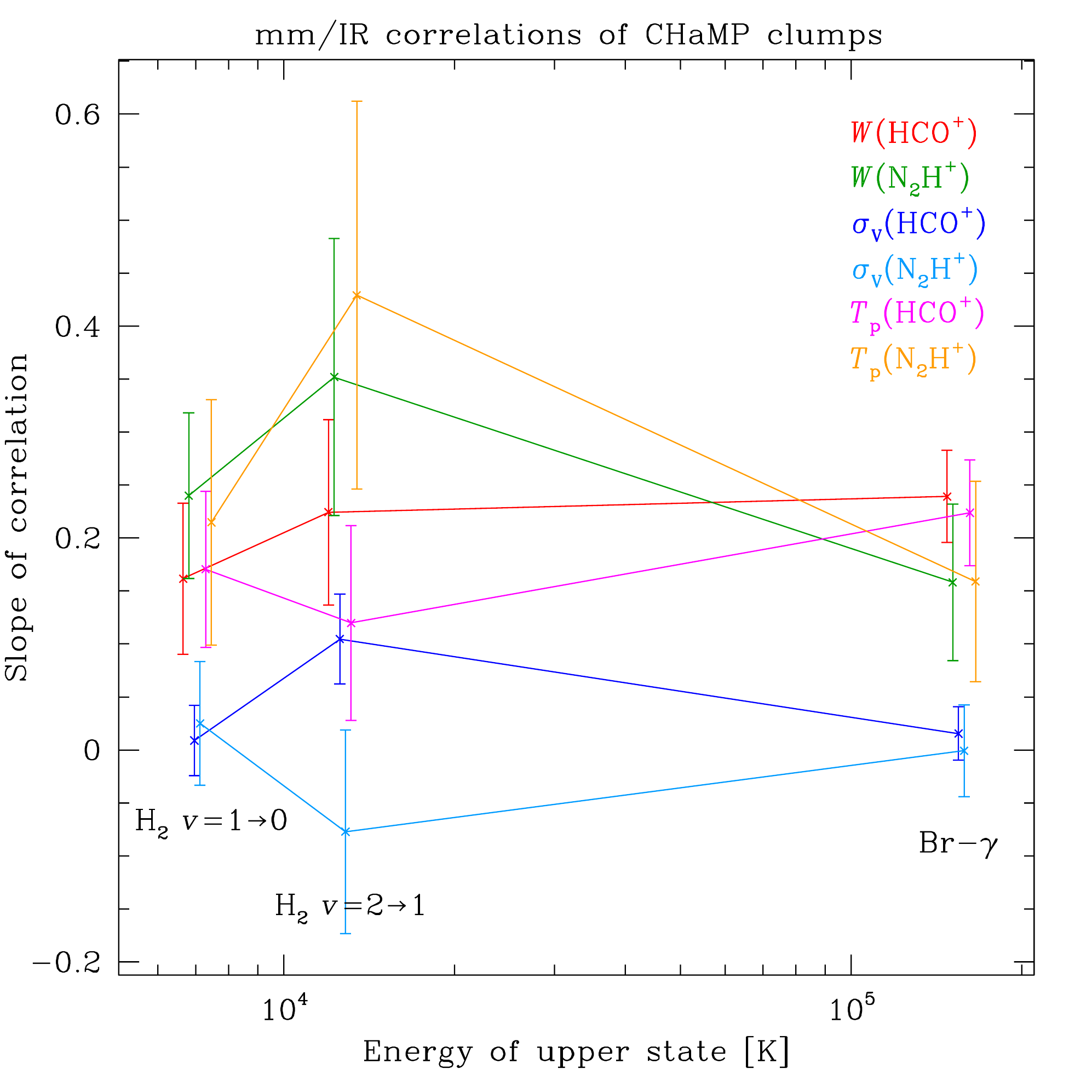}
(b)\hspace{-4mm}\includegraphics[angle=0,scale=0.44]{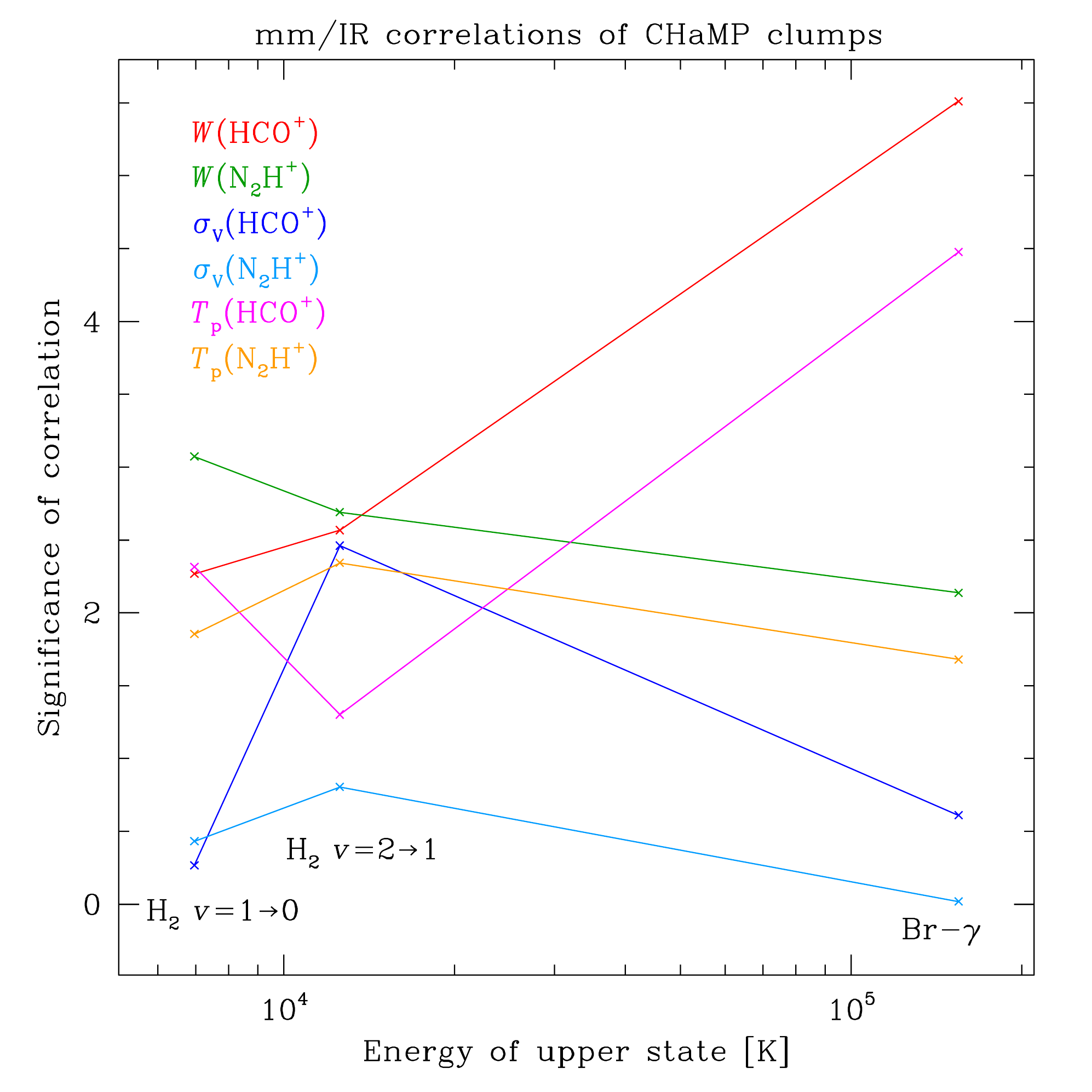}
\caption{Summary plots of all power-law correlations from Figs.\,\ref{hcopbrg}--\ref{hteffbrg}.  
$(a)$ Fitted slope of all power laws, with error bars indicating the least-squares uncertainty in the slope.  The points are colour-coded for the species (\hcop\ or \nnh) and whether the fit was to the integrated intensity, linewidth only, or peak brightness temperature only.  For clarity, each point's abscissa was offset slightly from its nominal value.  
$(b)$  Significance of the fits shown in panel ($a$), defined as the slope's value divided by the uncertainty.  One can consider this parameter the ``signal-to-noise'' of the fitted correlation, {\color{black} and follows well the trend in squared correlation coefficients $r^2$ (not shown, but given in the caption to each figure).  Whether from $r^2$ or this plot,} we see that the most significant correlation among the ones shown is between $W$(\hcop) and \brg\ at $>$5$\sigma$, even though this relation does not have the steepest fitted slope.}
\label{summary}
\end{figure}

Given the relationships revealed by Figures \ref{mmlines}--\ref{hteffbrg}, especially Figures \ref{hcopbrg}, \ref{n2hpbrg}, and \ref{hteffbrg}. it seems hard to avoid the conclusion that the brightness of the \hcop\ emission (= \hcop\ column density) is being largely driven by the same source(s) that produce the heating and ionisation signified by the \brg\ nebulae.  In other words, molecular clouds with \hcop-enhanced gas seem to signpost the physical interface where massive pre-main sequence stars first begin to heat and drive off their nascent molecular cocoons.  Such relationships between \hii\ regions and molecular clouds have, in a gross sense, been seen before \cite[e.g.,][]{bfs82}, but these have typically been in areas where the dispersal of gas has progressed further, and the young stellar population is less embedded, than in this sample.

Thus, this is the first direct correlation of which we are aware between a standard molecular ``dense gas tracer'' (\hcop) and the evolution of the cloud as traced by {\em ongoing} massive star formation activity (\brg\ emission).  Such a correlation perhaps ought to be expected, but until now there has been little {\em systematic} information correlating the physical conditions in the dense gas of massive clumps, as traced by their molecular emission alone, with their stellar progeny.  By this we mean that, while a good number of IR studies of individual regions, or small collections of regions, have examined (e.g.) \htwo\ emission in outflows \citep{CGN06}, there has been little to tie this information to the overall gas properties, or even the gross mm line emission.  Similarly, while a number of larger dense molecular gas surveys are now available \cite[e.g.,][]{L07,w10,s11}, they typically do not relate their data directly to any embedded protostellar content.

What is unexpected is that, as a fairly ``standard'' dense gas tracer, \hcop\ may turn out to actually trace not mass or density as such, but more likely a combination of gas column density and excitation from embedded massive young stars.  On the larger scales of galaxies or GMC complexes in our own Milky Way, ``dense gas tracers'' are typically thought to represent a fairly homogeneous state of the molecular ISM.  Theoretically, this has been challenged by studies such as those of \citet{kth07} or \cite{ncs08}, which show that radiative transfer through an entire population of clouds (with a range of properties) must be considered in order to explain (for example) the Kennicutt-Schmidt relationships as traced by these species.  Our results suggest that it may be more accurate to think of these tracers as representing a convolution of a larger population of colder, fainter, quiescent clouds with a rarer, brighter cohort that is being actively altered by their newly formed, massive protostellar products.  Such a picture is supported by recent observations of dense-gas tracers on large ($\sim$100\,pc) galactic scales \cite[e.g., in the nuclear bar of Maffei 2;][]{mt12}.  The mm line ratio $\eta$, in particular, is strongly tied to the presence of star formation and photon-dominated regions (PDRs), in contrast to the ratio of several other pairs of molecular species.  Thus, we claim that the case for $\eta$ being a significant signpost of molecular cloud evolution in the presence of active star formation is now strong, and could be utilised more widely in future studies.

In Paper I we found that massive dense clumps that are \hcop-bright ($W$ $>$ 12\,K\kms) comprise only $\sim$5\% of the population of such clumps: 12 of the 15 clumps from Paper I that fall into this category are already included here.  Therefore, most of the IR imaging needed to complete this survey will be of the more numerous, fainter population.  If the trend in Figure \ref{hcopbrg} is confirmed in these other clumps, it would tend to strengthen the case for long-lived massive clumps put forward in Paper I, and supported by the results of (e.g.) \cite{ncs08}.

What is the physical basis for the significance of $\eta$?  \nnh\ is known to trace better the cold, quiescent, pre-stellar dense gas in low-mass cores \cite[e.g.,][]{CWZ02} than do other dense-gas tracers like \hcop, CS, or HCN.  This happens because gas-phase CO is a net destroyer of \nnh, while the \hcop\ abundance is closely tied to the CO.  Where CO freezes out onto grains in the coldest parts of molecular clumps and cores, this production of \hcop\ and destruction of \nnh\ is hindered.  Whether this also happens in massive star-forming clumps has hitherto been less clear, although some hint of this has previously emerged \cite[P. Caselli, in prep.]{ZCP09,s11}.  

\begin{figure*}
(a)\hspace{-4mm}\includegraphics[angle=0,scale=0.44]{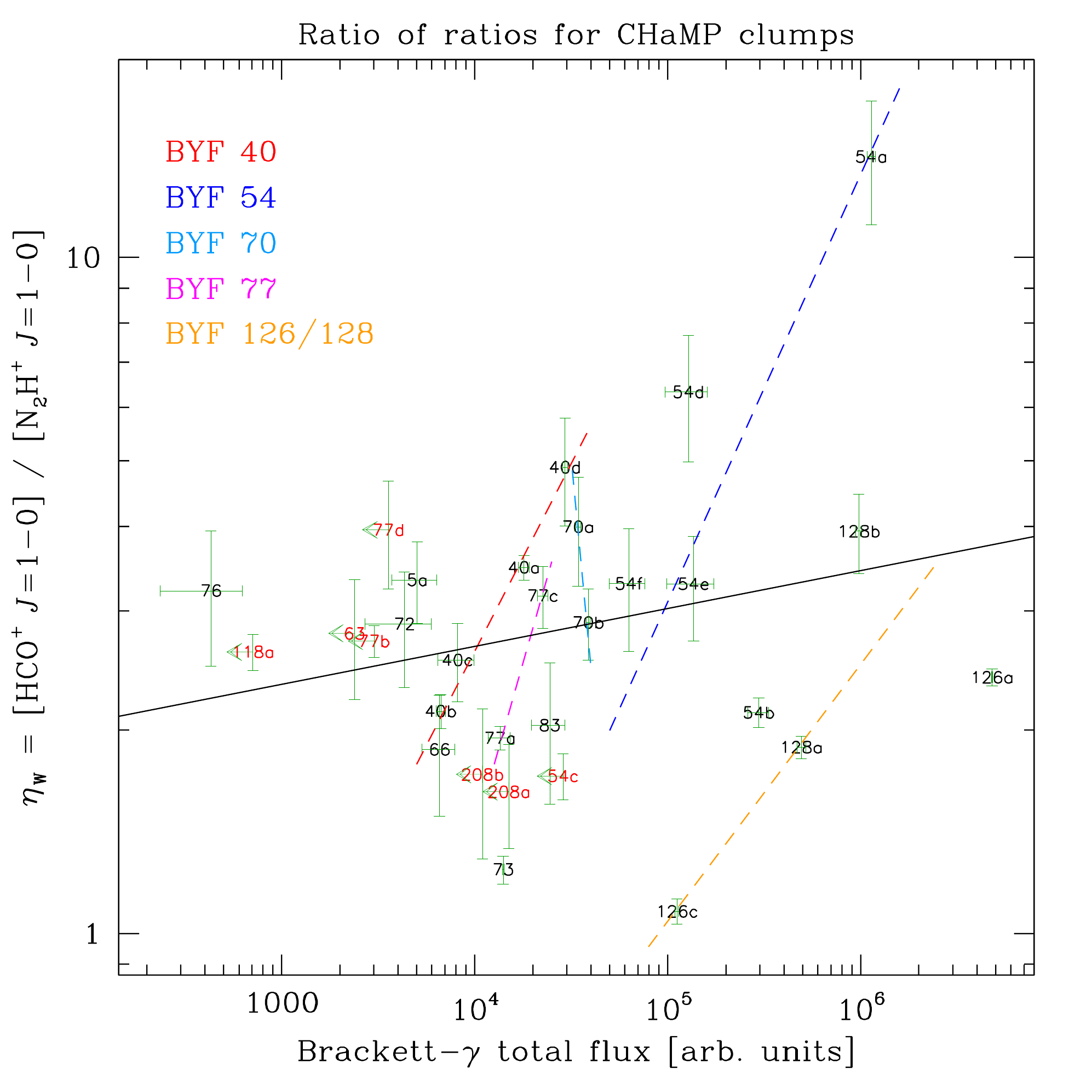}\hspace{-2mm}
(b)\hspace{-4mm}\includegraphics[angle=0,scale=0.44]{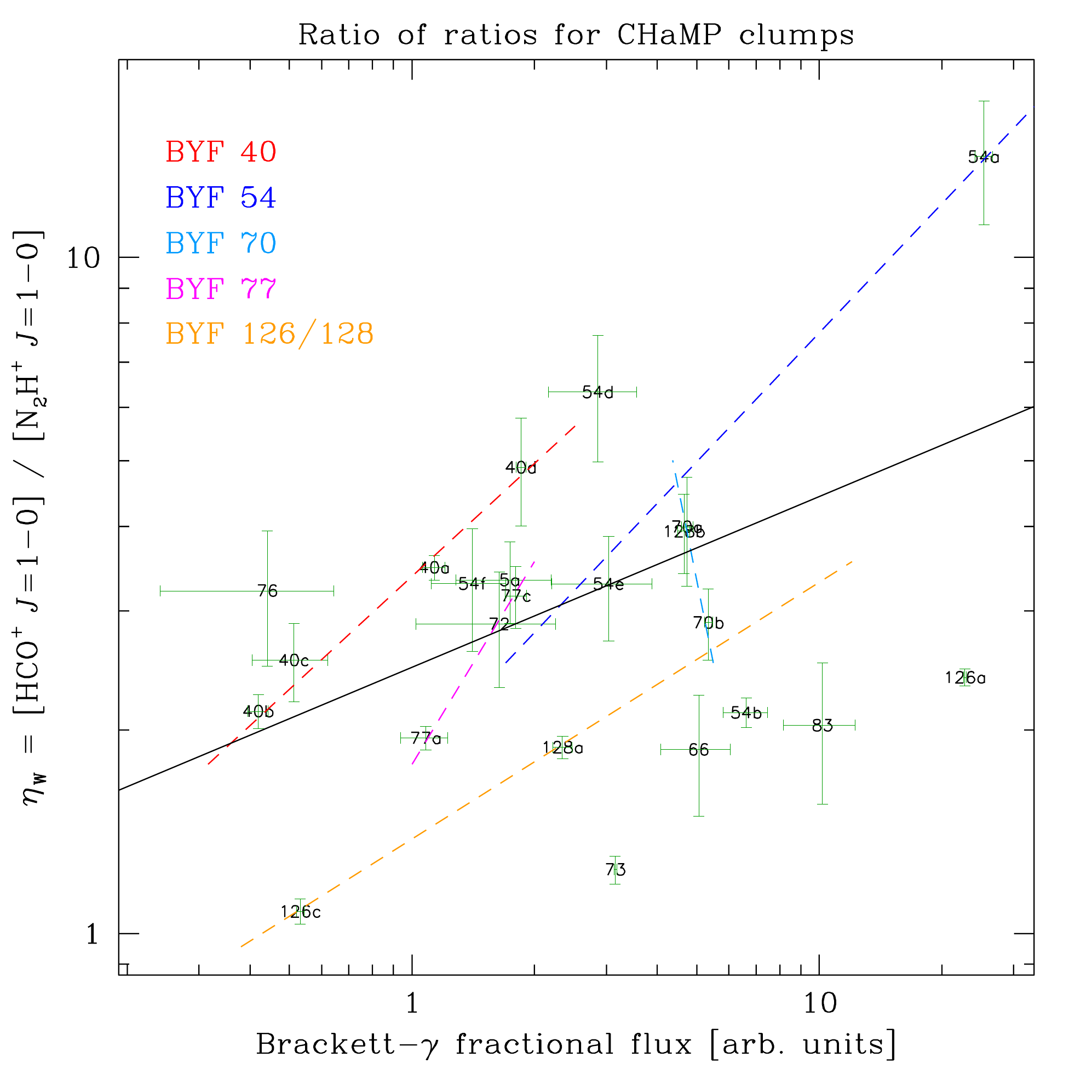}
\caption{
($a$) The mm-line ratio $\eta$ as a function of the \brg\ flux, labelled by clump number.  Other details are as in Fig.\,\ref{hcopbrg}.  Here the least-squares fit to the well-measured points (black labels and line) has slope $m=0.06\pm0.05$ and correlation coefficient $r^2=0.05$, with an rms scatter about this trend of a factor of 1.8 in the ordinate.  However, the slopes of fits to individual complexes seem much steeper, as shown in the various colours.  
($b$) Well-measured points from panel $a$, but now with their \brg\ flux normalised to the total bolometric flux of the relevant complex \citep[from][]{mtb13}.  Note the different abscissa scales in these two panels.  Here the least-squares fit is only to those 17 points that are part of the labelled complexes.  Thus, points for BYF\,5a, 66, 72, 73, 76, and 83 were not included in the fit, but nevertheless lie within a factor of $\sim$2.5 of this fit, with an rms scatter of 0.27 in log\,$y$.  The fit itself has $m=0.25\pm0.10$ and $r^2=0.28$, with an rms scatter about this trend of only 17\% in the ordinate (or 0.067 in log\,$y$).
}
\label{etaWbrg}
\end{figure*}

Some mm-line studies of dense clumps claim an ``evolutionary progression'' when only a few examples of the different behaviour of \nnh\ (or other species) are given.  Our larger sample of clumps with both mm-spectroscopic and IR-line data may allow us to establish a more precise version of such a paradigm.  We show in Figure \ref{etaWbrg}$a$ a combination of the data from Figures \ref{mmlines}--\ref{n2hpbrg}.  
While this plot exhibits a fairly weak correlation overall ($\sim$1.3$\sigma$), a more subtle but striking trend becomes evident upon closer examination.  {\color{black} We now subdivide the clumps into complexes as shown in colour in Figure \ref{etaWbrg}$a$, and seek a relation within each complex.  This is not to claim statistical robustness of any such intra-complex relations, but to test if we can discern a unifying {\em inter-complex} principle in these relations.}  Thus, {\em for clumps within a given complex}, in several instances $\eta$ appears to be strongly correlated with the \brg\ flux: the mean slope of these (admittedly poorly-determined) correlations is $m$$\sim$0.8 ({\em cf.} $m$ = 0.26 from Fig.\,\ref{hcopbrg}).  In other words, without necessarily claiming causality in either direction, the ratio $\eta$ may possibly be a locally strong function of the ionising flux among the denser clumps within a molecular cloud complex.

Next, we note that in Figure \ref{etaWbrg}$a$, if granting the possibility of a $\eta$--\brg\ relation {\em within} complexes, there seems to be a different normalisation of this relation {\em between} complexes, spanning $\sim$2 orders of magnitude in \brg\ flux between the extremes of the BYF\,40 and 126/128 complexes.  If real, to some extent this might be due to (e.g.) IMF sampling variations, which could strongly affect the EUV and FUV fluxes from a given star cluster of given stellar mass, yet we do not have this information at hand.  Nevertheless, we examined several other clump parameters that might hint at a simple physical origin of such a normalisation, such as clump mass, distance, etc.  We found that the total bolometric flux from a complex of clumps produces an interesting result, as shown in Figure \ref{etaWbrg}$b$.  This particular normalisation \cite[using data from a companion study to this, of the SEDs of all the CHaMP clumps;][]{mtb13} also has the advantage of making Figure \ref{etaWbrg}$b$ distance-independent.  {\color{black} With this normalisation, we now fit {\em all} the clumps within the labelled complexes (i.e., 17 points) with a single least-squares fit.  We find the significance of the overall fit to be much higher than those to individual complexes} (slope $0.25\pm0.10$, or 2.4$\sigma$), although there may be other parameters such as IMF sampling as above, or 3D projection effects, or the geometry of the ionising source distribution relative to any internal extinction, that contribute to the remaining dispersion.  {\color{black} In addition, 6 more normalised points not included in this fit have a small scatter about it.}

{\color{black} While perhaps somewhat speculative,} we believe this is quite remarkable: it suggests that the relationship between $\eta$ and \brg\ in Figure \ref{etaWbrg} is much stronger than the relationship in Figure \ref{hcopbrg} of \brg\ to \hcop\ alone.  If correct, this would mean that, as the ionising flux rises, the \hcop\ column density is rising {\em at the same time as} the \nnh\ column density is falling.  Therefore it seems a natural hypothesis to suppose that $\eta$ and the ionisation state of a clump depend strongly on the {\em current, local} massive star formation activity among the clumps in question.  Together with a similar result from \cite{mt12}, we believe our results show, in a fairly direct and robust way, that this indicator (the \hcop--\nnh--\brg\ relationship) may be a reliable evolutionary signpost in massive star and cluster formation.

\subsection{The IR Lines\label{irdisc}}
\begin{figure}
\includegraphics[angle=0,scale=0.44]{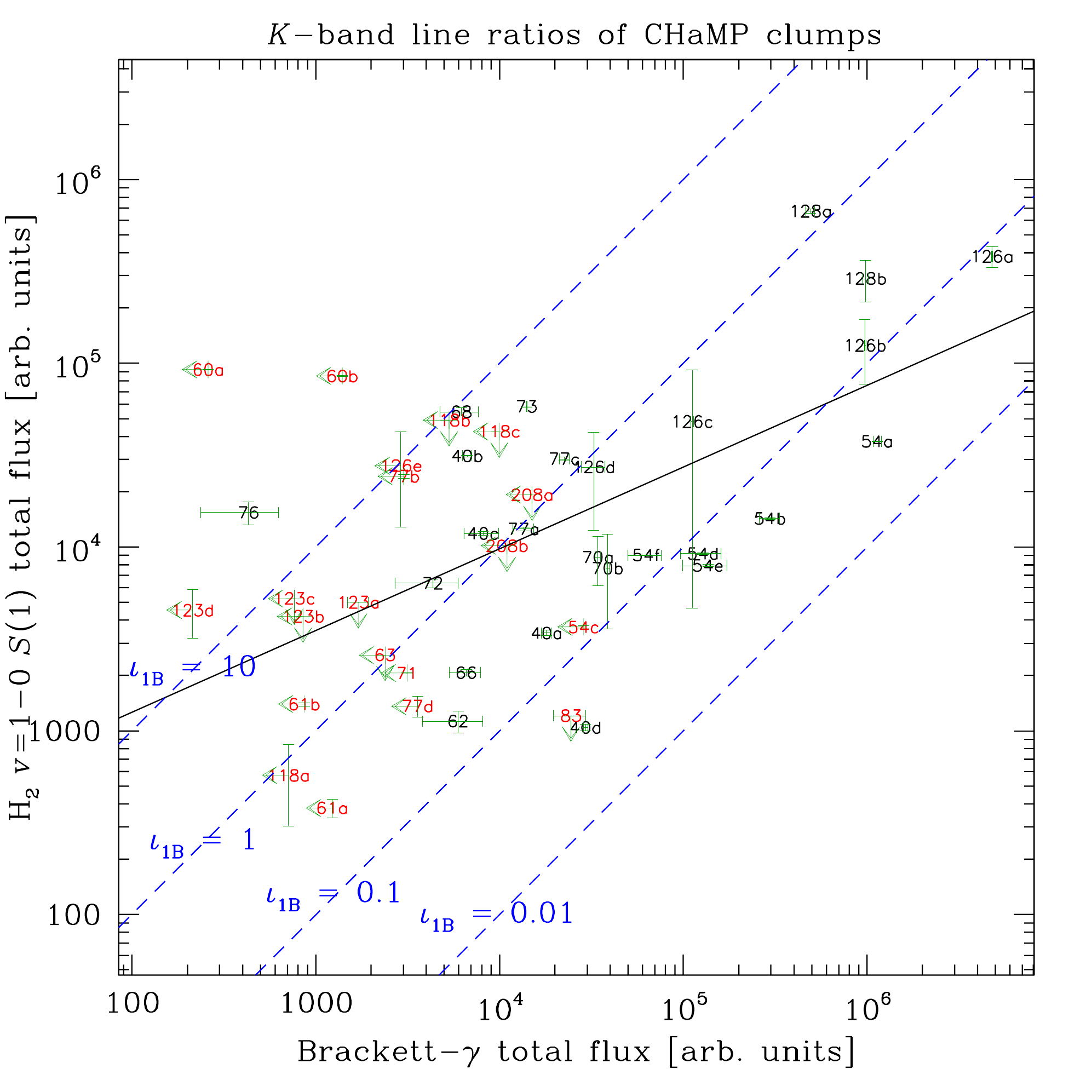}
\caption{\htwo\ \voz\ vs. \brg\ integrated fluxes, labelled by clump number.  Other details are as in Fig.\,\ref{hcopbrg}.  Here the unweighted least-squares fit to the well-measured points (black labels and line) has slope $0.44\pm0.13$, with an rms scatter about this trend of a factor of 4.1 in the ordinate. 
We also show contours of constant line ratio $\iota_{1B}$ = [\voz]/[\brg] in blue.}
\label{brghoz}
\end{figure}

\begin{figure}
\includegraphics[angle=0,scale=0.44]{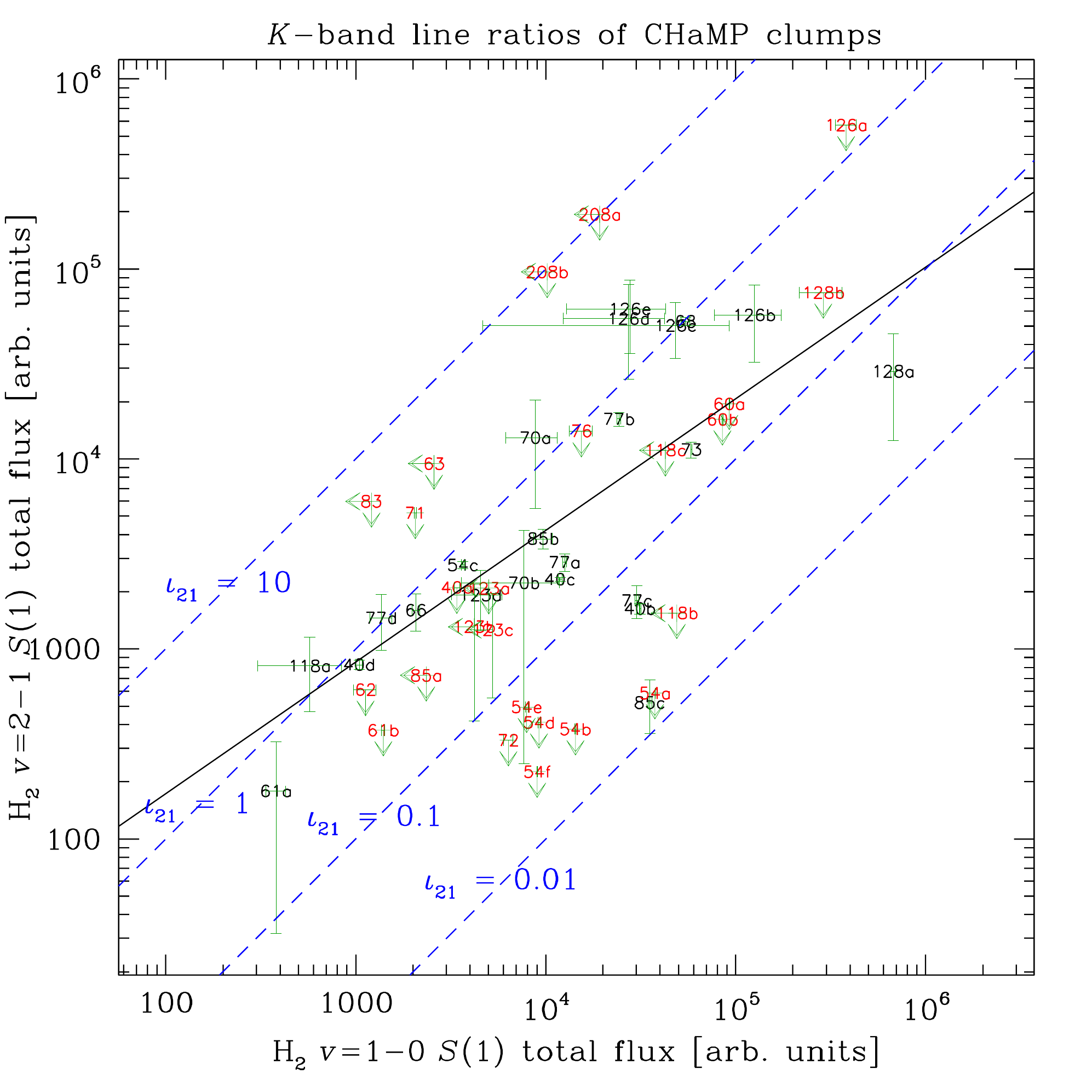}
\caption{\htwo\ \vto\ vs. \voz\ integrated fluxes, labelled by clump number.  Other details are as in Fig.\,\ref{hcopbrg}.  Here the unweighted least-squares fit to the well-measured points (black labels and line) has slope $0.69\pm0.15$, with an rms scatter about this trend of a factor of 4.0 in the ordinate. 
We also show contours of constant line ratio $\iota_{21}$ = [\vto]/[\voz] in blue.}
\label{hozhto}
\end{figure}

In Figure \ref{brghoz} we see that the \brg\ and \htwo\ \voz\ fluxes are marginally well-correlated with each other, which is consistent with the findings above, e.g. that the \hcop\ is well-correlated with the \brg\ but less so with the \htwo.  A comparison of the \brg\ emission with the \htwo\ \vto\ (not shown) reveals a similar correlation but with a larger uncertainty, again not inconsistent with the mm-IR relationships described earlier, even while it is possible that IMF sampling variations or other factors could also contribute to a large dispersion of NIR line ratios.  While a higher flux of emission from excited \htwo\ lines being correlated with a higher flux of \brg\ emission is not unexpected, perhaps more illuminating is a direct comparison of the two \htwo\ lines (Fig.\,\ref{hozhto}).  Although the correlation between clumps is again marginal, the line ratio itself is diagnostic of different excitation conditions.

For example, \cite{r98} conducted a detailed study of the \htwo\ line ratio in the reflection nebula Parsamyan 18.  They showed that the \htwo\ emission in that source is driven primarily by fluorescence from a single UV-luminous star, coupled with some thermalisation of the \voz\ line in regions of higher \htwo\ density.  Fluorescence alone produces a line ratio $\iota_{21}\sim0.6$, while thermalisation of the lower-$v$ levels can preferentially enhance the \voz\ line emission, thus depressing the $\iota_{21}$ ratio below its fluorescent value, sometimes significantly ($\iota_{21}<0.1$).  Such small values of $\iota_{21}$ can be obtained when the temperature is fairly low ($T_{\rm ex}$ $\lapp$ 2000\,K) and the density high enough to allow the \htwo\ to self-shield in the presence of a strong UV field ($n>40$$G_0$\,cm$^{-3}$, where $G_0=1$ for the average interstellar radiation field).  Since $G_0$ can exceed 10$^4$ near massive Young Stellar Objects (YSOs), significant thermalisation of the \htwo\ lines implies the presence of at least some gas at densities $>$10$^{5-6}$.  The purely thermal $\iota_{21}$ ratio will actually depend on the temperature in the \htwo\ gas, e.g. when the \htwo\ excitation temperature $T_{\rm ex}$ = 2000\,K, $\iota_{21}$ = 0.083, while at $T_{\rm ex}$ = 4000\,K, $\iota_{21}$ = 0.33 (T.\,Geballe, 1995)\footnote{ Quoted by T. Kerr (2004), http://www.jach.hawaii.edu/UKIRT/astronomy/calib/spec\_cal/h2\_s.html}, and at higher temperatures approaches the fluorescent value.  Therefore, we see that one requires a high density but low-excitation gas to observe very small $\iota_{21}$ ratios in the presence of a strong UV field.

In Figure \ref{hozhto} we see that $\iota_{21}$ is consistent (within the uncertainties) with fluorescence in most of our clumps, with a smaller number of sources showing \htwo\ emission that is more thermal in nature.  Here fluorescence is the preferred interpretation for $\iota_{21}\sim0.6$ over high-temperature thermalisation, since the average molecular density in our clumps as determined in Paper I from the \hcop\ line data ($n\sim10^{3-4}$ over the 40$''$ Mopra beam) is likely too low to afford self-shielding to the \htwo\ when near a massive YSO.  But where some clumps have lower $\iota_{21}$, the presence of some higher density molecular gas would seem to be indicated.  In each case, the emission morphology tends to match expectations for the physical origin of these line ratios.  For example, in BYF\,70a, 73, or 77b where $\iota_{21}$ is close to 0.6, the emission structure is clearly that of \hii\ regions with associated PDRs, and so fluorescence would be expected to drive the \htwo\ emission.  Conversely, in BYF\,62 or 72 the \vto\ is hard to detect at all, while the morphology resembles narrow \htwo\ jets from low-mass protostars which are dominated by thermal emission from shocks.  Curiously, while the BYF 54 clumps (see Fig.\,\ref{sample}) show very strong \brg\ and \htwo\ \voz\ emission, the \vto\ emission is virtually non-existent and $\iota_{21}$ $\ll$ 1, implying that the \voz\ nebulosity we see is dominated by thermal emission at low temperatures (i.e., $\sim$ 1000\,K), despite their projected proximity to what seem like PDRs adjacent to \hii\ regions.  Therefore, it may be that this complex has, e.g., unusually strong wind shocks exciting the \htwo\ emission, instead of the (apparently) more usual fluorescence, and may be worth deeper investigation on this basis.  


{\color{black} Finally, we reiterate that the calibration dispersion could not create a spurious correlation of the form in Figure \ref{hcopbrg}; indeed it would only weaken the true correlation.  In particular, Figure \ref{etaWbrg} would be unaffected by any variations in transparency, as all complexes within a clump were observed simultaneously.  Similarly in Figures \ref{brghoz} and \ref{hozhto}, since the \brg\ and both \htwo\ lines were nearly always observed within $\sim$20$^m$ of each other, it is likely that each experienced similar atmospheric conditions, changes in which would only move any clump parallel to the blue dashed lines, but not otherwise smear out any underlying correlation.}

\section{Conclusions\label{concl}}

We have presented new results of mm-wave and near-IR emission-line imaging of 60 massive, dense, star-forming molecular clumps, the first $\sim$20\% of the CHaMP sample (Paper I).  From these, we see the first systematic variations in line ratios with clump evolutionary state.  Our findings include the following.

1. Among the clumps showing bright molecular emission in \hcop\ and \nnh, there is a wide variation in the intensity ratio $\eta$ of these lines, from $\sim$1--14.  Among the fainter clumps, $\eta$ is somewhat more uniform, $\sim$2--5.

2. The \hcop\ intensity of the clumps is strongly correlated with the \brg\ integrated flux associated with the clump, while the \nnh\ intensity appears relatively uncorrelated with the \brg\ emission, when measured across all clumps.  There are also weaker correlations between either molecular species and the associated \htwo\ IR emission lines.  The $W$(\hcop)-\brg\ relation arises mostly in the line brightness, rather than the linewidth, suggesting that the dominant effect of the energy output of massive YSOs on the clump gas is radiative, rather than mechanical.  This is consistent with the fluorescent interpretation for the \htwo\ line emission, below.

3. The mm-line ratio $\eta$ = [\hcop]/[\nnh] is even more strongly correlated with bright \brg\ emission among clumps {\bf {\em within}} a given molecular complex, and is at least partly normalised {\bf {\em between}} complexes by the total bolometric luminosity of the complex.  This situation occurs in the minority of all our clumps, i.e. those with a significant nearby UV field from embedded massive YSOs.  The majority of our clumps show only a mild, apparently random variation in $\eta$ $\sim$ 2--5.

4. This suggests a common trend among massive dense clumps in the presence of active massive star and star cluster formation.  Before significant massive star formation has occurred in a clump, the \hcop/\nnh\ column density ratio $\eta$ is constant to within a factor of $\sim$2.  As the ionising flux impinging upon a clump rises, the \hcop\ column density rises while the \nnh\ column density simultaneously falls.

5. We propose that the $\eta$-\brg\ relation is a useful new diagnostic of the progress in a massive molecular clump towards the formation of massive clusters with stars of sufficient luminosity to ionise and potentially disperse the clump.

6. In most cases, the \htwo\ emission is consistent with fluorescent excitation in PDRs near \hii\ regions, and the emission morphology often conforms to this origin.  In a few cases, the \htwo\ emission seems mostly thermally excited, and again the morphology in such instances often resembles that from protostellar jets driving shocks into the surrounding gas.  BYF\,54 seems to be a counterexample to both these groups, exhibiting thermal \htwo\ line ratios in the presence of a strong UV field.

These and other results suggest that the \hcop/\nnh\ intensity ratio $\eta$, together with the \brg\ flux, may provide useful signposts of massive star formation, and allow us to tie observations more closely to certain theoretical models of the evolution of massive, dense molecular clumps, contributing to a more complete picture of the formation of massive stars and star clusters.

\vspace{-4mm}
\section*{Acknowledgments}

We thank the staff members of the ATNF and AAO for their support of the observations.  PJB gratefully acknowledges support through NSF grant AST-0645412 to JCT at the University of Florida, a Professional Development Leave award in 2011 from the University of Florida, and receipt of an AAO Distinguished Visitorship in 2011.  SNO and LEA were partially supported by an REU supplement to JCT's NSF grant.  We also thank the referee for helpful suggestions and comments which led to several improvements in the paper.

{\em FACILITIES:} {Mopra (MOPS), AAT (IRIS2)} 

\section*{Supporting Information}
Additional Supporting Information may be found in the online version of this article:

{\bf Appendix A:} Narrow-band NIR Images.

\appendix

\begin{figure*}
\section{Narrow-band Near-Infrared Images\label{app}}
(a)\hspace{-3mm}\includegraphics[angle=-90,scale=0.43]{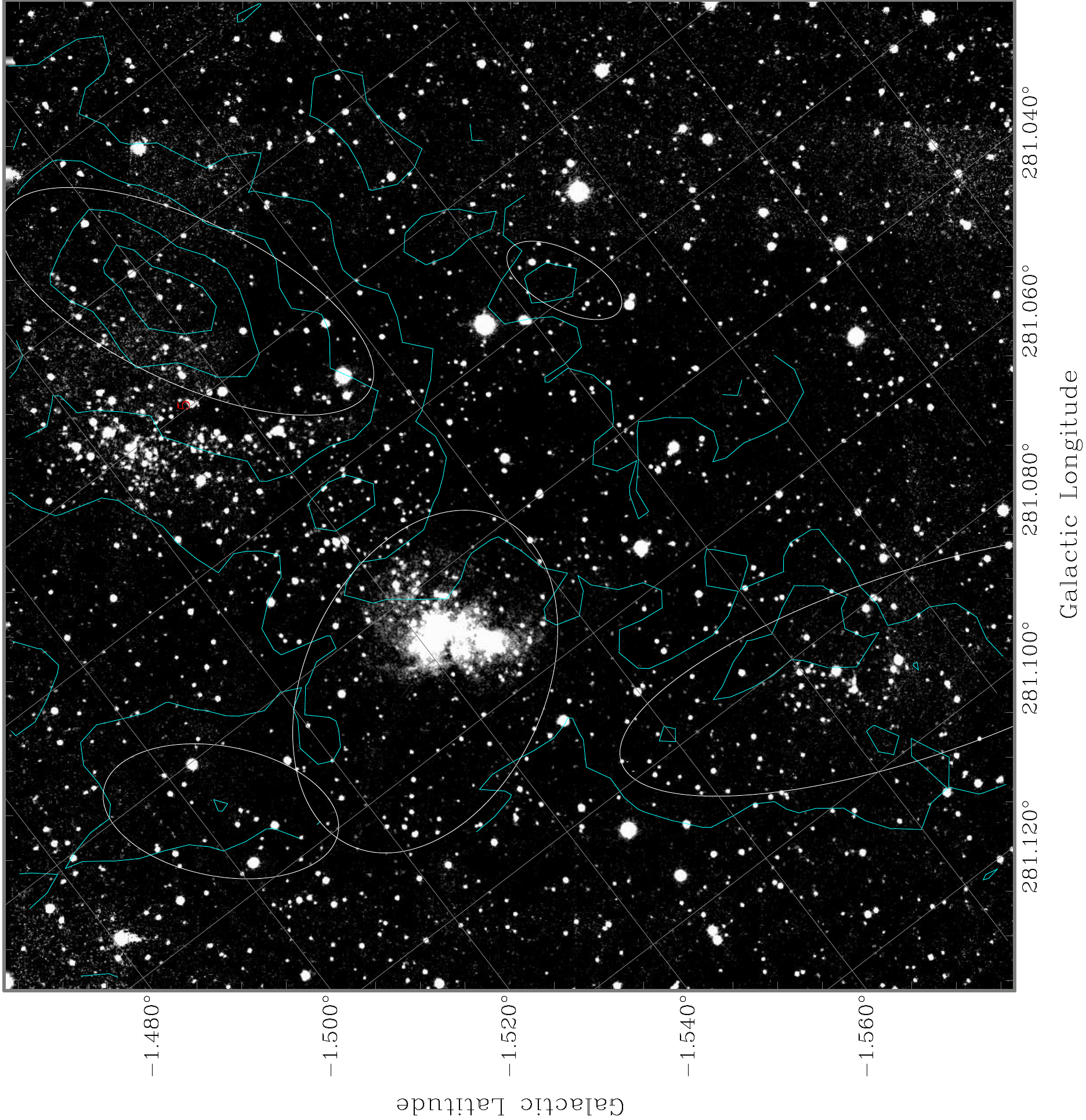}\hspace{3mm}
(b)\hspace{-3mm}\includegraphics[angle=-90,scale=0.43]{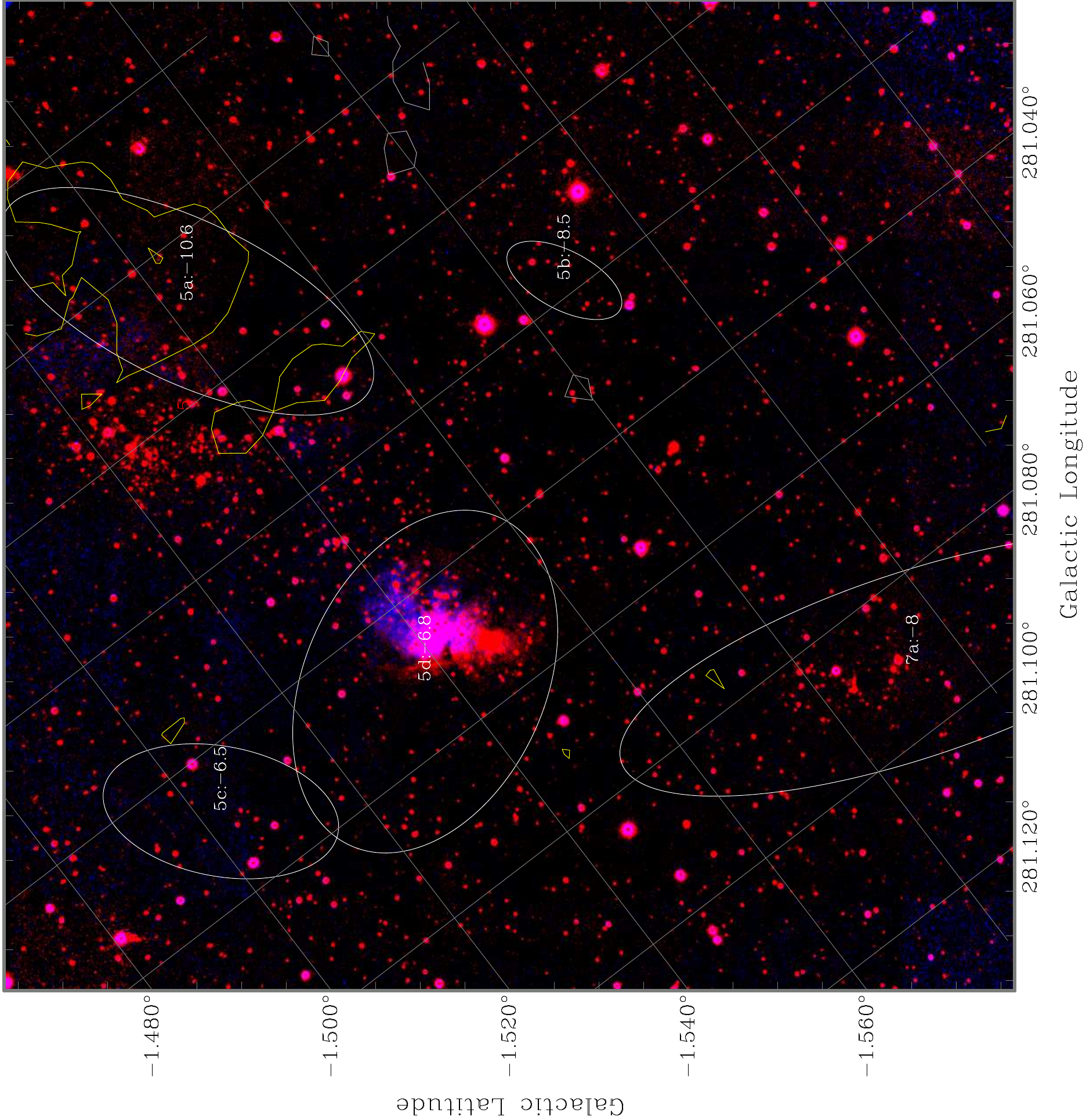}
\caption{(a) $K$-band line-free continuum image of BYF\,5 (part of Region 1 from Paper I) on a linear brightness scale.  Contours are overlaid from Mopra \hcop\ integrated intensity (cyan levels at 4, 8, 12, and 16 times the rms level of 0.318\,K\kms), while the white ellipses represent gaussian fits to the \hcop\ emission.  (b) Colour-composite image of continuum-subtracted \brg\ (blue) with the same image as in ($a$) (red) on a linear brightness scale.  (Poor weather prevented us from obtaining the corresponding images for this field in the other filters.)  Overlaid here are contours of Mopra \nnh\ integrated intensity (grey levels at --3 and yellow at 3 times the rms of 0.274\,K\kms) with the same ellipses as in (a).  At a distance of 3.2\,kpc, the scale is 40$''$ = 0.621\,pc or 1\,pc = 0\fdg0229 = 64\farcs5.  In this and subsequent figures, images not otherwise indicated are from observations in 2011 with shorter integration times.  Images from deeper 2006--07 data are so noted.}
\label{byf5}
\end{figure*}

\vspace*{-6mm}
\begin{figure*}
(a)\hspace{-3mm}\includegraphics[angle=-90,scale=0.41]{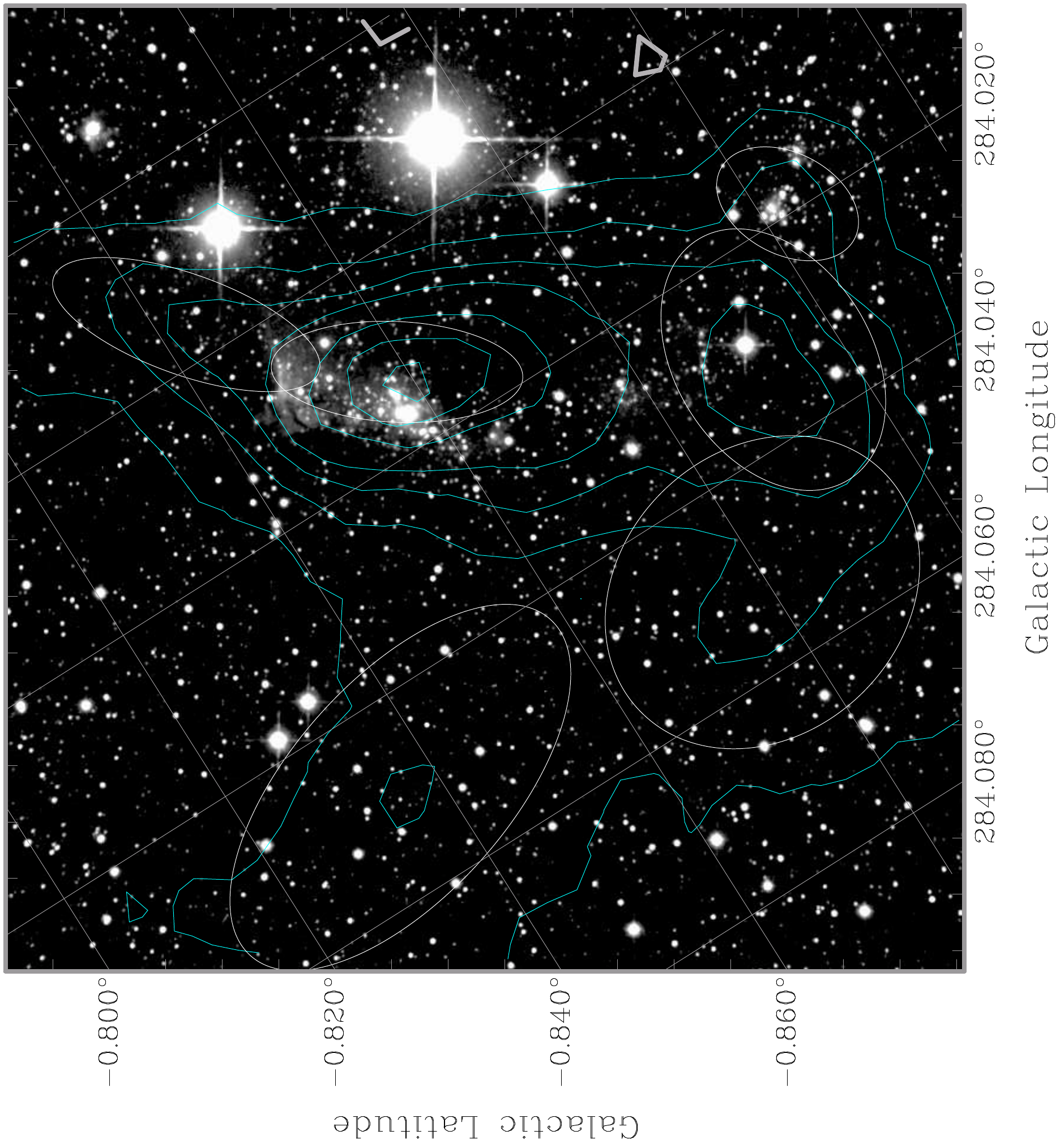}\hspace{3mm}
(b)\hspace{-3mm}\includegraphics[angle=-90,scale=0.41]{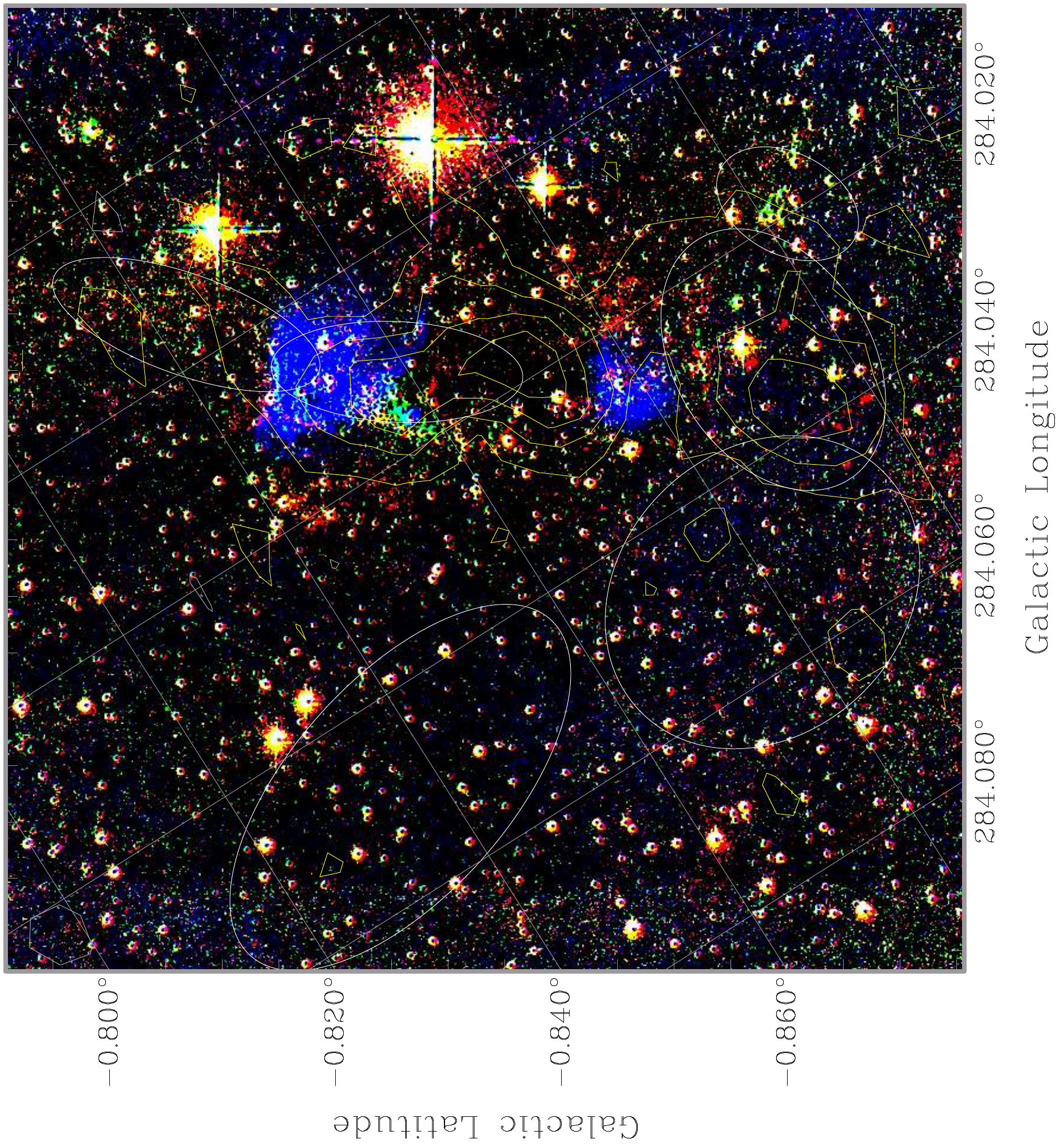}
\caption{Same as Fig.\,\ref{sample}, but for BYF\,40 (part of Region 6 from Paper I) using more sensitive 2006 data.  (a) $K$-band line-free continuum image with \hcop\ contours (heavy grey at 0, cyan at 8, 16, 24, 32, 48, 64, and 80 times the rms level of 0.312\,K\kms).  (b) RGB-pseudo-colour image of the continuum-subtracted $K$-band spectral lines \nnh\ contours (grey at --2.5, yellow at 2.5, 5, 7.5, and 10 times the rms of 0.588\,K\kms).  White ellipses show gaussian fits to the \hcop\ emission in both panels.  At a distance of 2.5\,kpc, the scale is 40$''$ = 0.485\,pc or 1\,pc = 0\fdg0229 = 82\farcs5.
\label{byf40}}
\end{figure*}

\begin{figure*}
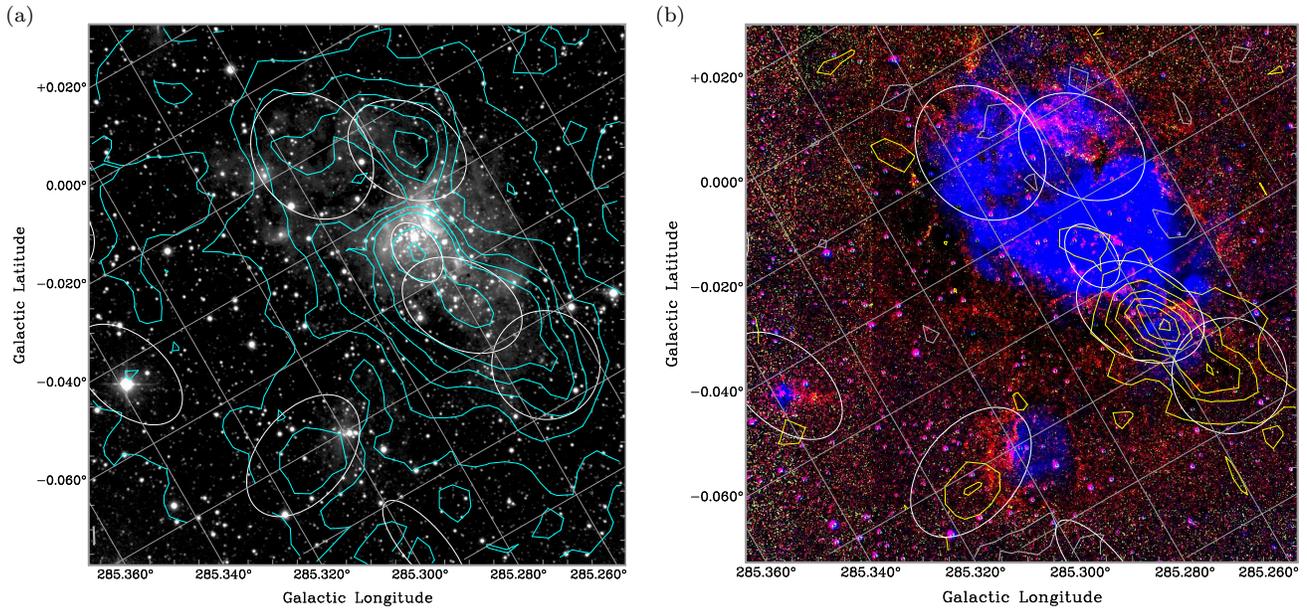

(a)\hspace{-3mm}\includegraphics[angle=-90,scale=0.41]{byf54Kcont}\hspace{3mm}
(b)\hspace{-3mm}\includegraphics[angle=-90,scale=0.41]{byf54rgbM} 
\caption{Same as Fig.\,\ref{sample}, but for BYF\,54 (part of Region 8 from Paper I) using more sensitive 2006 data.  (a) $K$-band line-free continuum image with \hcop\ contours (heavy grey at 0, cyan at 4, 8, 12, 16, 20, 24, 40, and 56 times the rms level of 0.401\,K\kms).  (b) RGB-pseudo-colour image of the continuum-subtracted $K$-band spectral lines with \nnh\ contours (grey at --2, yellow at 2, 4, 6, 8, 10, and 12 times the rms level of 0.603\,K\kms).  White ellipses show gaussian fits to the \hcop\ emission in both panels.  At a distance of 2.5\,kpc, the scale is 40$''$ = 0.485\,pc or 1\,pc = 0\fdg0229 = 82\farcs5.
\label{byf54}}
\end{figure*}

\begin{figure*}
(a)\hspace{-3mm}\includegraphics[angle=-90,scale=0.41]{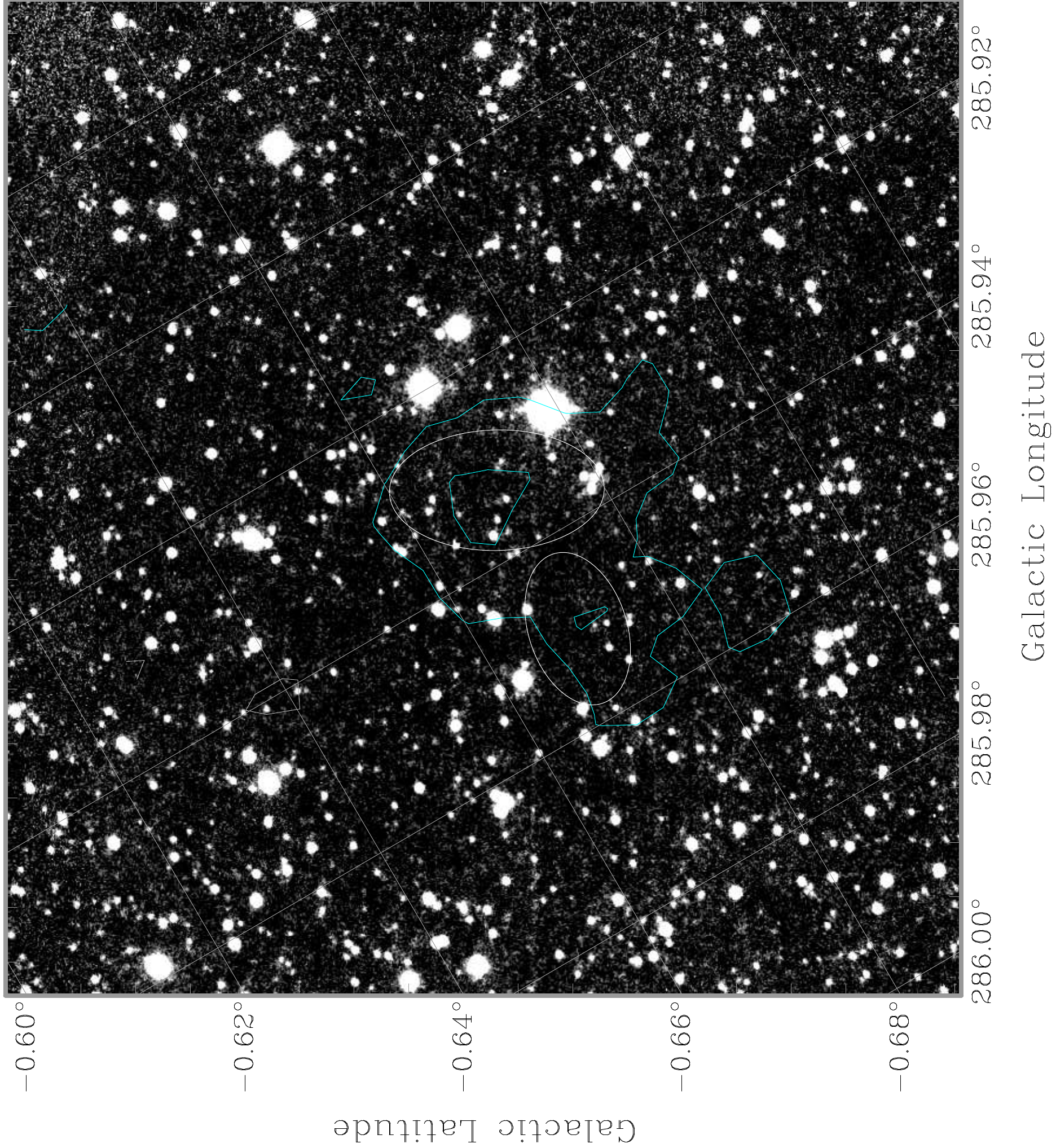}\hspace{3mm}
(b)\hspace{-3mm}\includegraphics[angle=-90,scale=0.41]{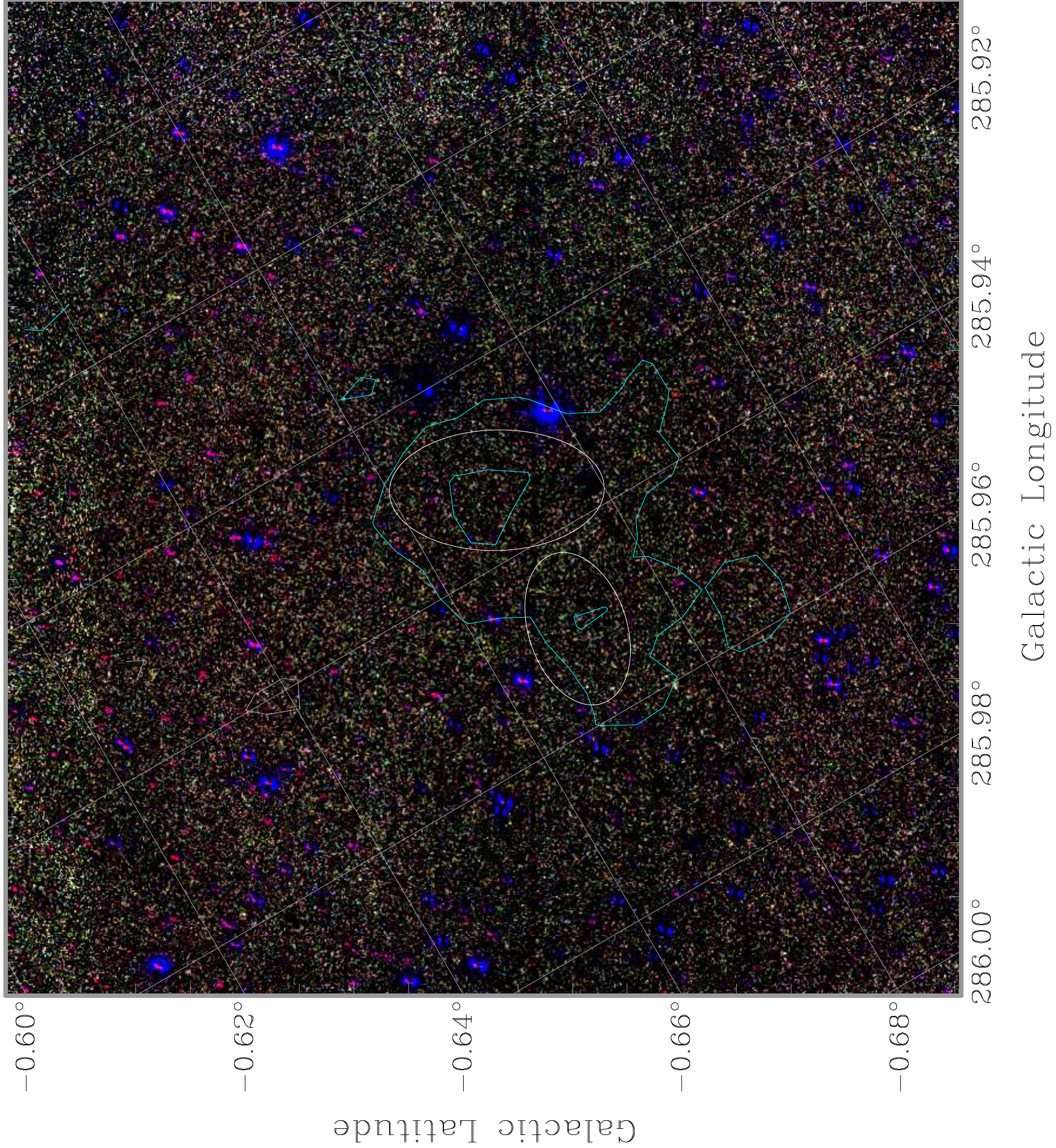}
\caption{Same as Fig.\,\ref{sample}, but for BYF\,60 (part of Region 9 from Paper I).  (a) $K$-band line-free continuum image with \hcop\ contours (grey at --2.5 and cyan at 2.5 and 5 times the rms level of 0.295\,K\kms).  (b) RGB-pseudo-colour image of the continuum-subtracted $K$-band spectral lines with the same contours as in (a) (\nnh\ was not mapped in this field).  White ellipses show gaussian fits to the \hcop\ emission in both panels.  At a distance of 5.3\,kpc, the scale is 40$''$ = 1.03\,pc or 1\,pc = 0\fdg0108 = 38\farcs9.}
\label{byf60}
\end{figure*}

\begin{figure*}
(a)\hspace{-3mm}\includegraphics[angle=-90,scale=0.41]{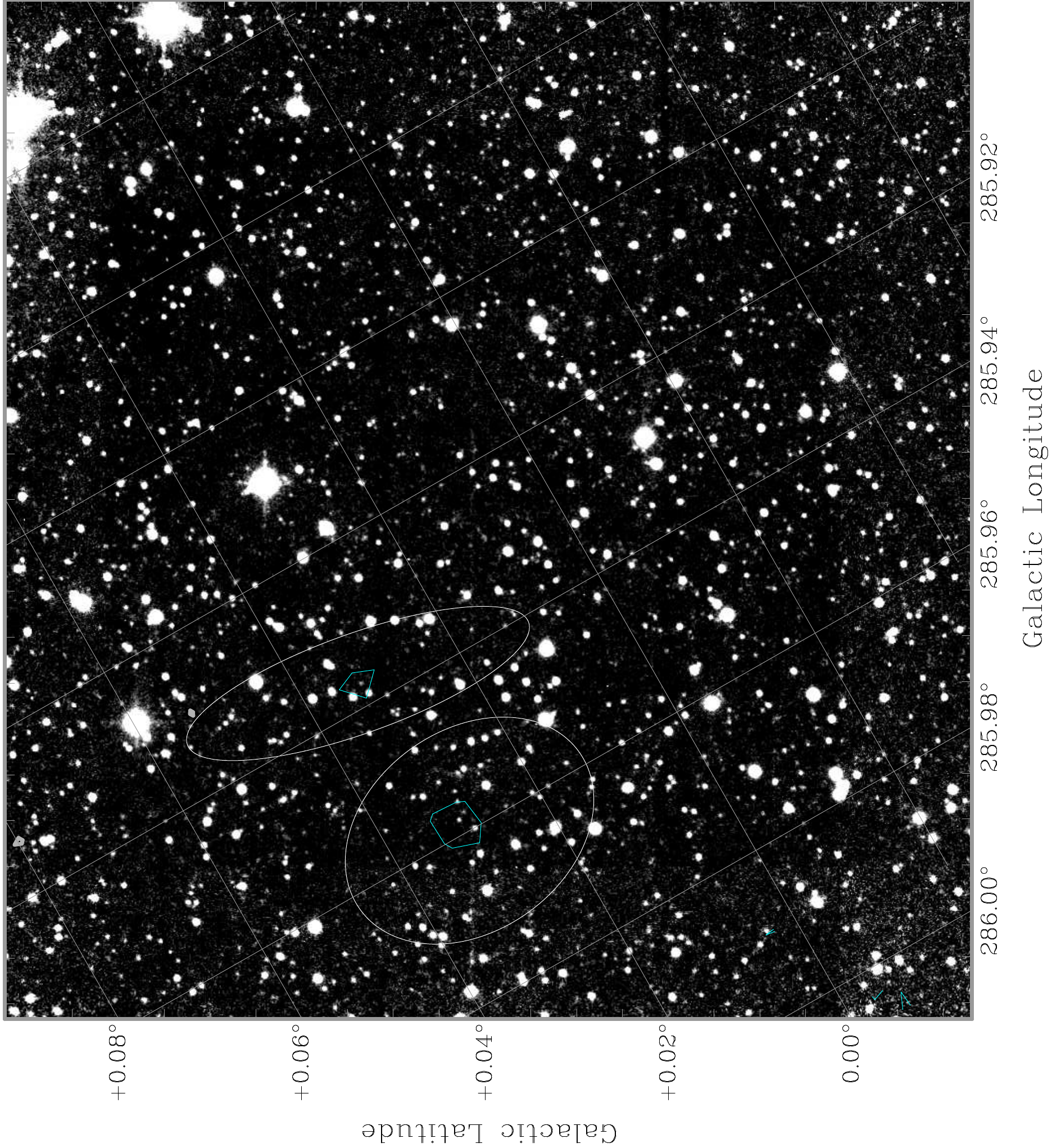}\hspace{3mm}
(b)\hspace{-3mm}\includegraphics[angle=-90,scale=0.41]{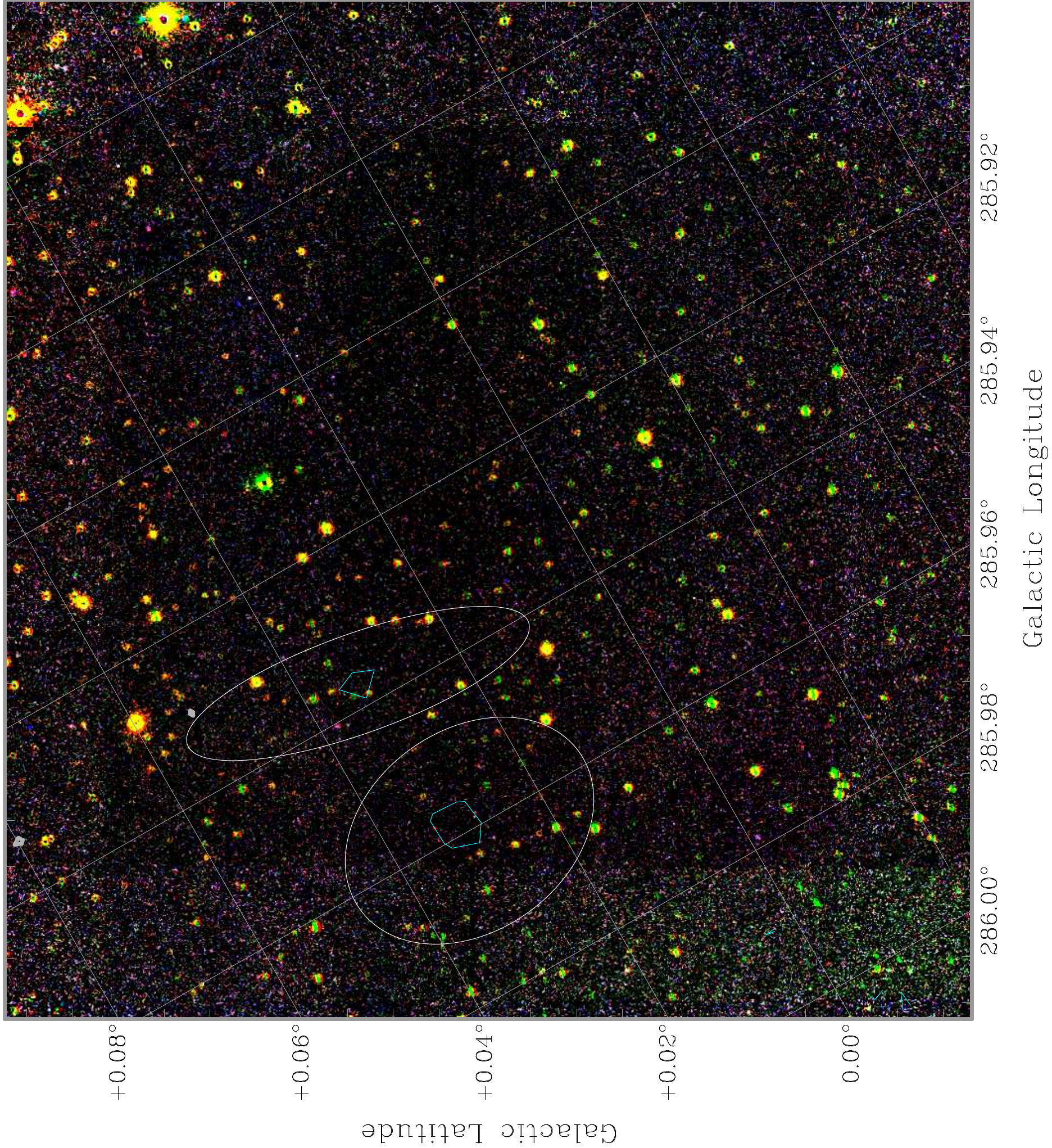}
\caption{Same as Fig.\,\ref{sample}, but for BYF\,61 (part of Region 9 from Paper I, see Fig.\,\ref{reg9}).  (a) $K$-band line-free continuum image with both \hcop\ contours (heavy grey at 0 and cyan at 6, 10, and 14 times the rms level of 0.404\,K\kms) and \nnh\ contours (grey at --3 and yellow at 3 times the rms level of 0.253\,K\kms).  (b) RGB-pseudo-colour image of the continuum-subtracted $K$-band spectral lines with the same contours as in (a).  White ellipses show gaussian fits to the \hcop\ emission in both panels.  At a distance of 2.5\,kpc, the scale is 40$''$ = 0.485\,pc or 1\,pc = 0\fdg0229 = 82\farcs5.}
\label{byf61}
\end{figure*}

\begin{figure*}
(a)\hspace{-3mm}\includegraphics[angle=-90,scale=0.42]{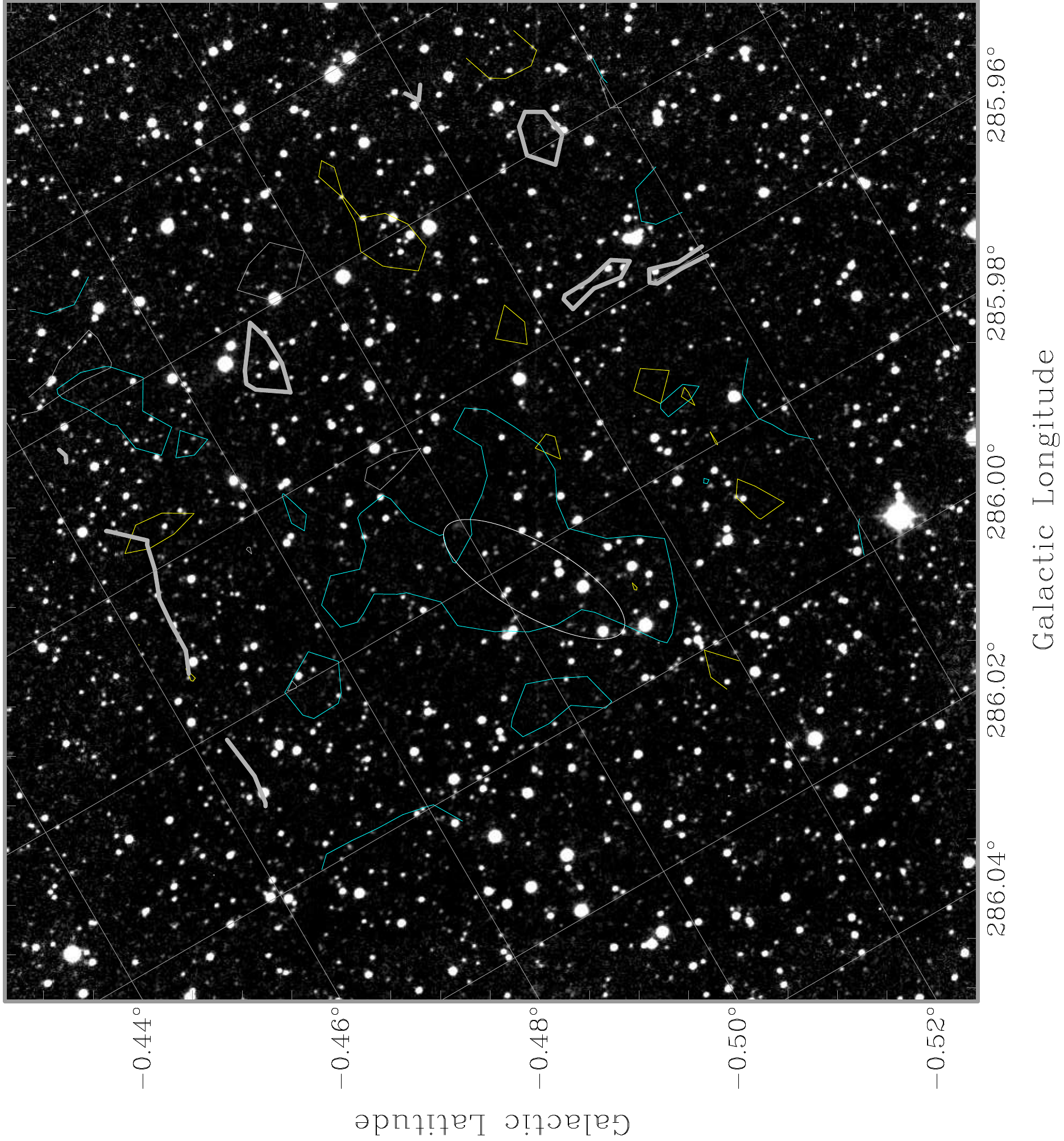}\hspace{3mm}
(b)\hspace{-3mm}\includegraphics[angle=-90,scale=0.42]{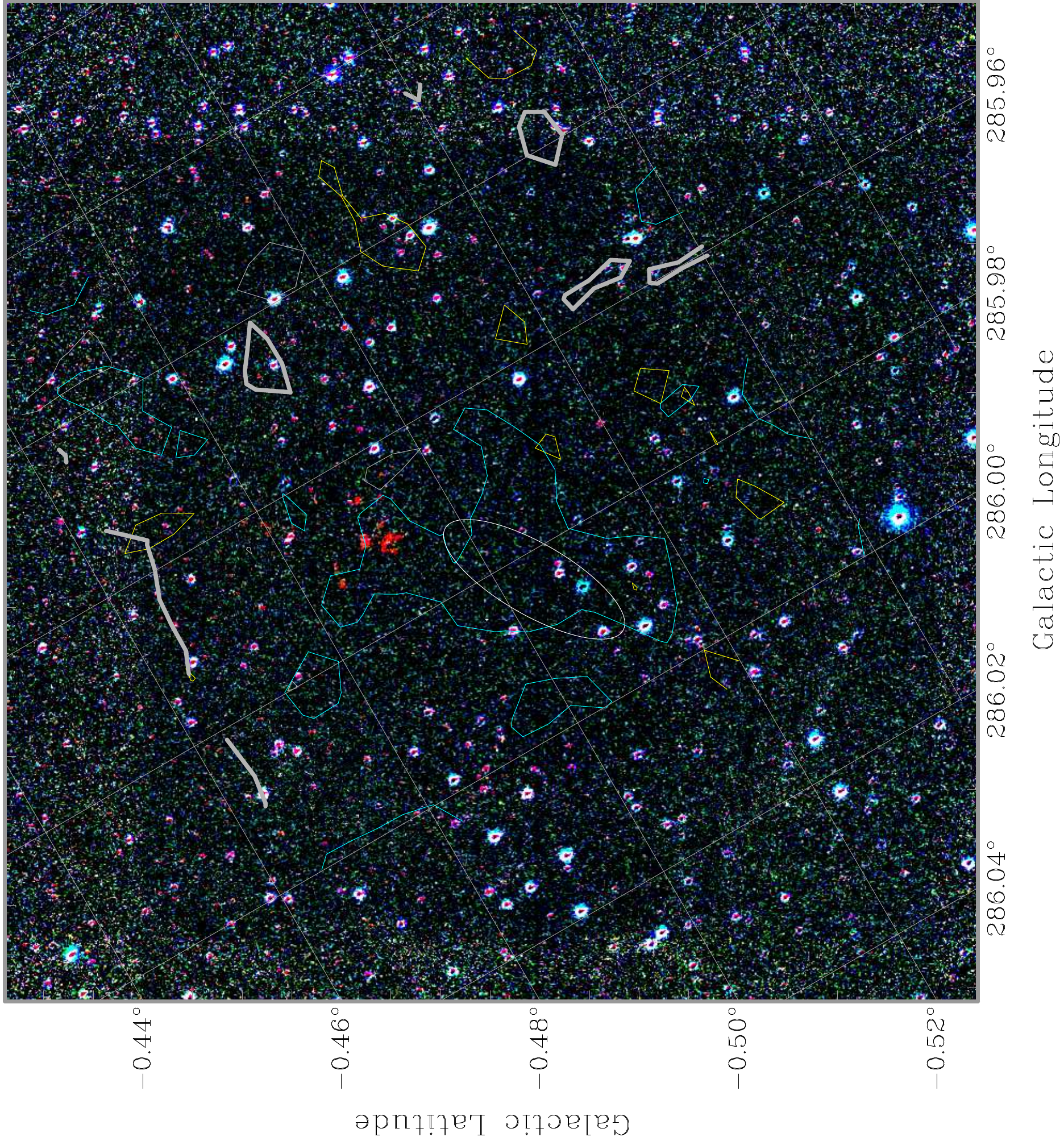}
\caption{Same as Fig.\,\ref{sample}, but for BYF\,62 (part of Region 9 from Paper I, see Fig.\,\ref{reg9}).  (a) $K$-band line-free continuum image with both \hcop\ contours (heavy grey at 0 and cyan at 6, 10, and 14 times the rms level of 0.404\,K\kms) and \nnh\ contours (grey at --3 and yellow at 3 times the rms level of 0.253\,K\kms).  (b) RGB-pseudo-colour image of the continuum-subtracted $K$-band spectral lines with the same contours as in (a).  The white ellipse shows a gaussian fit to the \hcop\ emission in both panels.  At a distance of 2.5\,kpc, the scale is 40$''$ = 0.485\,pc or 1\,pc = 0\fdg0229 = 82\farcs5.}
\label{byf62}
\end{figure*}

\begin{figure*}
(a)\hspace{-3mm}\includegraphics[angle=-90,scale=0.41]{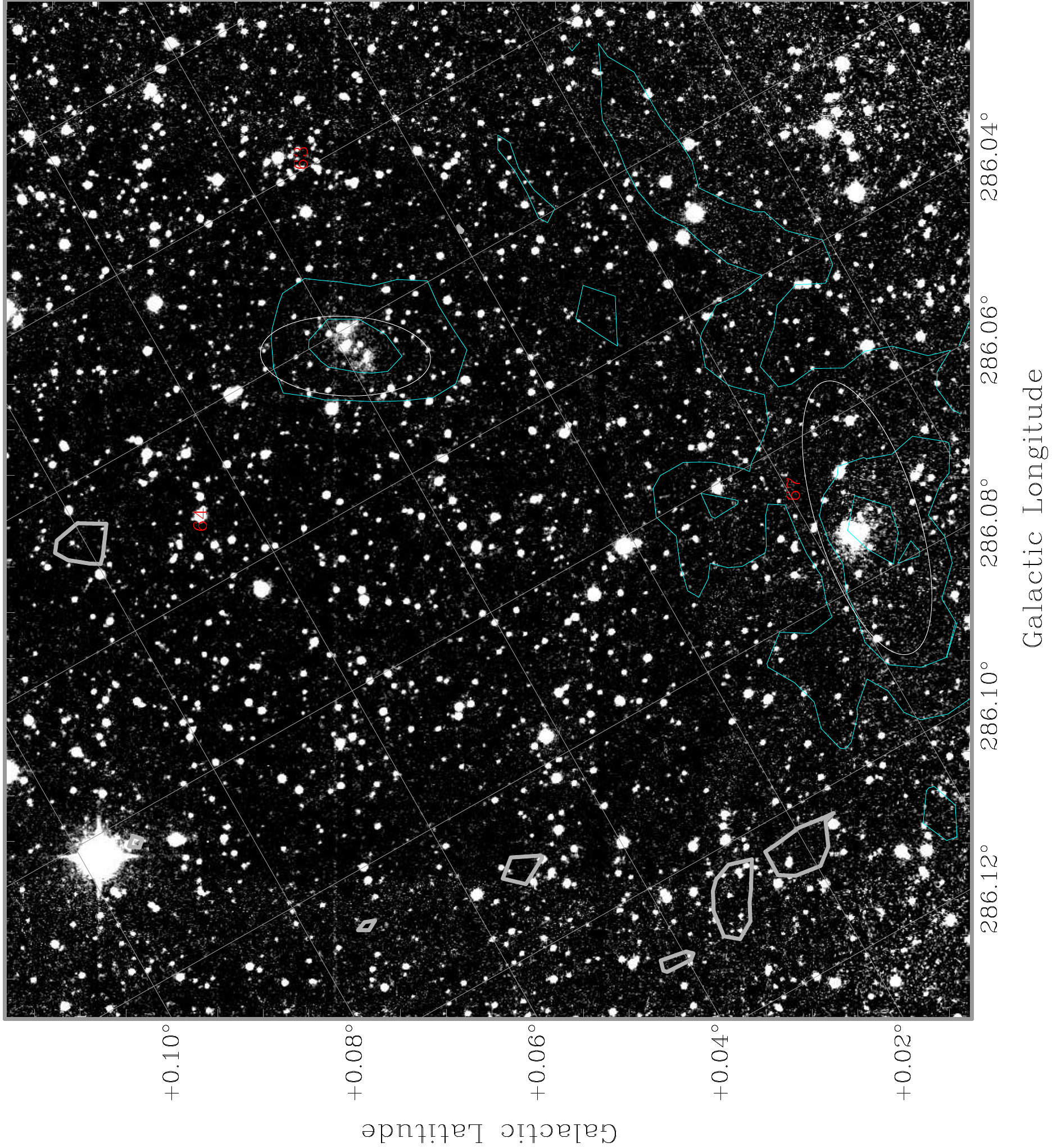}\hspace{3mm}
(b)\hspace{-3mm}\includegraphics[angle=-90,scale=0.41]{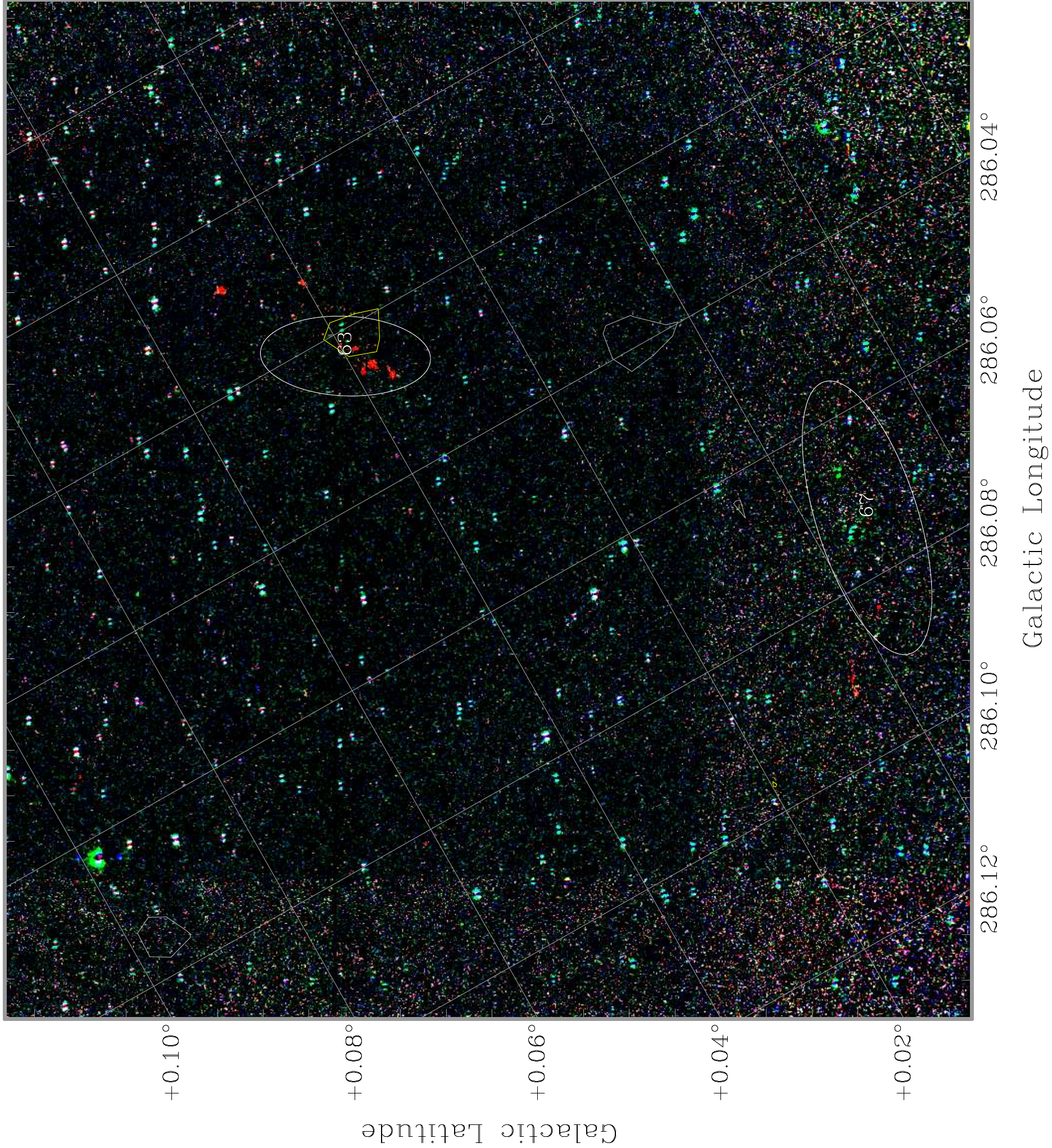}
\caption{Same as Fig.\,\ref{sample}, but for BYF\,63 and 67 (part of Region 9 from Paper I, see Fig.\,\ref{reg9}).  (a) $K$-band line-free continuum image with \hcop\ contours (heavy grey at 0 and cyan at 6, 10, and 14 times the rms level of 0.404\,K\kms).  (b) RGB-pseudo-colour image of the continuum-subtracted $K$-band spectral lines with \nnh\ contours (grey at --3 and yellow at 3 times the rms level of 0.253\,K\kms).  White ellipses show gaussian fits to the \hcop\ emission in both panels.  At a distance of 2.5\,kpc, the scale is 40$''$ = 0.485\,pc or 1\,pc = 0\fdg0229 = 82\farcs5.}
\label{byf63}
\end{figure*}


\begin{figure*}
(a)\hspace{-3mm}\includegraphics[angle=-90,scale=0.40]{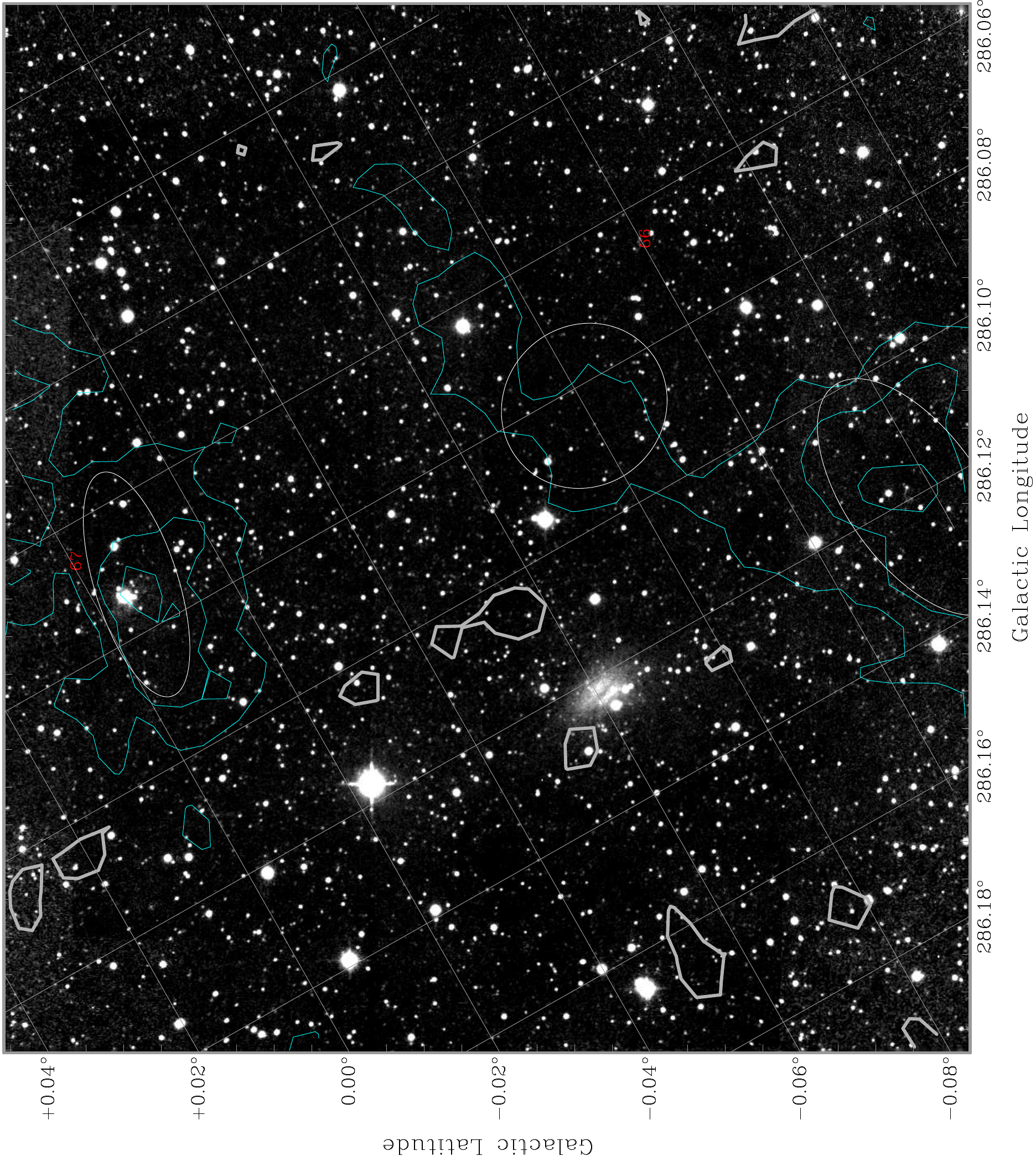}\hspace{3mm}
(b)\hspace{-3mm}\includegraphics[angle=-90,scale=0.40]{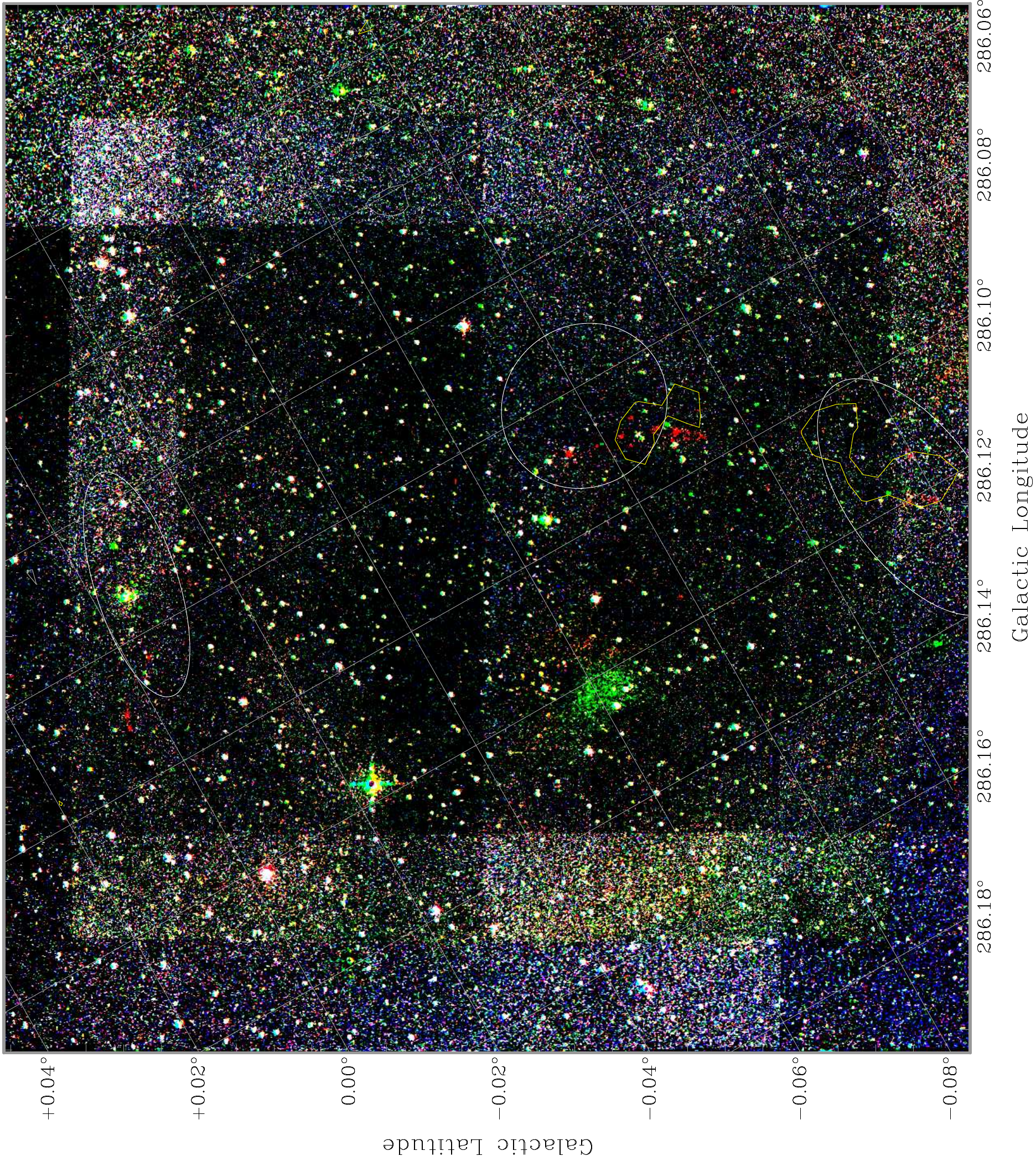}
\caption{Same as Fig.\,\ref{sample}, but for BYF\,67, 66, and 69 (part of Region 9 from Paper I, see Fig.\,\ref{reg9}).  (a) $K$-band line-free continuum image with \hcop\ contours (heavy grey at 0 and cyan at 6, 10, and 14 times the rms level of 0.404\,K\kms).  (b) RGB-pseudo-colour image of the continuum-subtracted $K$-band spectral lines with \nnh\ contours (grey at --3 and yellow at 3 times the rms level of 0.253\,K\kms).  White ellipses show gaussian fits to the \hcop\ emission in both panels.  At a distance of 2.5\,kpc, the scale is 40$''$ = 0.485\,pc or 1\,pc = 0\fdg0229 = 82\farcs5.  Note the $\sim$arcmin-sized background galaxy at $l$ = 286\fdg12, $b$ = --0\fdg16.}
\label{byf67}
\end{figure*}

\begin{figure*}
(a)\hspace{-3mm}\includegraphics[angle=-90,scale=0.42]{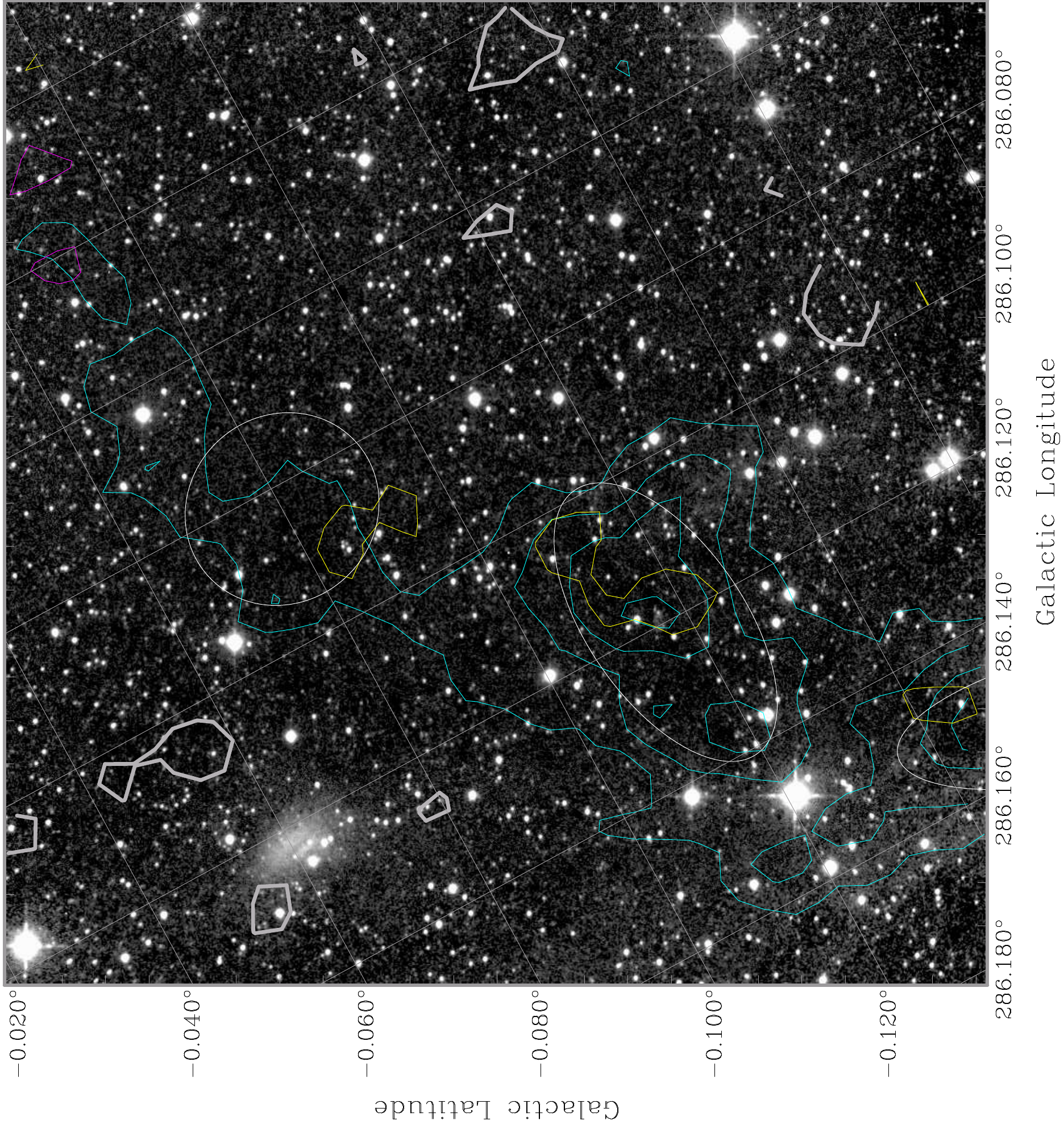}\hspace{3mm}
(b)\hspace{-3mm}\includegraphics[angle=-90,scale=0.42]{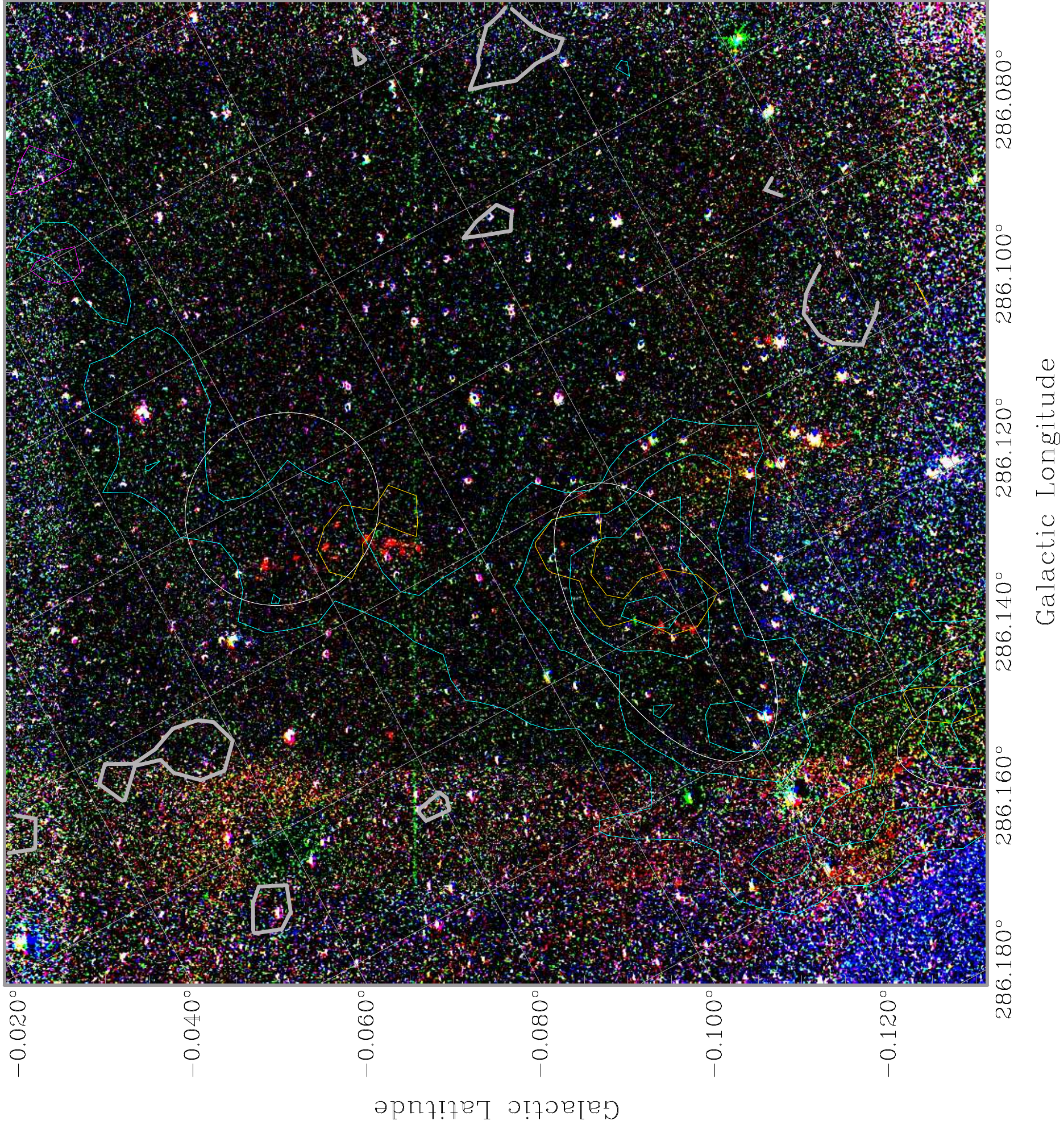}
\caption{Same as Fig.\,\ref{sample}, but for BYF\,66 and 69 (part of Region 9 from Paper I, see Fig.\,\ref{reg9}) using more sensitive 2007 data.  (a) $K$-band line-free continuum image with both \hcop\ contours (heavy grey at 0, cyan at 6, 9, 12, and 15 times the rms level of 0.404\,K\kms) and \nnh\ contours (yellow at --3, gold at +3 times the rms level of 0.253\,K\kms).  (b) RGB-pseudo-colour image of the continuum-subtracted $K$-band spectral lines with the same contours as in (a).  White ellipses show gaussian fits to the \hcop\ emission in both panels.  At a distance of 2.5\,kpc, the scale is 40$''$ = 0.485\,pc or 1\,pc = 0\fdg0229 = 82\farcs5.  Note the $\sim$arcmin-sized background galaxy at $l$ = 286\fdg12, $b$ = --0\fdg16.}
\label{byf66} 
\end{figure*}

\begin{figure*}
(a)\hspace{-3mm}\includegraphics[angle=-90,scale=0.40]{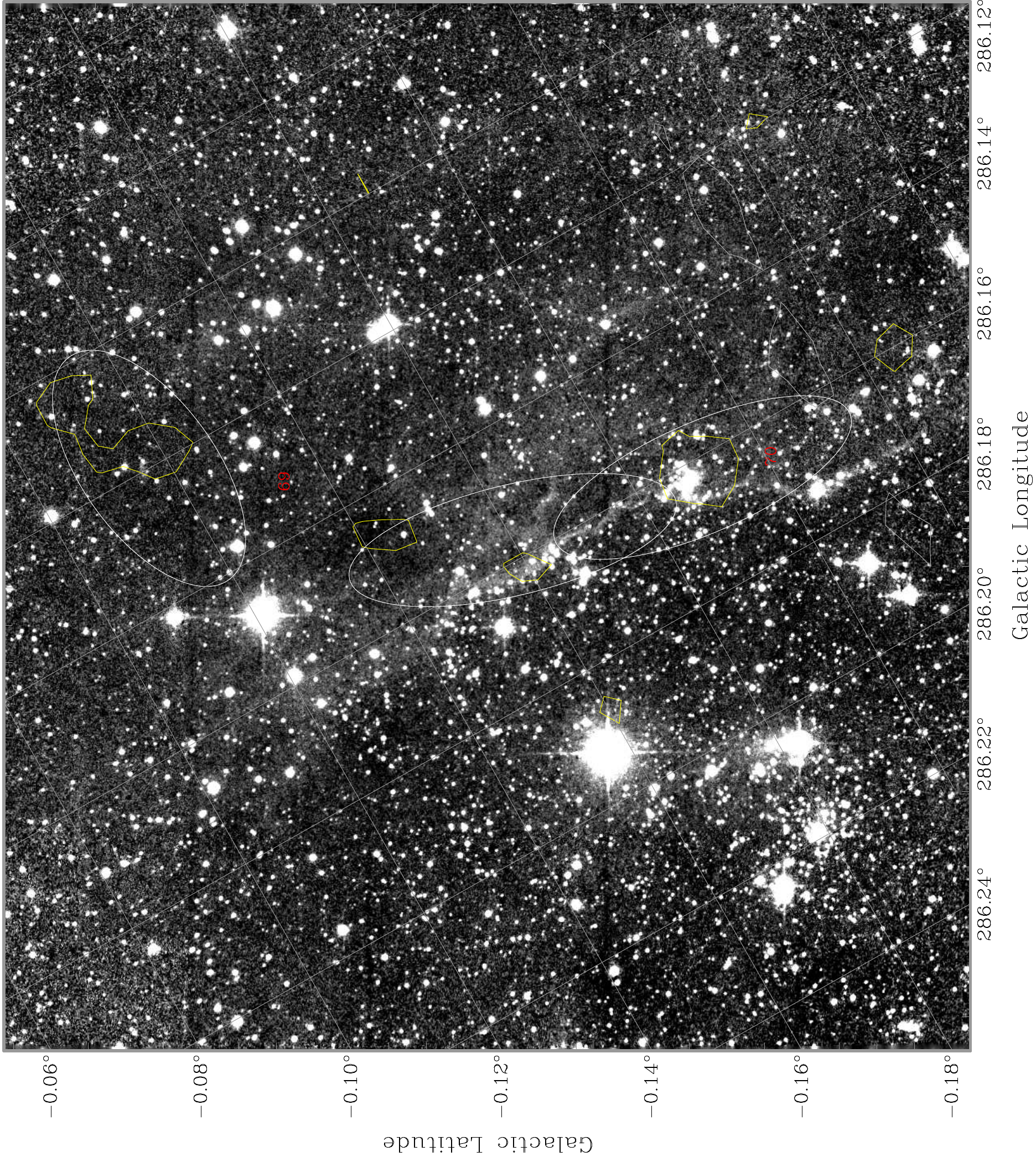}\hspace{3mm}
(b)\hspace{-3mm}\includegraphics[angle=-90,scale=0.40]{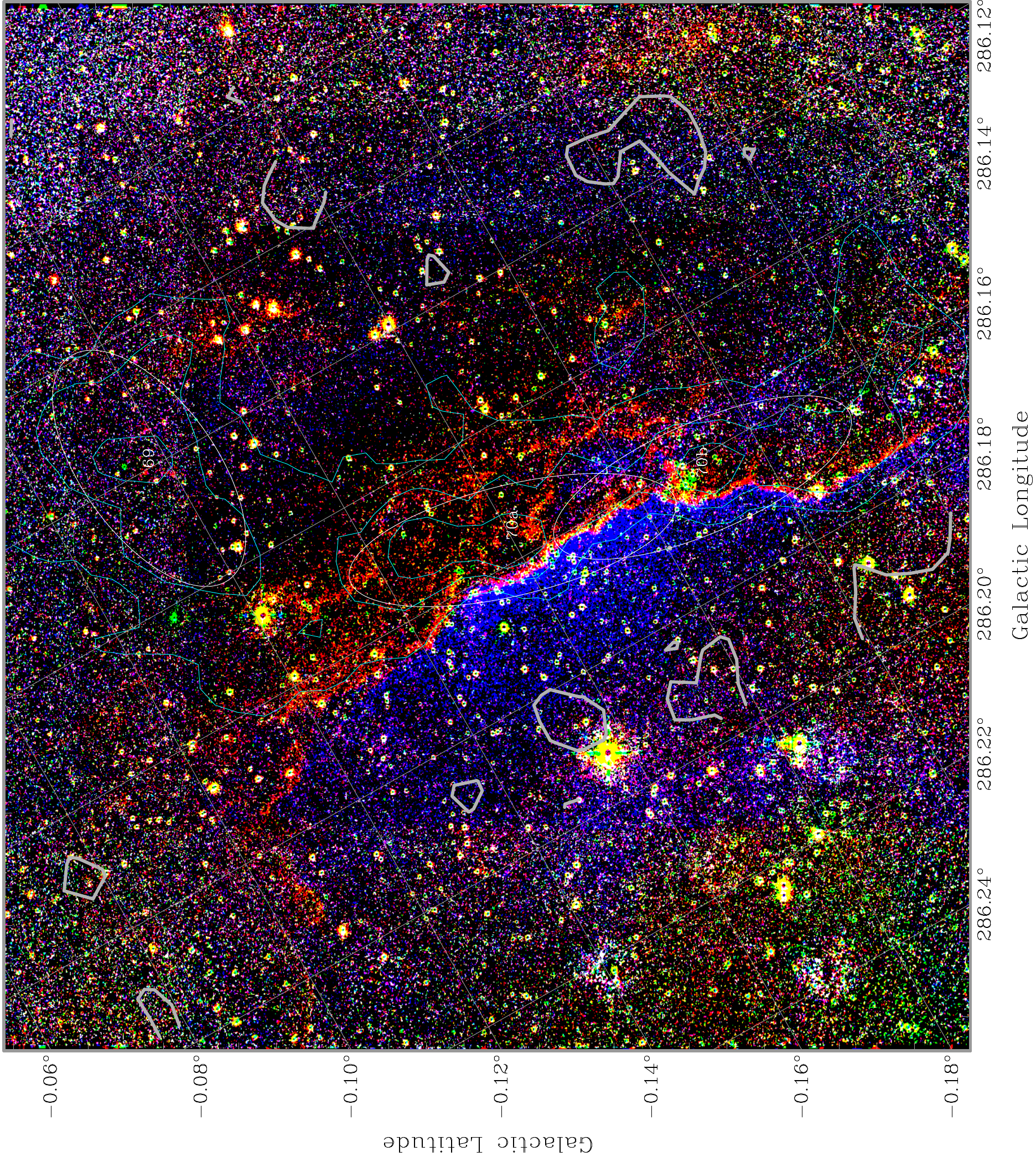}
\caption{Same as Fig.\,\ref{sample}, but for BYF\,69 and 70 (part of NGC 3324 and Region 9 from Paper I, see Fig.\,\ref{reg9}).  (a) $K$-band line-free continuum image with \nnh\ contours (grey at --3 and yellow at 3 times the rms level of 0.253\,K\kms).  (b) RGB-pseudo-colour image of the continuum-subtracted $K$-band spectral lines with \hcop\ contours (heavy grey at 0 and cyan at 6, 10, and 14 times the rms level of 0.404\,K\kms).  White ellipses show gaussian fits to the \hcop\ emission in both panels.  At a distance of 2.5\,kpc, the scale is 40$''$ = 0.485\,pc or 1\,pc = 0\fdg0229 = 82\farcs5.}
\label{byf70} 
\end{figure*}

\begin{figure*}
(a)\hspace{-3mm}\includegraphics[angle=-90,scale=0.42]{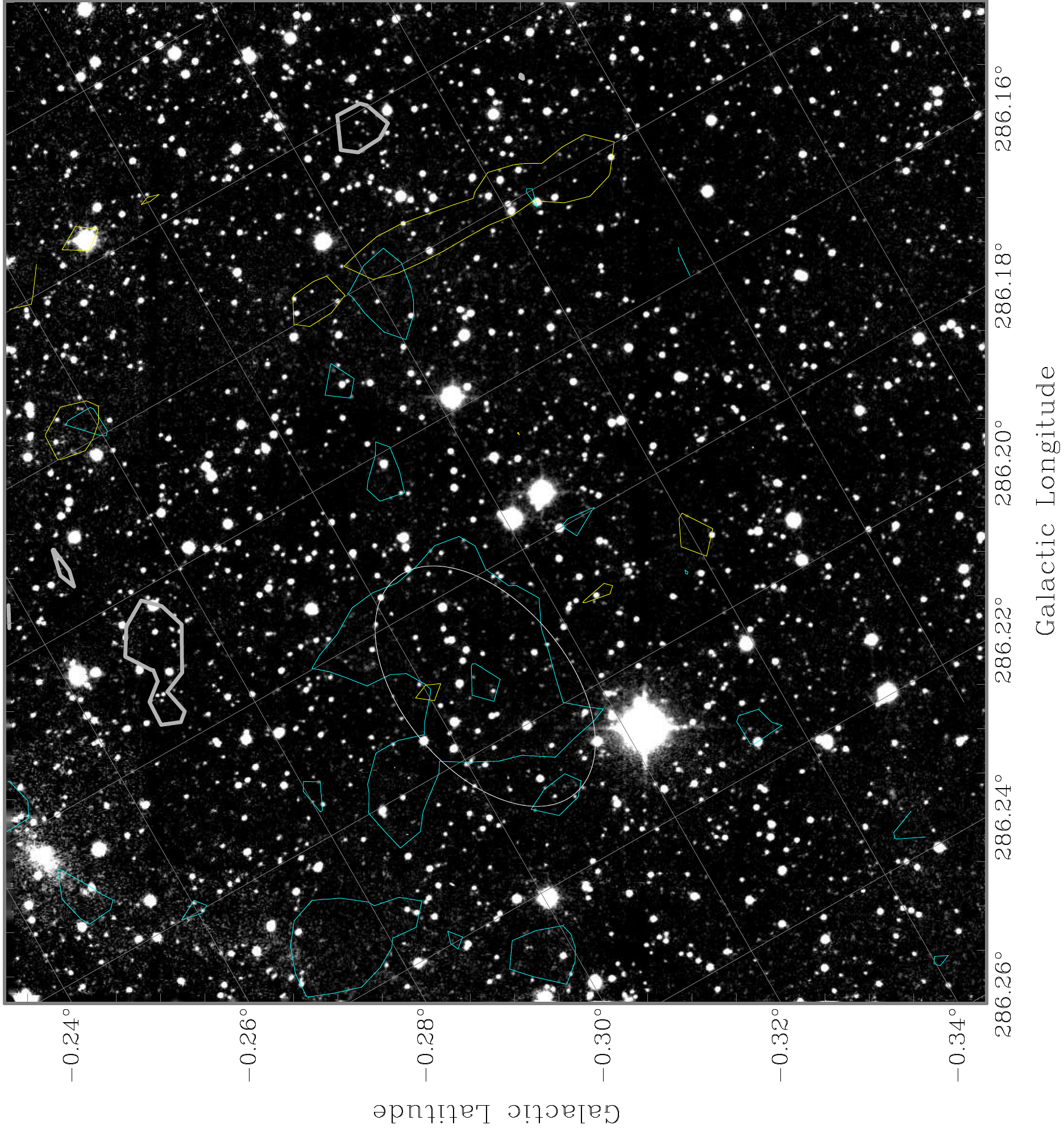}\hspace{3mm}
(b)\hspace{-3mm}\includegraphics[angle=-90,scale=0.42]{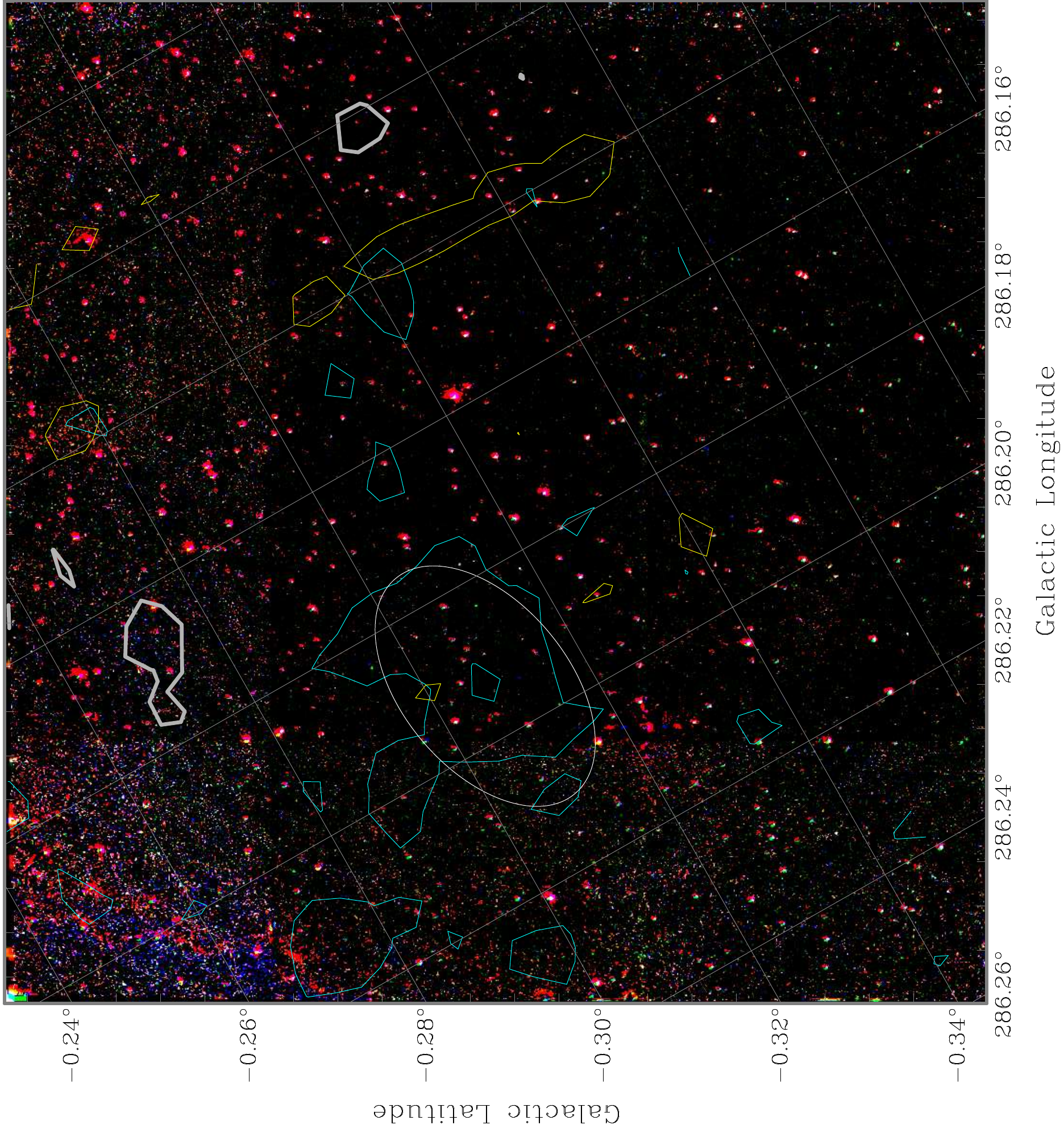}
\caption{Same as Fig.\,\ref{sample}, but for BYF\,71 (part of Region 9 from Paper I, see Fig.\,\ref{reg9}).  (a) $K$-band line-free continuum image with \hcop\ contours (heavy grey at 0, cyan at 6 and 10 times the rms level of 0.404\,K\kms) and \nnh\ contours (grey at --3 and yellow at 3 times the rms level of 0.253\,K\kms).  (b) RGB-pseudo-colour image of the continuum-subtracted $K$-band spectral lines with the same contours as in (a).  The white ellipse shows a gaussian fit to the \hcop\ emission in both panels.  At a distance of 2.5\,kpc, the scale is 40$''$ = 0.485\,pc or 1\,pc = 0\fdg0229 = 82\farcs5.}
\label{byf71}
\end{figure*}

\begin{figure*}
(a)\hspace{-3mm}\includegraphics[angle=-90,scale=0.40]{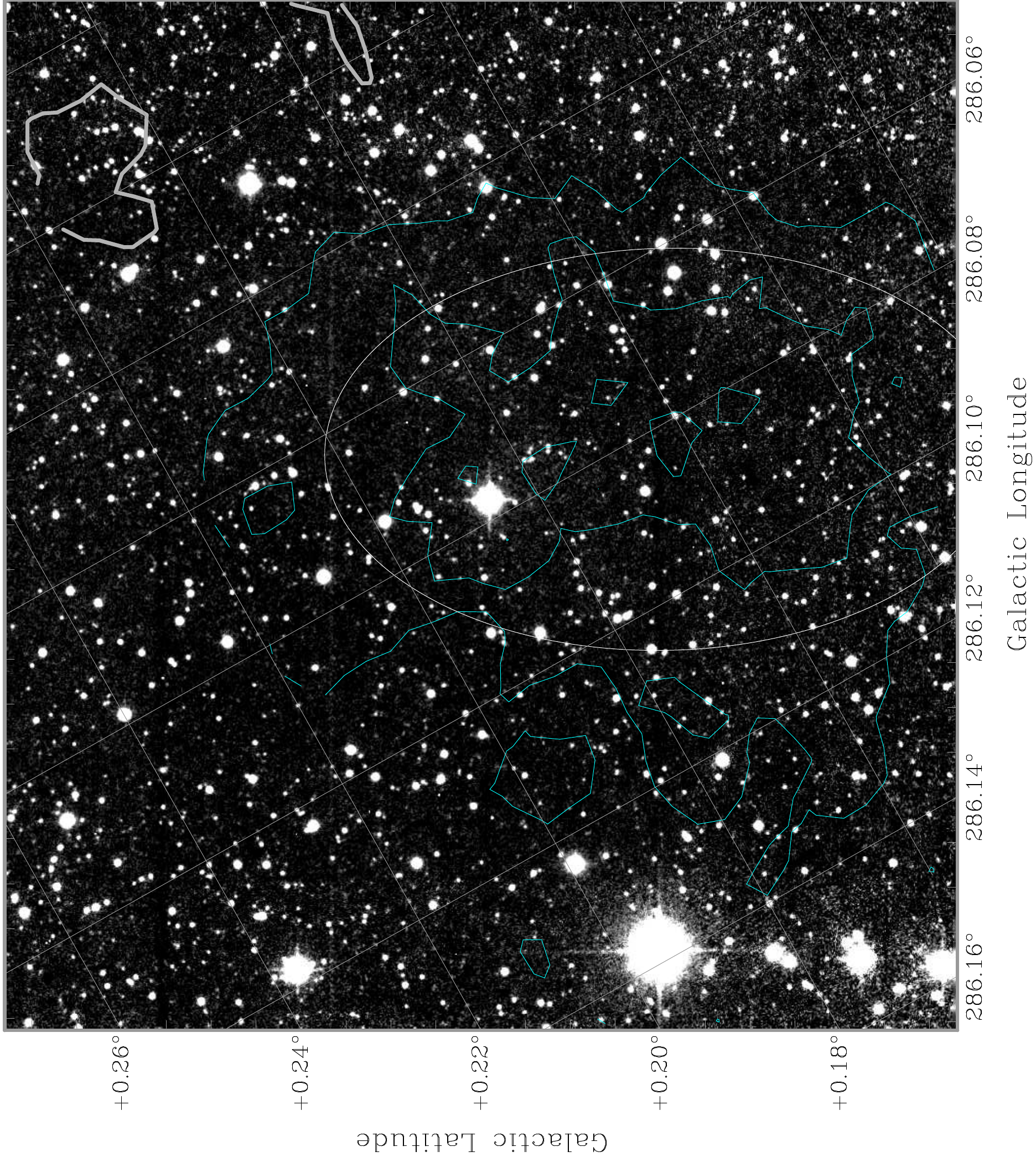}\hspace{3mm}
(b)\hspace{-3mm}\includegraphics[angle=-90,scale=0.40]{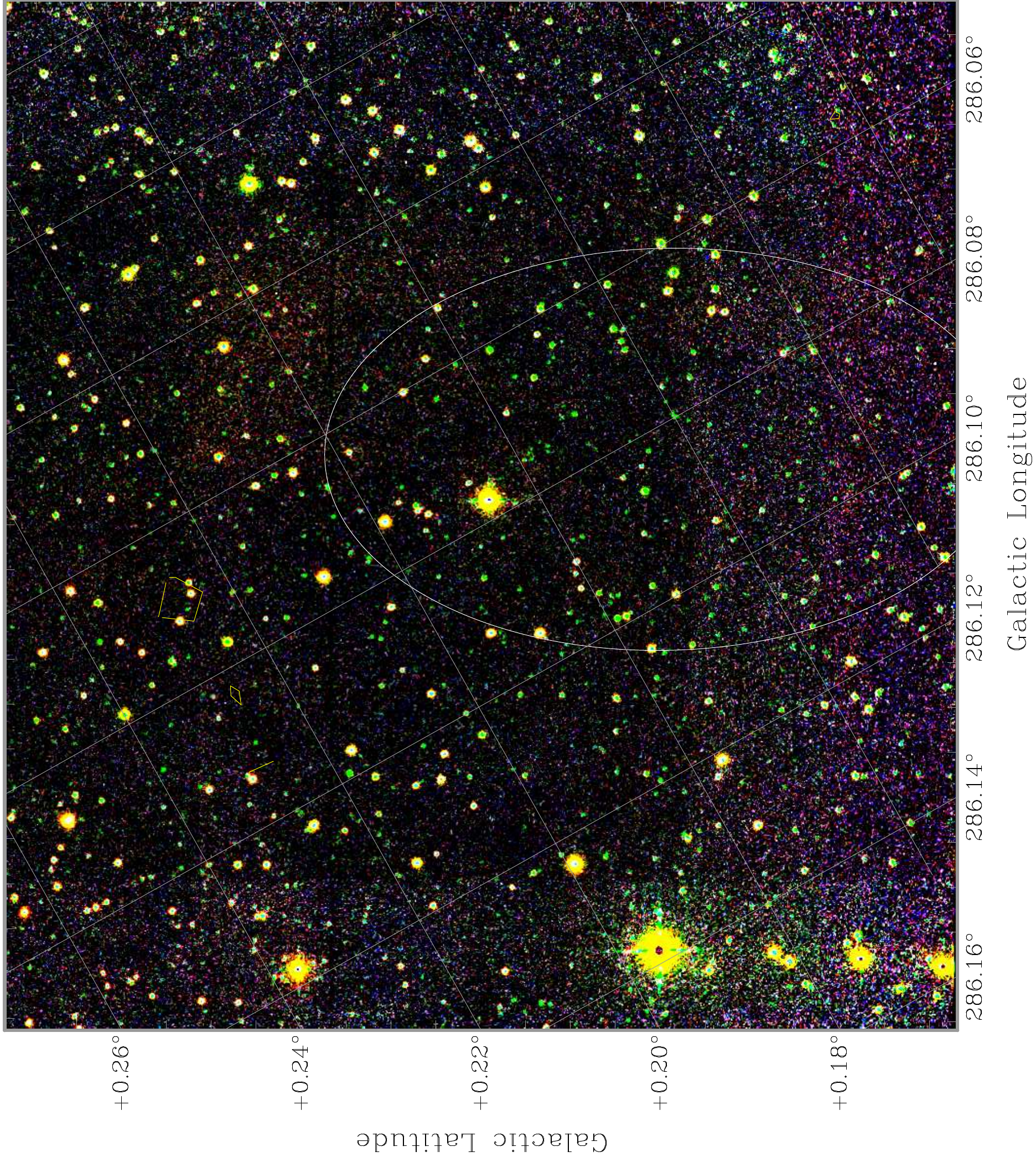}
\caption{Same as Fig.\,\ref{sample}, but for BYF\,68 (part of Region 9 from Paper I, see Fig.\,\ref{reg9}).  (a) $K$-band line-free continuum image with \hcop\ contours (heavy grey at 0 and cyan at 6, 10, and 14 times the rms level of 0.404\,K\kms).  (b) RGB-pseudo-colour image of the continuum-subtracted $K$-band spectral lines with \nnh\ contours (grey at --3 and yellow at 3 times the rms level of 0.253\,K\kms).  The white ellipse shows a gaussian fit to the \hcop\ emission in both panels.  At a distance of 2.5\,kpc, the scale is 40$''$ = 0.485\,pc or 1\,pc = 0\fdg0229 = 82\farcs5.}
\label{byf68}
\end{figure*}


\clearpage

\begin{figure*}
(a)\hspace{-3mm}\includegraphics[angle=-90,scale=0.40]{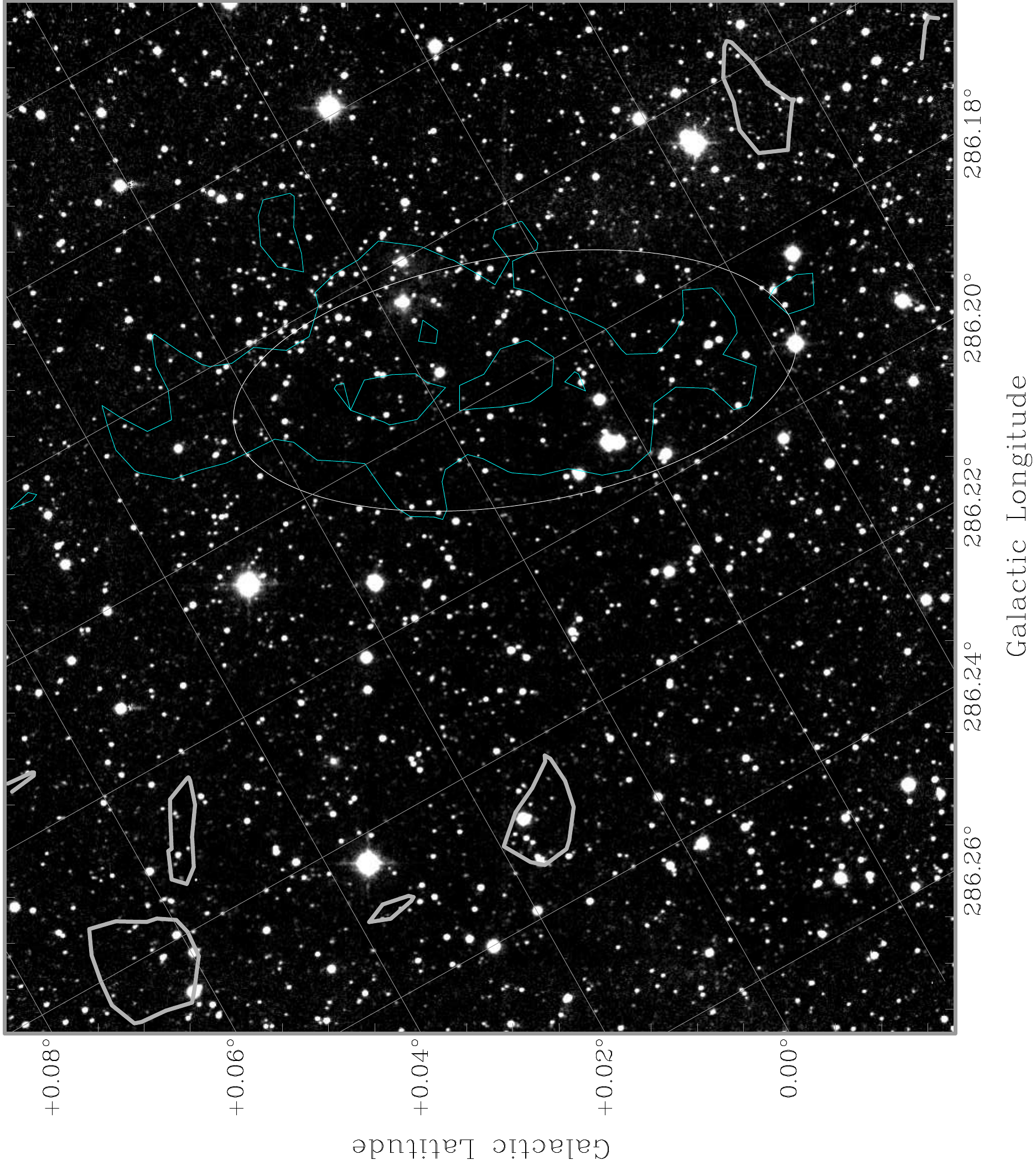}\hspace{3mm}
(b)\hspace{-3mm}\includegraphics[angle=-90,scale=0.40]{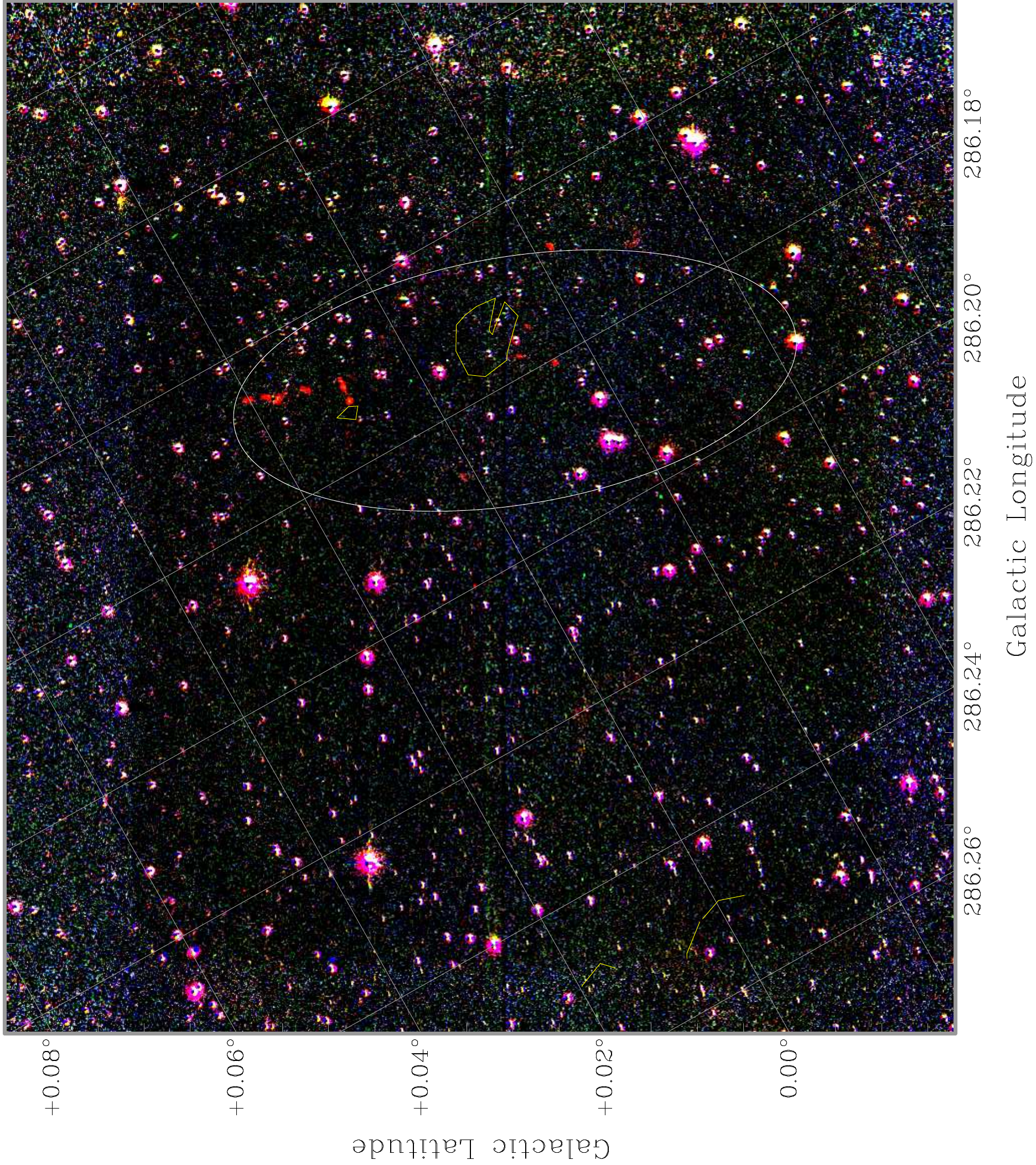}
\caption{Same as Fig.\,\ref{sample}, but for BYF\,72 (part of Region 9 from Paper I, see Fig.\,\ref{reg9}).  (a) $K$-band line-free continuum image with \hcop\ contours (heavy grey at 0 and cyan at 6, 10, and 14 times the rms level of 0.404\,K\kms).  (b) RGB-pseudo-colour image of the continuum-subtracted $K$-band spectral lines with \nnh\ contours (grey at --3 and yellow at 3 times the rms level of 0.253\,K\kms).  The white ellipse shows a gaussian fit to the \hcop\ emission in both panels.  At a distance of 2.5\,kpc, the scale is 40$''$ = 0.485\,pc or 1\,pc = 0\fdg0229 = 82\farcs5.}
\label{byf72}
\end{figure*}

\begin{figure*}
(a)\hspace{-3mm}\includegraphics[angle=-90,scale=0.41]{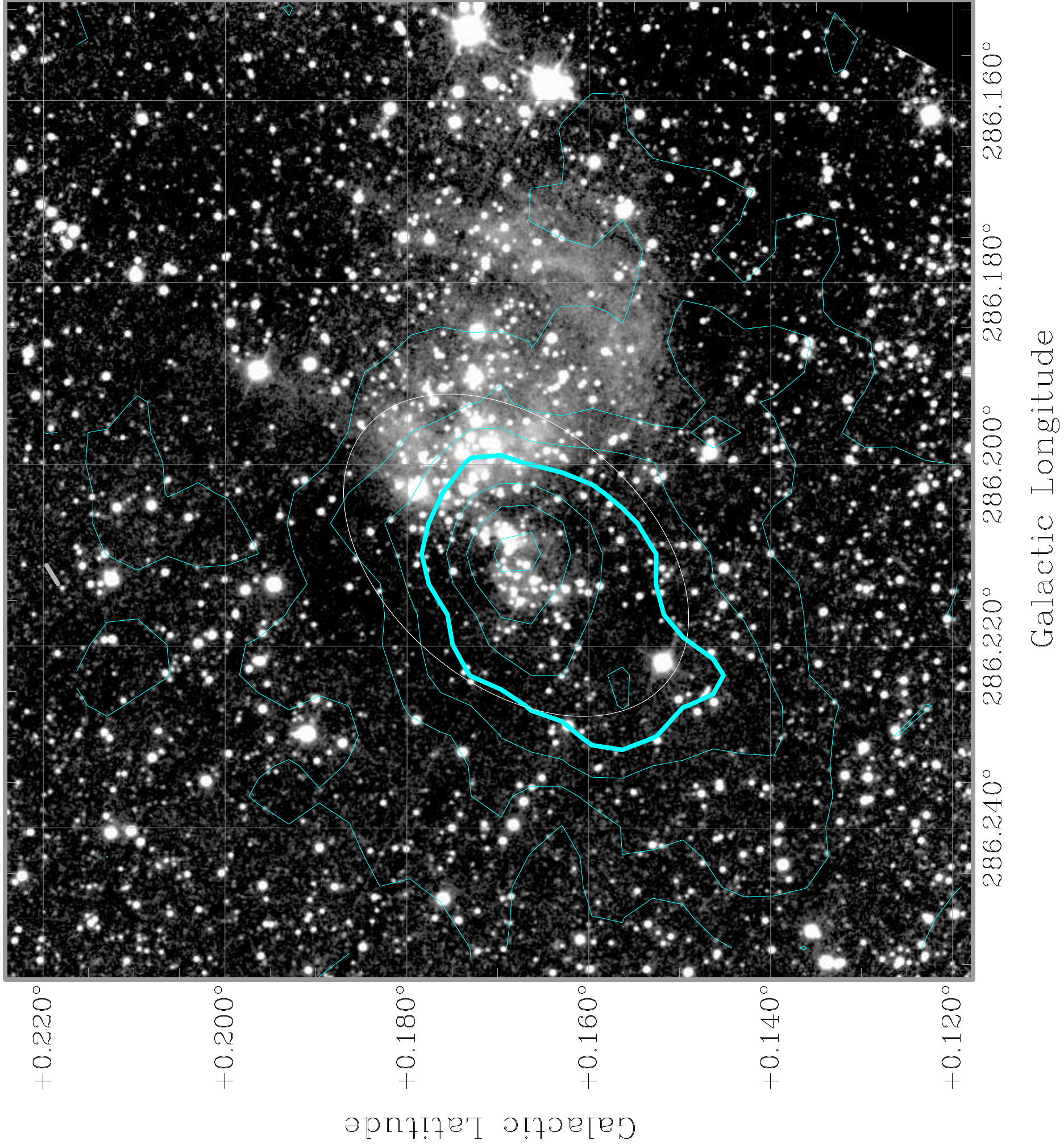}\hspace{3mm}
(b)\hspace{-3mm}\includegraphics[angle=-90,scale=0.41]{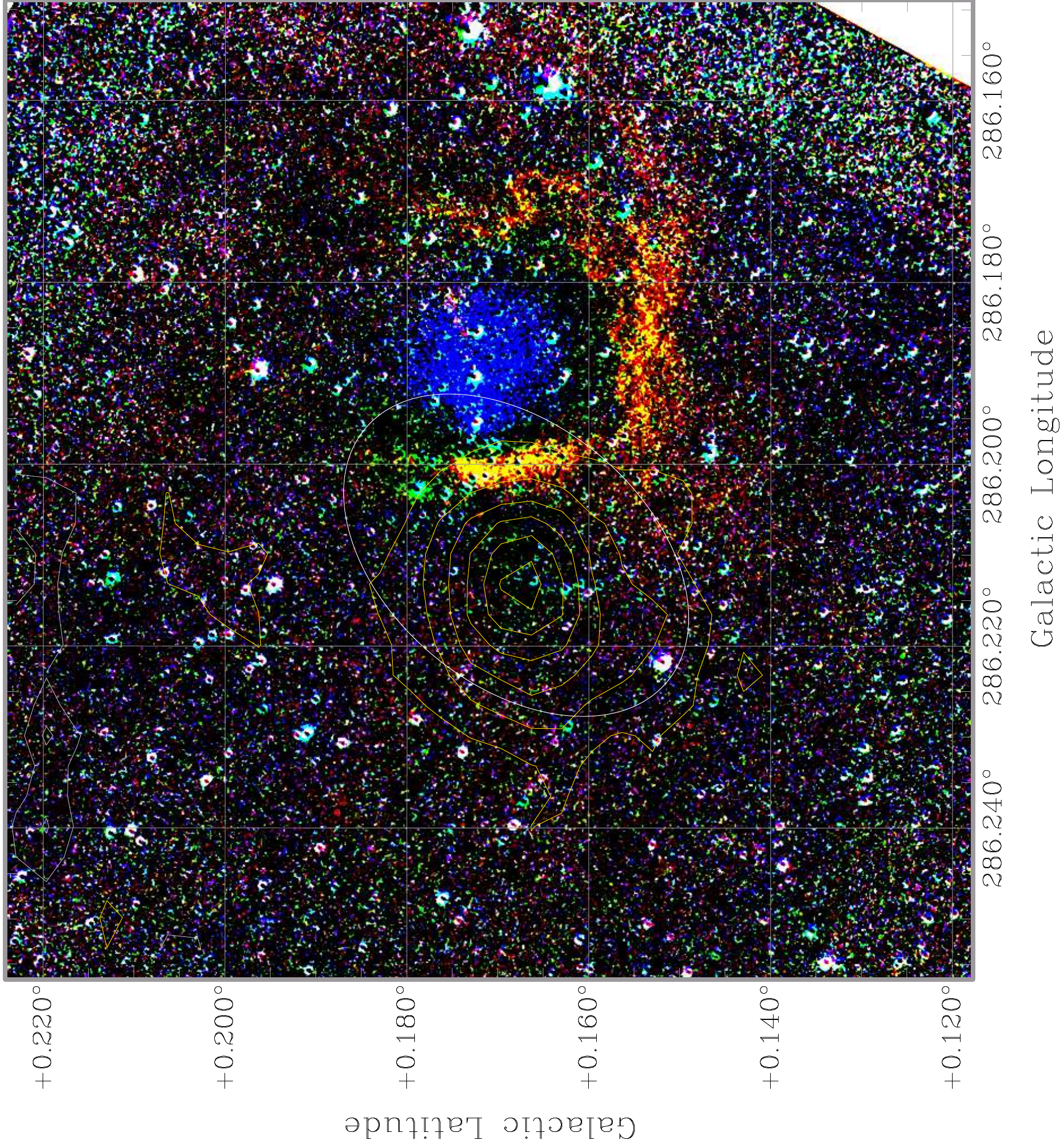}
\caption{Same as Fig.\,\ref{sample}, but for BYF\,73 (part of Region 9 from Paper I, see Fig.\,\ref{reg9}) using more sensitive 2006 data.  (a) $K$-band line-free continuum image with \hcop\ contours (heavy grey at 0, cyan at 6, 9, 12, 15(heavy), 19, 22.5, and 26 times the rms level of 0.404\,K\kms).  (b) RGB-pseudo-colour image of the continuum-subtracted $K$-band spectral lines with \nnh\ contours (grey at --3 and --6, yellow at 3, 6, ..., 15, and 18 times the rms level of 0.253\,K\kms).  The white ellipse shows a gaussian fit to the \hcop\ emission in both panels.  At a distance of 2.5\,kpc, the scale is 40$''$ = 0.485\,pc or 1\,pc = 0\fdg0229 = 82\farcs5.  Note that Fig.\,7a in \citet{BYR10} has the colour assignments of the \htwo\ lines switched; this error is corrected in (b).
\label{byf73}}
\end{figure*}

\begin{figure*}
(a)\hspace{-3mm}\includegraphics[angle=-90,scale=0.42]{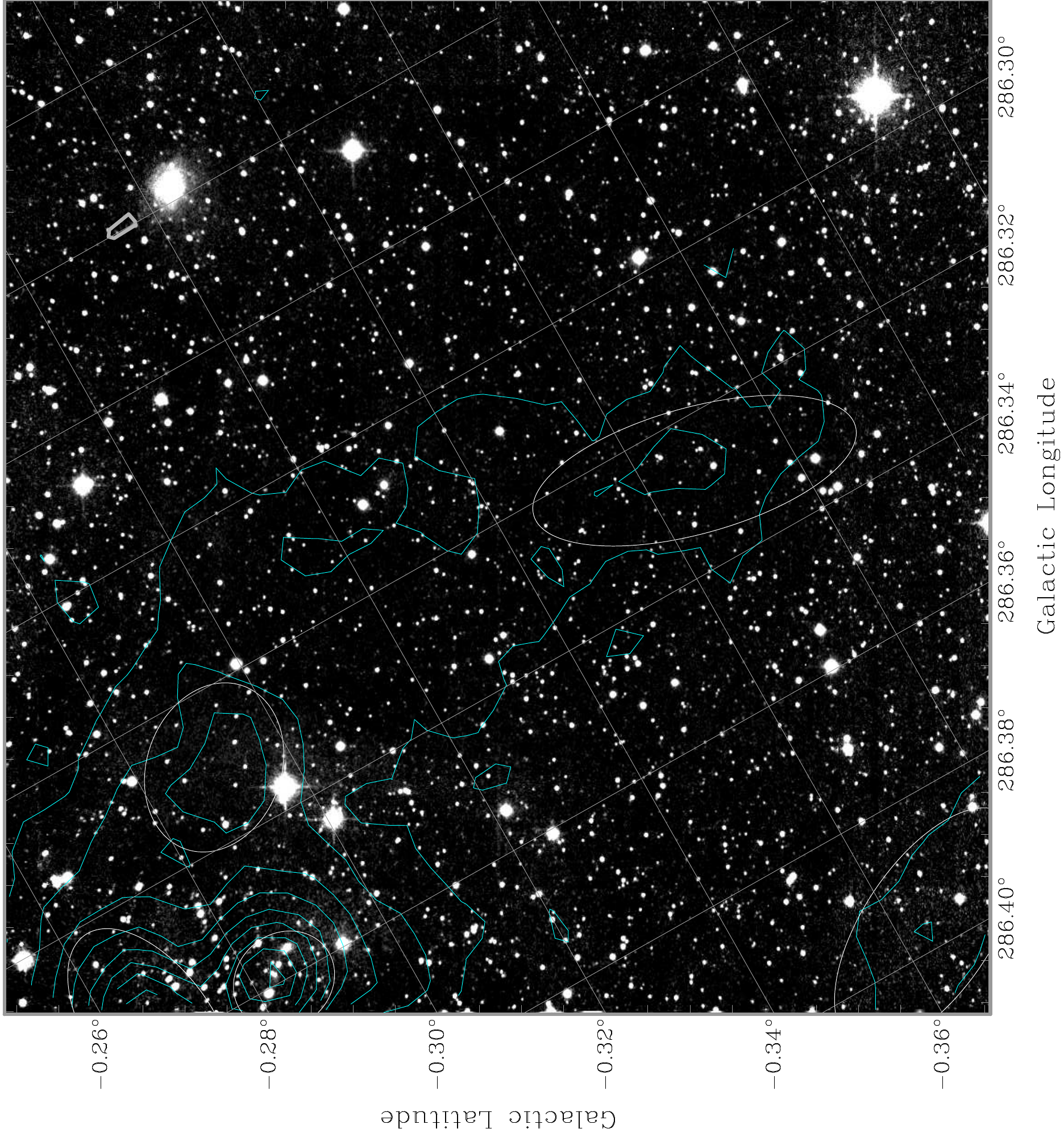}\hspace{3mm}
(b)\hspace{-3mm}\includegraphics[angle=-90,scale=0.42]{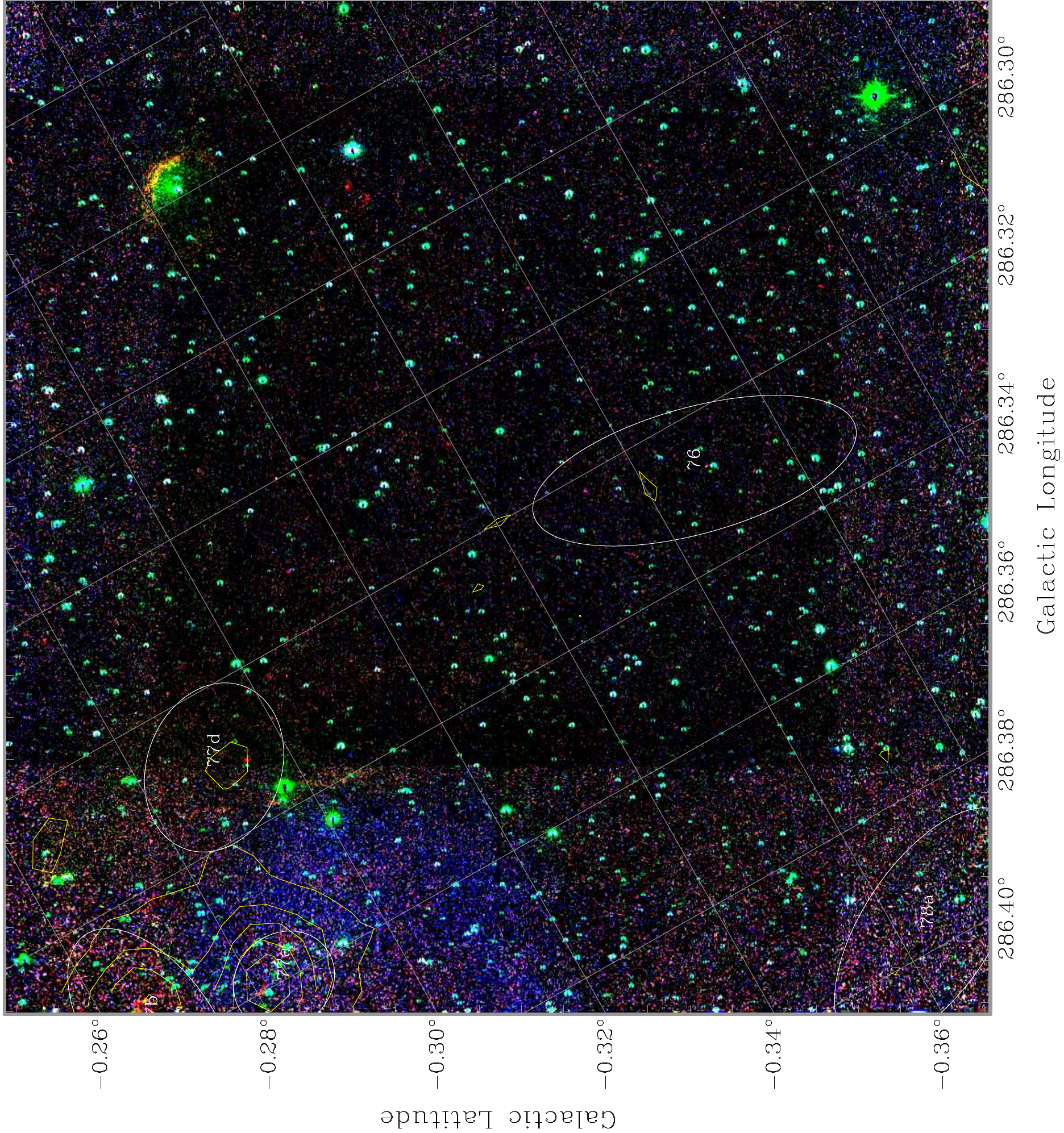}
\caption{Same as Fig.\,\ref{sample}, but for BYF\,76--78 (part of Region 9 from Paper I, see Fig.\,\ref{reg9}).  (a) $K$-band line-free continuum image with \hcop\ contours (heavy grey at 0 and cyan at 6, 10, and 14 times the rms level of 0.404\,K\kms).  (b) RGB-pseudo-colour image of the continuum-subtracted $K$-band spectral lines with \nnh\ contours (grey at --3 and yellow at 3 times the rms level of 0.253\,K\kms).  White ellipses show gaussian fits to the \hcop\ emission in both panels.  At a distance of 2.5\,kpc, the scale is 40$''$ = 0.485\,pc or 1\,pc = 0\fdg0229 = 82\farcs5.}
\label{byf76}
\end{figure*}

\begin{figure*}
(a)\hspace{-3mm}\includegraphics[angle=-90,scale=0.42]{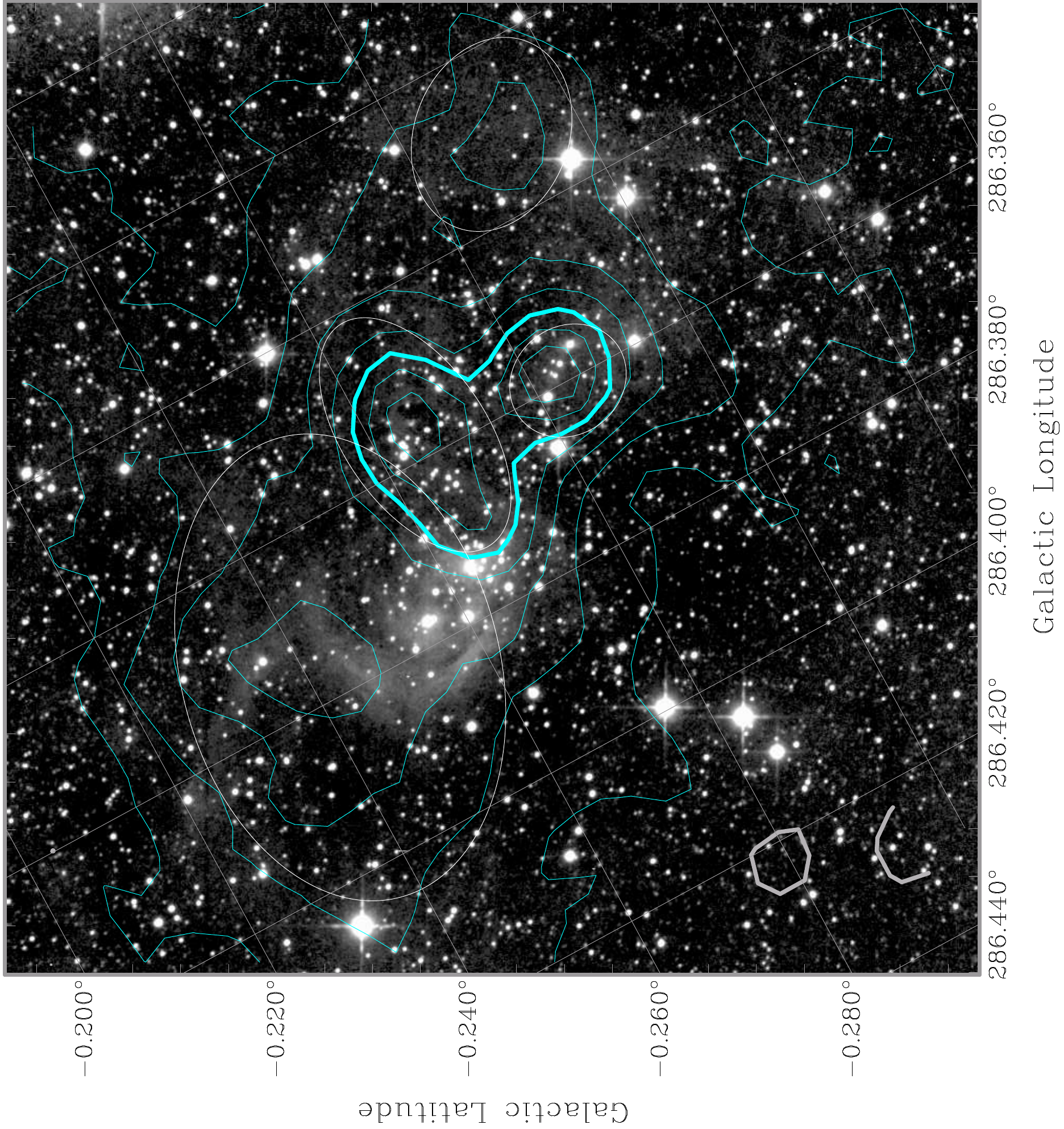}\hspace{3mm}
(b)\hspace{-3mm}\includegraphics[angle=-90,scale=0.42]{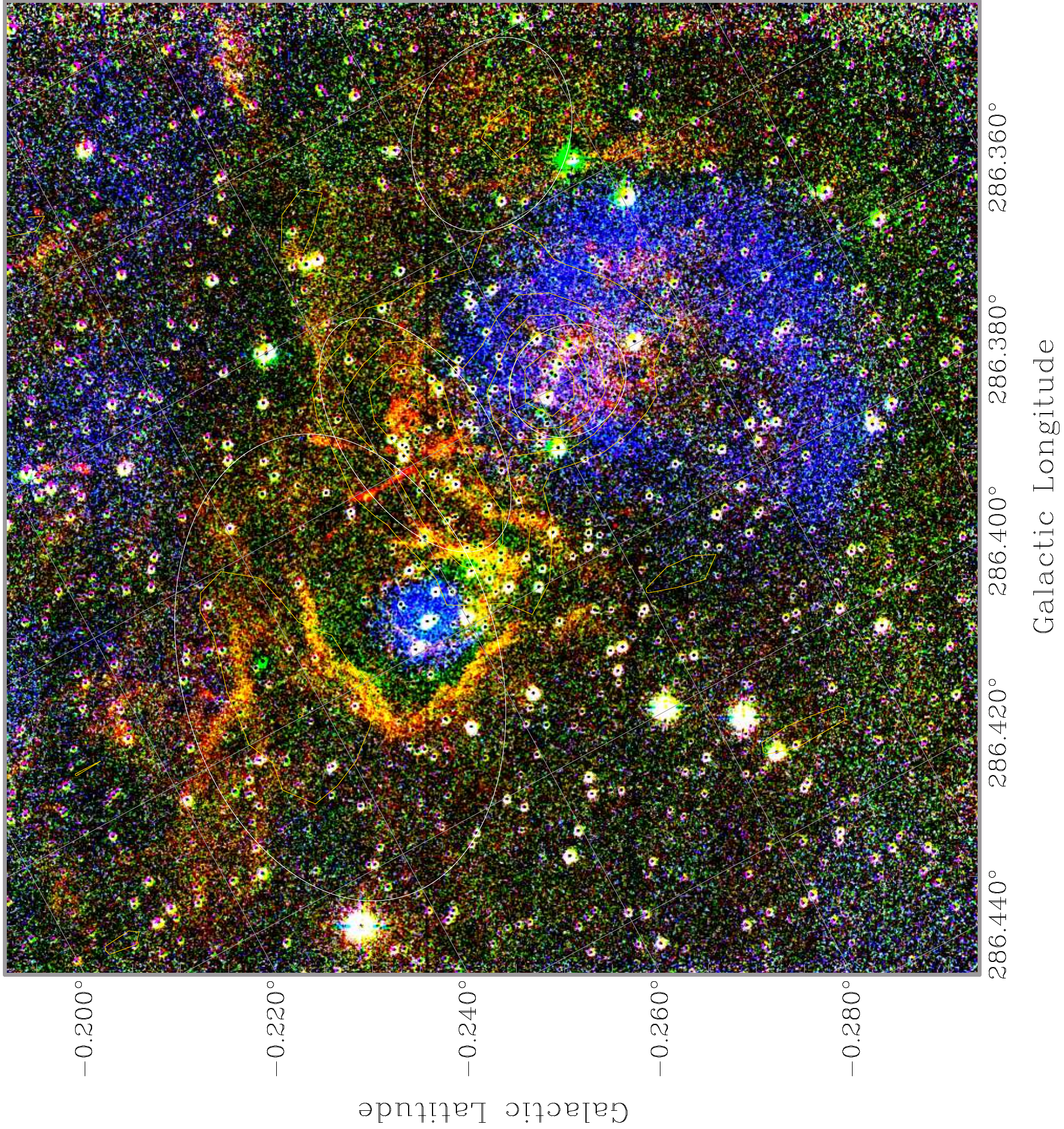}
\caption{Same as Fig.\,\ref{sample}, but for BYF\,77 (part of Region 9 from Paper I, see Fig.\,\ref{reg9}) using more sensitive 2006 data.  (a) $K$-band line-free continuum image with \hcop\ contours (heavy grey at 0, cyan at 5, 10, 15, 20, 25(heavy), 30, and 35 times the rms level of 0.404\,K\kms).  (b) RGB-pseudo-colour image of the continuum-subtracted $K$-band spectral lines with \nnh\ contours (grey at --3 and --6, yellow at 3, 6, 9, 12, and 15 times the rms level of 0.253\,K\kms).  White ellipses show gaussian fits to the \hcop\ emission in both panels.  At a distance of 2.5\,kpc, the scale is 40$''$ = 0.485\,pc or 1\,pc = 0\fdg0229 = 82\farcs5.
\label{byf77}}
\end{figure*}

\begin{figure*}
(a)\hspace{-3mm}\includegraphics[angle=-90,scale=0.43]{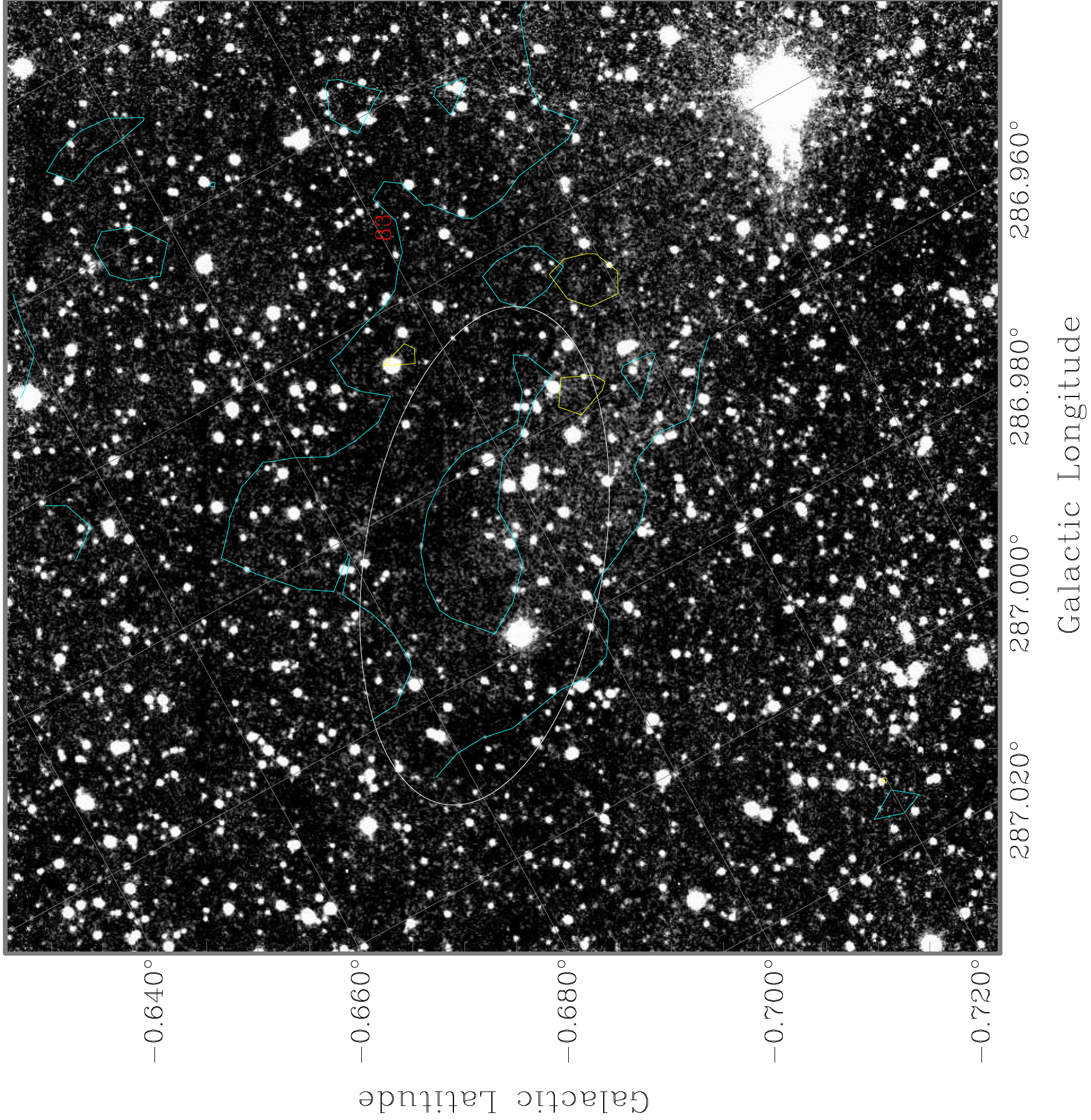}\hspace{3mm}
(b)\hspace{-3mm}\includegraphics[angle=-90,scale=0.43]{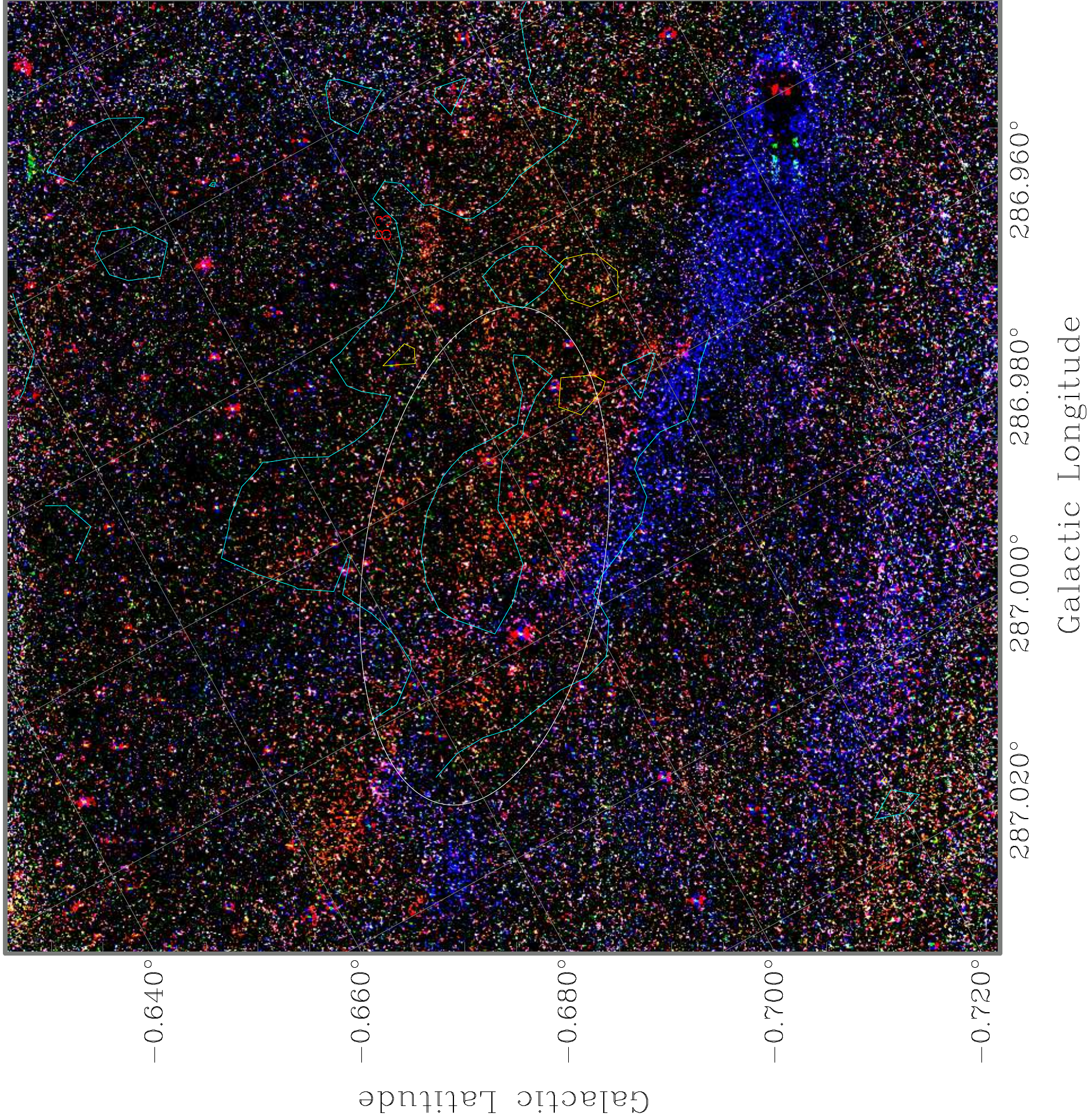}
\caption{Same as Fig.\,\ref{sample}, but for BYF\,83 (part of Region 10 from Paper I).  (a) $K$-band line-free continuum image with both \hcop\ contours (cyan at 4 and 8 times the rms level of 0.373\,K\kms) and \nnh\ contours (yellow at 3 times the rms level of 0.378\,K\kms).  (b) RGB-pseudo-colour image of the continuum-subtracted $K$-band spectral lines with the same contours as in (a).  The white ellipse shows a gaussian fit to the \hcop\ emission in both panels.  At a distance of 2.5\,kpc, the scale is 40$''$ = 0.485\,pc or 1\,pc = 0\fdg0229 = 82\farcs5.}
\label{byf83}
\end{figure*}

\begin{figure*}
(a)\hspace{-3mm}\includegraphics[angle=-90,scale=0.42]{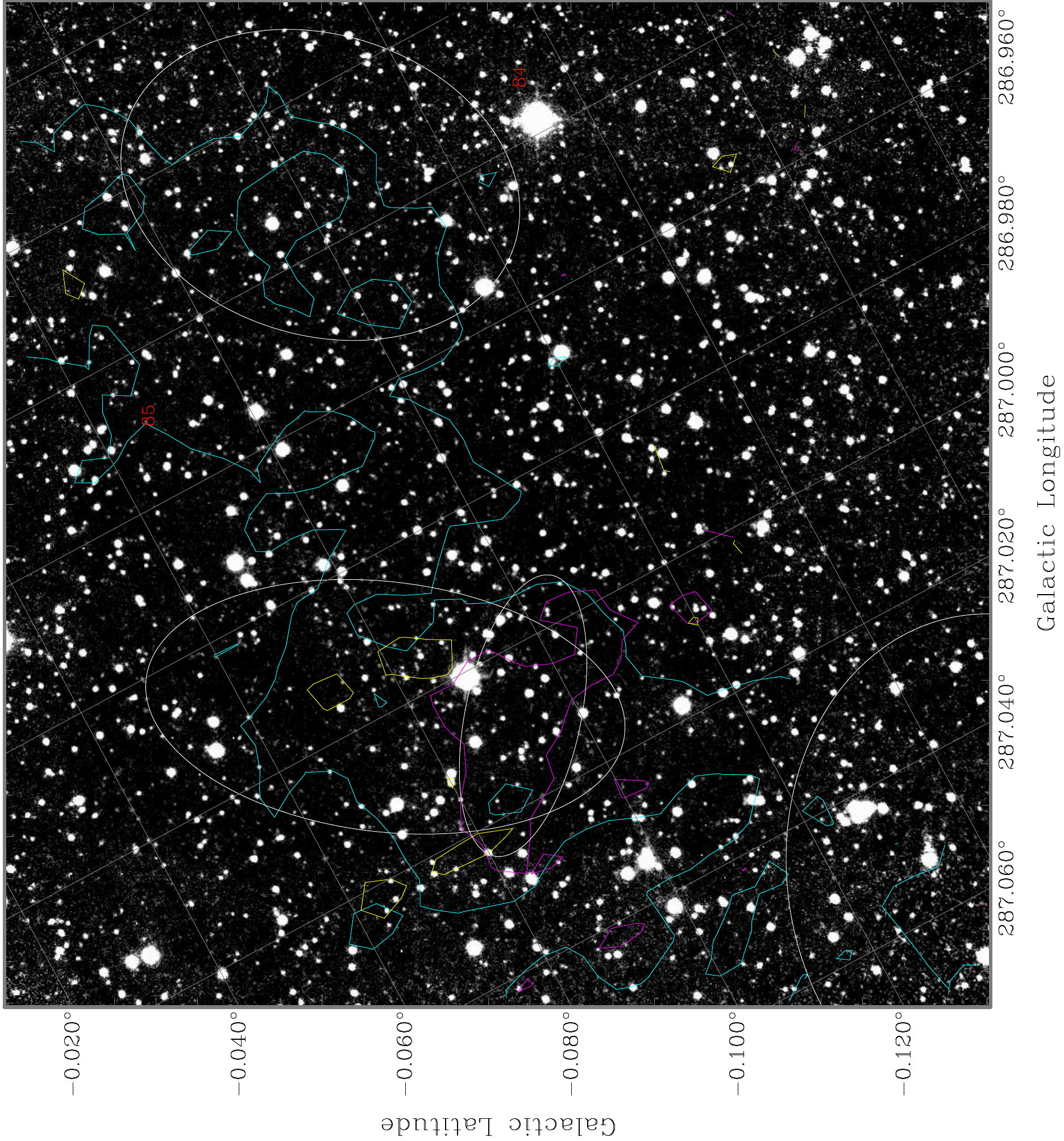}\hspace{3mm}
(b)\hspace{-3mm}\includegraphics[angle=-90,scale=0.42]{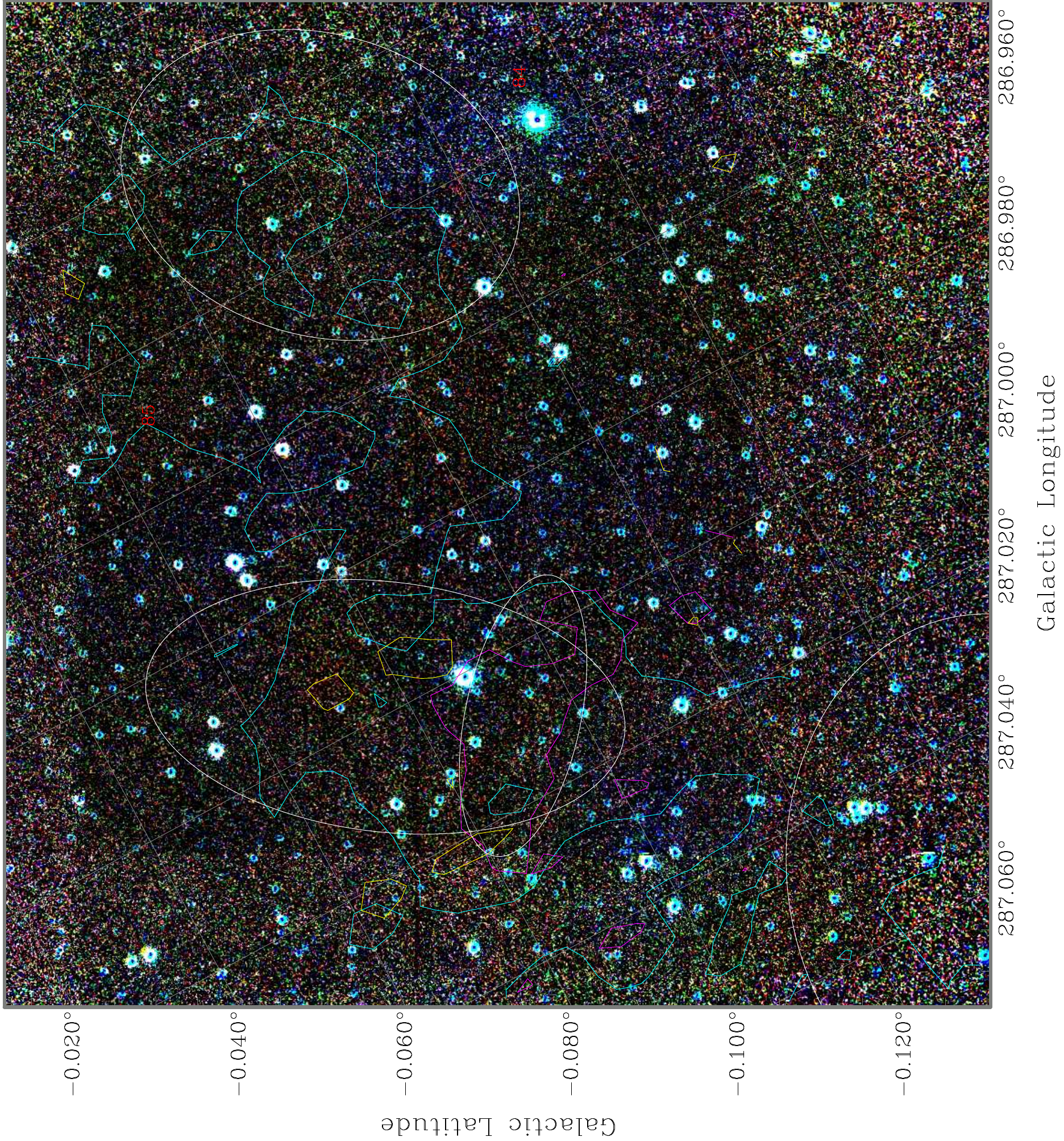}
\caption{Same as Fig.\,\ref{sample}, but for BYF\,85 (part of Region 10 from Paper I).  (a) $K$-band line-free continuum image.  In this field we show three sets of contours since there are two velocity components visible in the \hcop\ data.  At \vlsr = --20\kms\ are cyan \hcop\ contours (at 4 and 8 times the rms level of 0.329\,K\kms) and yellow \nnh\ contours (at 3 times the rms level of 0.332\,K\kms).  At \vlsr = --9\kms\ are magenta \hcop\ contours (at 3 times the rms level of 0.313\,K\kms).  (b) RGB-pseudo-colour image of the continuum-subtracted $K$-band spectral lines with the same contours as in (a).  White ellipses show gaussian fits to the \hcop\ emission in both panels.  At a distance of 2.5\,kpc, the scale is 40$''$ = 0.485\,pc or 1\,pc = 0\fdg0229 = 82\farcs5.}
\label{byf85}
\end{figure*}


\begin{figure*}
(a)\hspace{-3mm}\includegraphics[angle=-90,scale=0.41]{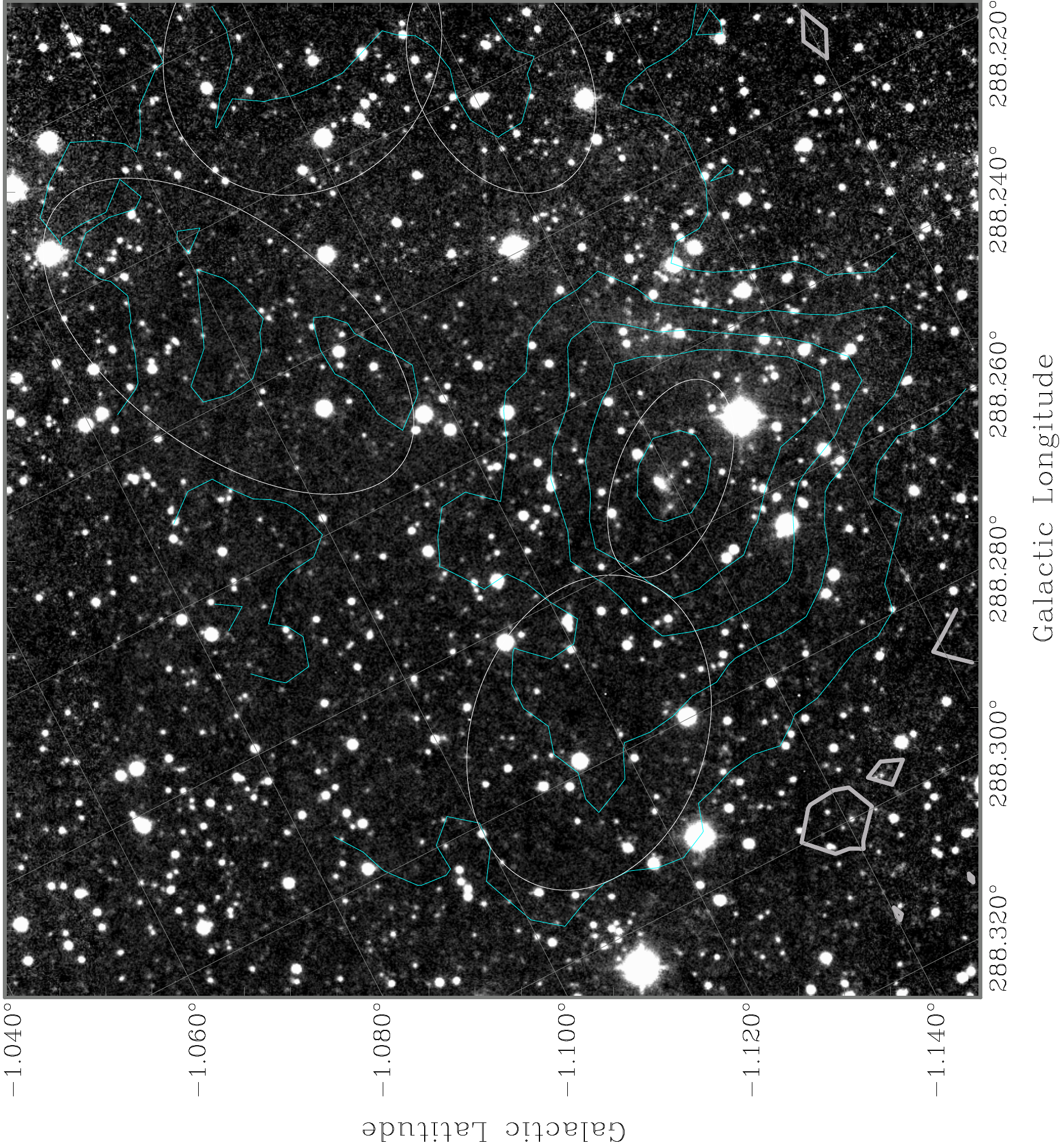}\hspace{3mm}
(b)\hspace{-3mm}\includegraphics[angle=-90,scale=0.41]{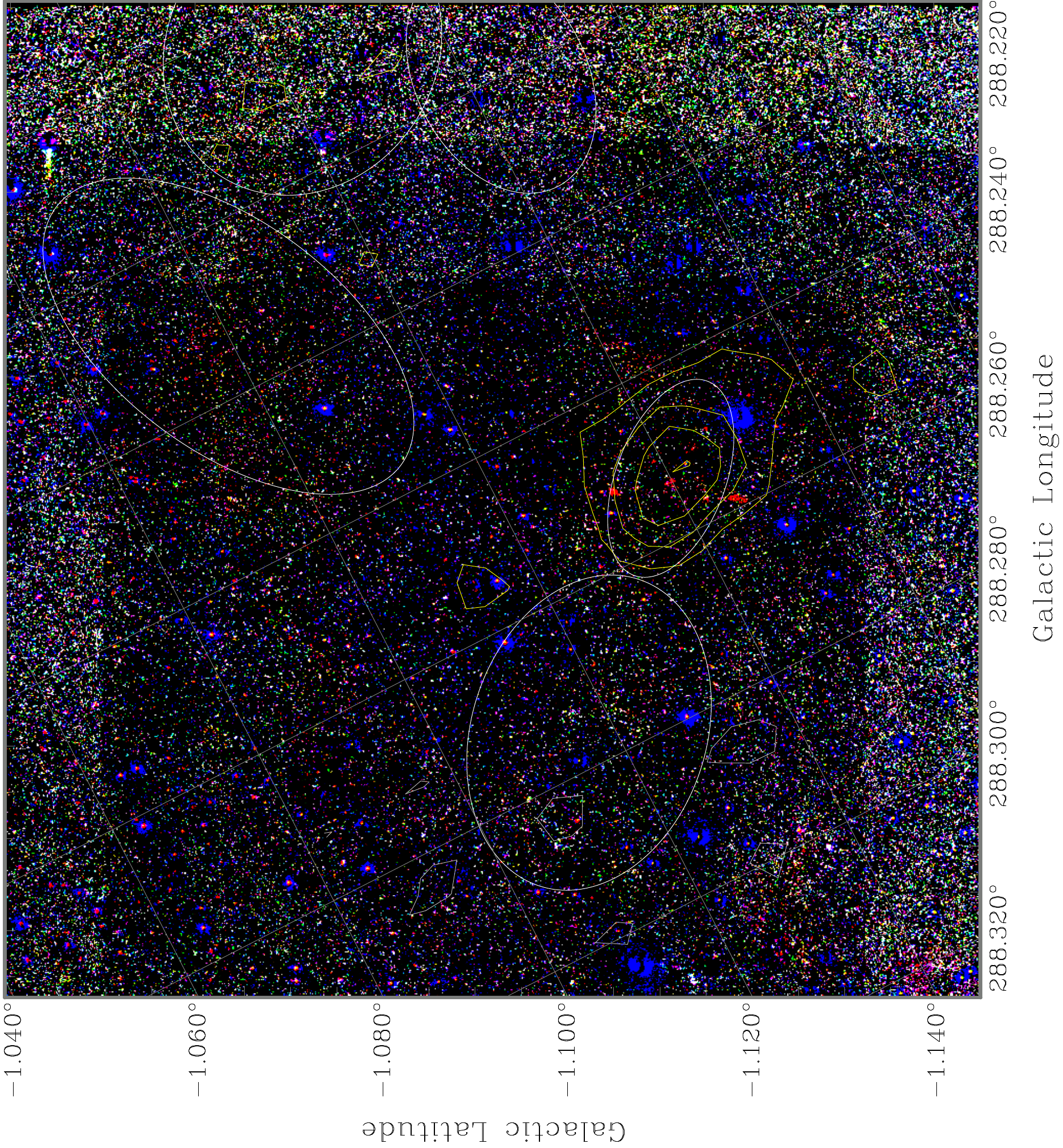}
\caption{Same as Fig.\,\ref{sample}, but for BYF\,118 (part of Region 11 from Paper I).  (a) $K$-band line-free continuum image with \hcop\ contours (heavy grey at 0, cyan at 5, 10 15, 20, and 40 times the rms level of 0.237\,K\kms).  (b) RGB-pseudo-colour image of the continuum-subtracted $K$-band spectral lines with \nnh\ contours (grey at --3, yellow at 3, 6, 9, and 12 times the rms level of 0.217\,K\kms).  White ellipses show gaussian fits to the \hcop\ emission in both panels.  At a distance of 2.5\,kpc, the scale is 40$''$ = 0.485\,pc or 1\,pc = 0\fdg0229 = 82\farcs5.}
\label{byf118}
\end{figure*}

\begin{figure*}
(a)\hspace{-3mm}\includegraphics[angle=-90,scale=0.38]{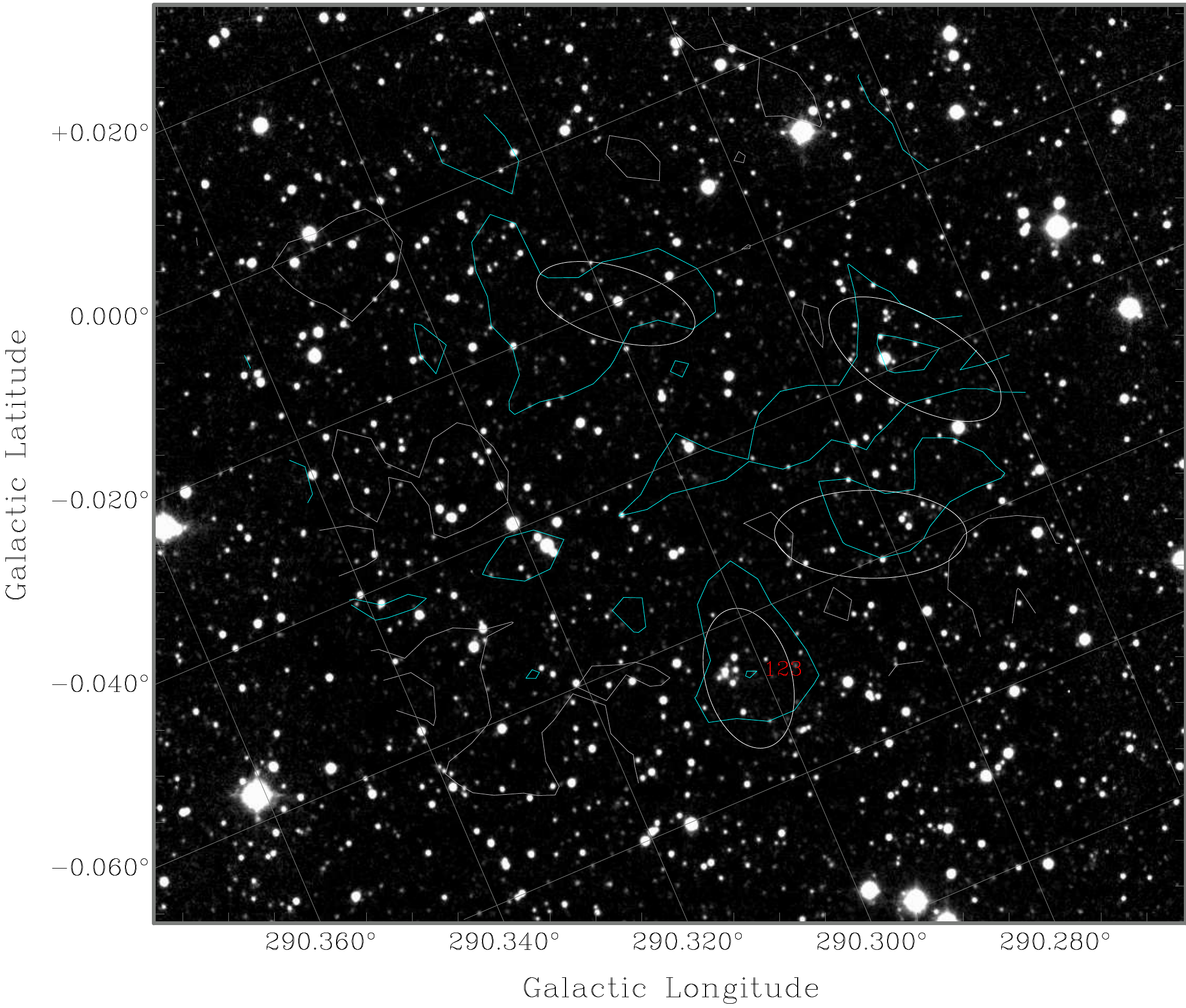}\hspace{3mm}
(b)\hspace{-3mm}\includegraphics[angle=-90,scale=0.38]{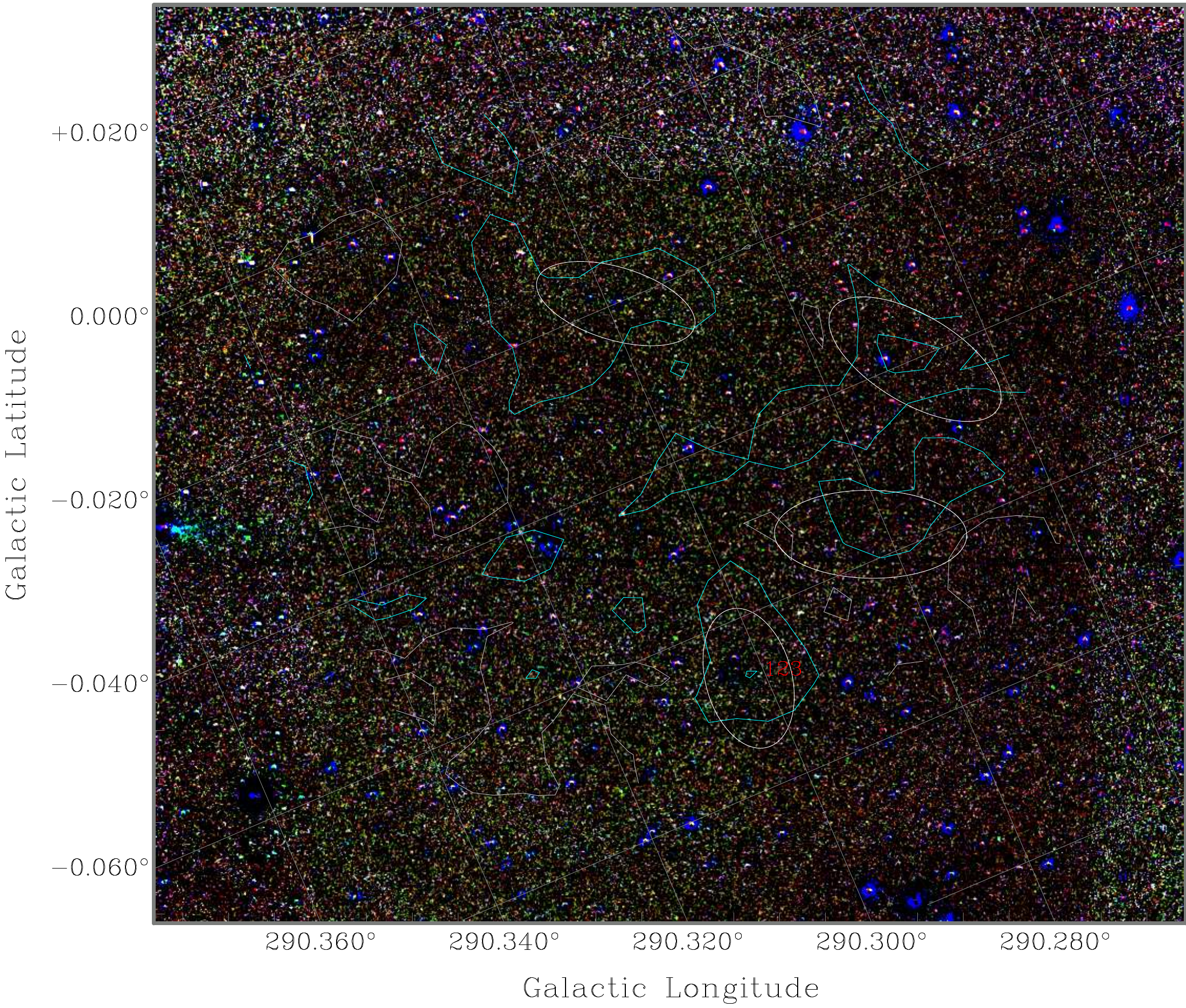}
\caption{Same as Fig.\,\ref{sample}, but for the isolated source BYF\,123 from Paper I.  (a) $K$-band line-free continuum image with \hcop\ contours (grey at 0, cyan at 2 and 4 times the rms level of 0.418\,K\kms).  (b) RGB-pseudo-colour image of the continuum-subtracted $K$-band spectral lines with the same contours as in (a).  (No \nnh\ data are available for this field.)  White ellipses show gaussian fits to the \hcop\ emission in both panels.  At a distance of 6.8\,kpc, the scale is 40$''$ = 1.32\,pc or 1\,pc = 0\fdg0084 = 30\farcs3.}
\label{byf123}
\end{figure*}

\begin{figure*}
(a)\hspace{-3mm}\includegraphics[angle=-90,scale=0.42]{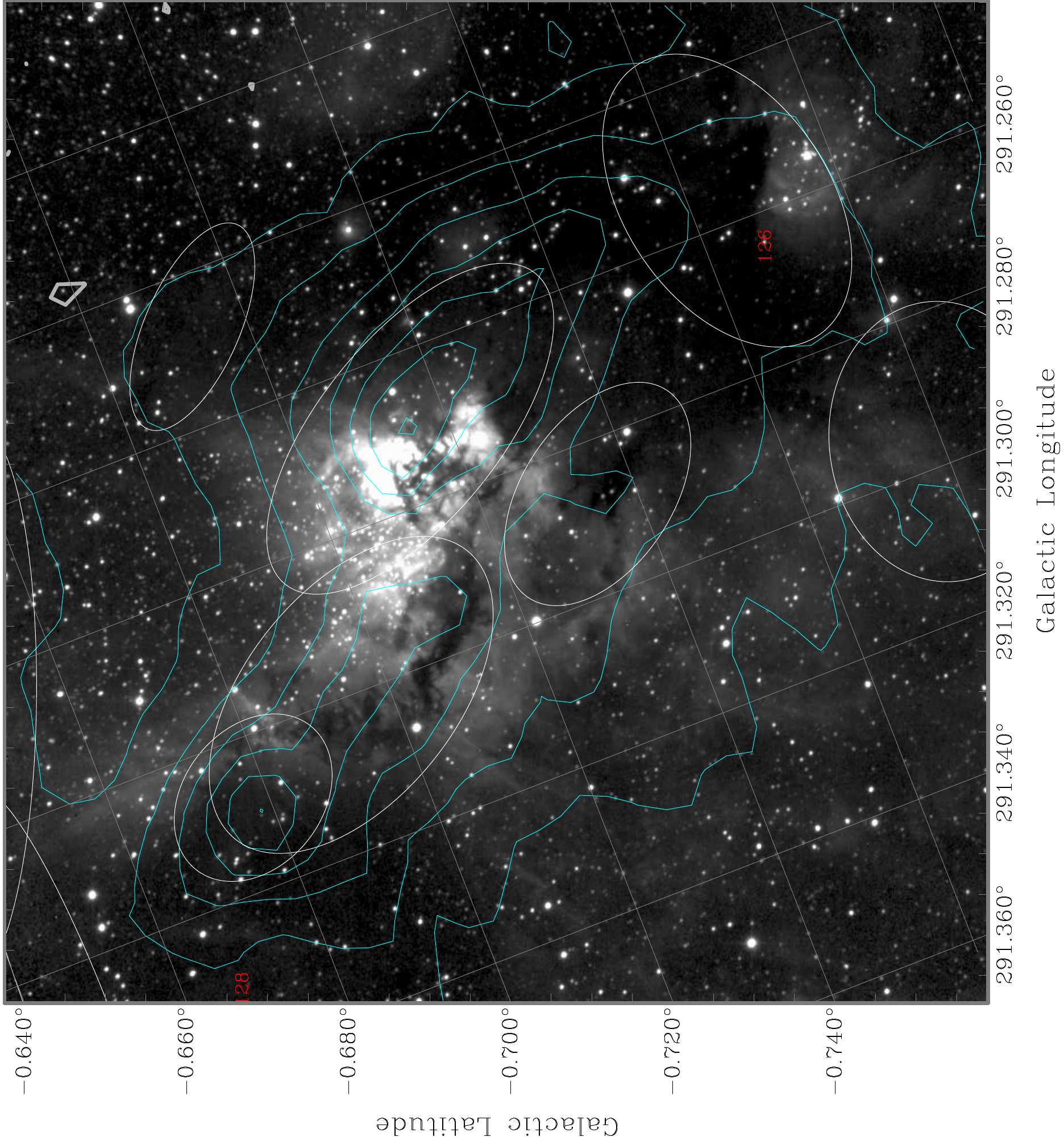}\hspace{3mm}
(b)\hspace{-3mm}\includegraphics[angle=-90,scale=0.42]{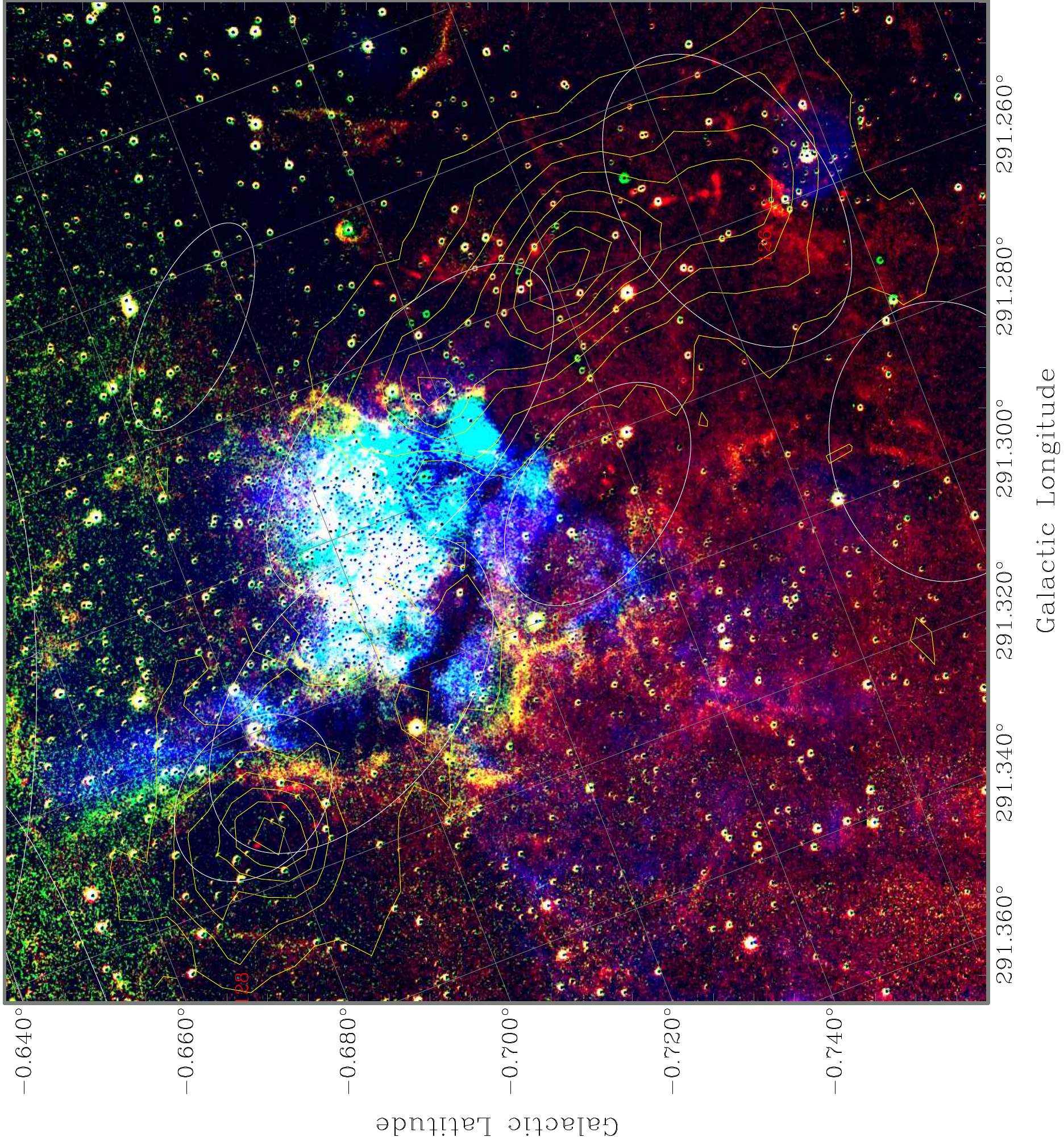}
\caption{Same as Fig.\,\ref{sample}, but for BYF\,126 and 128, also known as NGC 3576 (part of Region 13 from Paper I) 
(a) $K$-band line-free continuum image with \hcop\ contours (heavy grey at 0, cyan at 10, 20, ..., 60, and 70 times the rms level of 0.424\,K\kms).  (b) RGB-pseudo-colour image of the continuum-subtracted $K$-band spectral lines with \nnh\ contours (grey at --4, yellow at 4, 8, ..., 28, and 32 times the rms level of 0.386\,K\kms).  White ellipses show gaussian fits to the \hcop\ emission in both panels.  At a distance of 2.4\,kpc, the scale is 40$''$ = 0.465\,pc or 1\,pc = 0\fdg0239 = 85\farcs9.
\label{byf126}}
\end{figure*}

\begin{figure*}
(a)\hspace{-3mm}\includegraphics[angle=-90,scale=0.38]{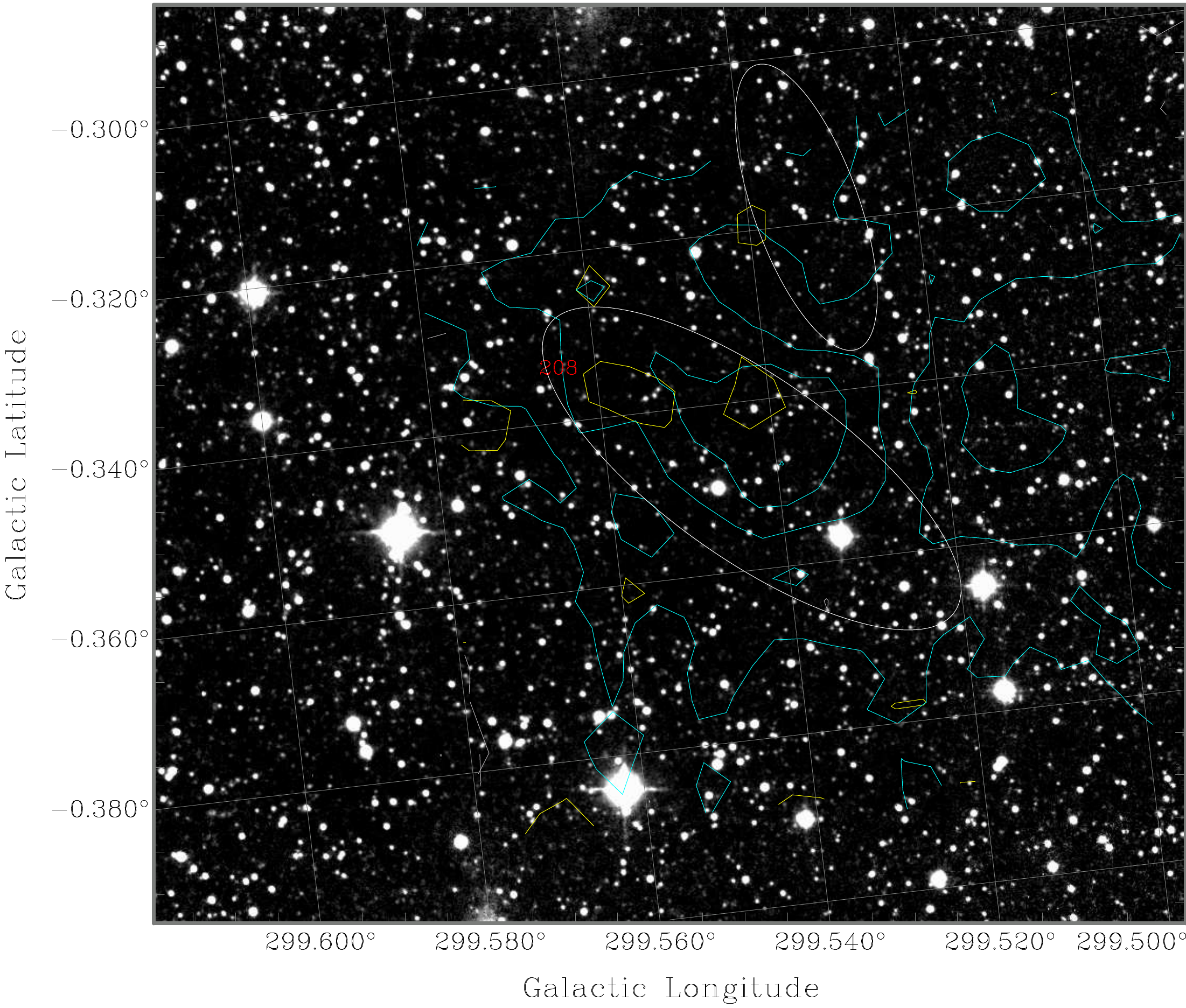}\hspace{3mm}
(b)\hspace{-3mm}\includegraphics[angle=-90,scale=0.38]{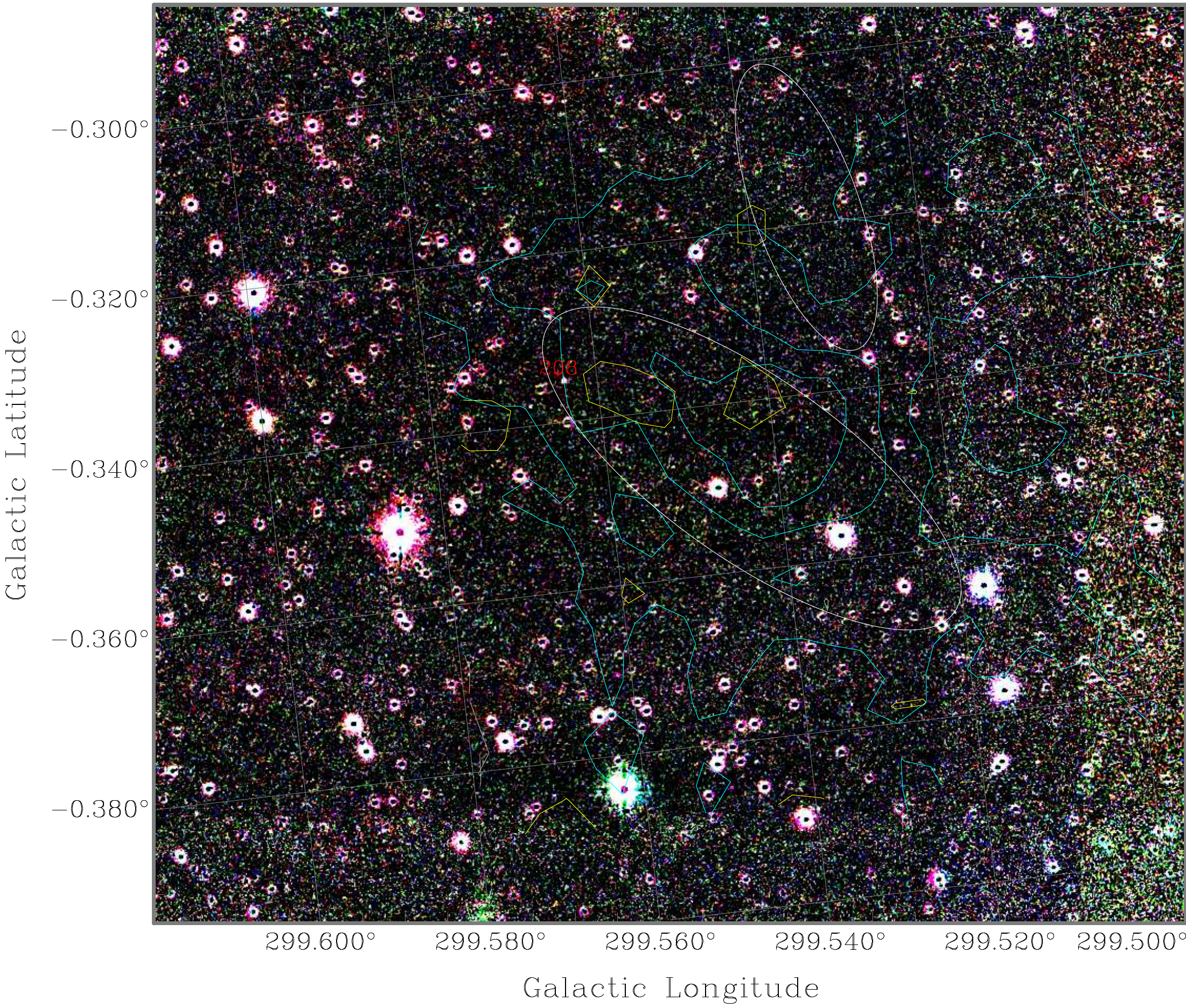}
\caption{Same as Fig.\,\ref{sample}, but for BYF\,208 (part of Region 26 from Paper I).  (a) $K$-band line-free continuum image with both \hcop\ contours (grey at --3, cyan at 3, 6, and 9 times the rms level of 0.230\,K\kms) and \nnh\ contours (grey at --2.5 and yellow at 2.5 times the rms level of 0.199\,K\kms).  (b) RGB-pseudo-colour image of the continuum-subtracted $K$-band spectral lines with the same contours as in (a).  White ellipses show gaussian fits to the \hcop\ emission in both panels.  At a distance of 4.7\,kpc, the scale is 40$''$ = 0.911\,pc or 1\,pc = 0\fdg0122 = 43\farcs9.}
\label{byf208}
\end{figure*}

\label{lastpage}

\end{document}

%% file: table2a.tex
  5a & 280.988 & --1.537 &  1.56(22) & --12.0,--2.0  & 0.62(21) & --7.6 (11) & 1.39 (9) & 1.26 (9) & 143(12) &   46(16) & 104(3) \\
  5b &     N   &     N   &   -- (38) & --12.0,--2.0  &  -- (37) &     --     &   --     &   --     &   --    &    --    &   --   \\
  5c & 281.034 & --1.505 &  0.85(21) & --12.0,--2.0  & 0.64(20) & --7.25(11) & 1.54 (3) & 1.42 (3) & 116(13) &   $<$46  &  11(5) \\
  5d & 281.048 & --1.540 &  0.80(21) & --12.0,--2.0  & 0.66(20) &     S      &   S      &   S      &  45(16) &   $<$46  & 143(5) \\
  7a & 281.074 & --1.570 &  0.84(23) & --12.0,--2.0  & 0.66(22) & --6.4 (10) & 0.80(25) & 0.55(25)\vspace{2mm}&64(14)&$<$47&108(5)\\
 40a & 284.017 & --0.871 &  4.20(24) &   +6.0,+12.0  & 1.53(30) &  +8.58(12) & 1.39(11) & 1.26(11) &  95(13) &   36(18) & 144(3) \\
 40b & 284.036 & --0.893 &  3.20(23) &   +6.0,+12.0  & 1.36(28) &  +8.88(15) & 1.19 (8) & 1.04 (8) &  74(14) &   45(16) &  90(3) \\
 40c & 284.019 & --0.899 &  1.41(24) &   +6.0,+12.0  & 1.06(29) &  +8.5 (6)  & 1.0 (6)  & 0.8 (6)  &  $<$52  &   $<$40  &    --  \\
 40d & 283.989 & --0.836 &  0.94(23) &   +6.0,+12.0  & 0.93(28) &  +9.2 (5)  & 2.06 (9) & 1.98 (9) &  $<$38  &   $<$30  & 124(5) \\
 40e &     N   &     N   &   -- (20) &   +6.0,+12.0  &  -- (24) &     --     &   --     &   --\vspace{2mm}& -- &   --   &   --   \\
 40f & 284.052 & --0.836 &  0.73(22) &   +6.0,+12.0  & 0.77(27) &  +8.2 (10) & 1.3 (5)  & 1.2 (5)  &  $<$38  &   $<$30  &  56(5) \\
 40g &     N   &     N   &   -- (24) &   +6.0,+12.0  &  -- (30) &     --     &   --     &   --     &   --    &    --    &   --   \\
 54a & 285.266 & --0.047 &  0.95(27) &  --1.0,+6.0   & 0.89(31) &  +3.25(23) & 2.06(10) & 1.98(10) &  $<$30  &   $<$30  &  72(5) \\
 54b & 285.257 & --0.073 &  4.87(30) &  --1.0,+6.0   & 1.93(35) &  +2.75(15) & 1.30(18) & 1.16(18) &  57(15) &   $<$30  &  31(3) \\
 54c & 285.251 & --0.090 &  3.25(30) &  --1.0,+6.0   & 1.40(34) &  +4.04(31) & 0.63(31) & 0.39(31)\vspace{2mm}&51(15)&51(15)&0(5)\\
 54d & 285.261 & --0.037 &  0.94(27) &  --1.0,+6.0   & 0.81(31) &  +3.6 (5)  & 1.7 (5)  & 1.6 (5)  &  57(15) &   $<$30  & 121(5) \\
 54e & 285.261 & --0.007 &  1.25(29) &  --1.0,+6.0   & 1.03(34) &  +4.10(25) & 0.41(34) & 0.39(34) &  $<$45  &   $<$30  &   0(5) \\
 54f & 285.307 & --0.077 &  1.08(29) &  --1.0,+6.0   & 0.93(33) &  +2.1 (5)  & 2.51 (5) & 2.44 (5) &  36(18) &   31(20) & 117(5) \\
 54g &     N   &     N   &   -- (46) &  --1.0,+6.0   &  -- (53) &     --     &   --     &   --     &   --    &    --    &   --   \\
 54h &     N   &     N   &   -- (26) &  --1.0,+6.0   &  -- (30) &     --     &   --     &--\vspace{2mm}& --  &    --    &   --   \\
 61a &     N   &     N   &   -- (20) & --25.0,--19.0 &  -- (25) &     --     &   --     &   --     &   --    &    --    &   --   \\
 61b &     N   &     N   &   -- (21) & --25.0,--19.0 &  -- (26) &     --     &   --     &   --     &   --    &    --    &   --   \\
 62  &     N   &     N   &   -- (36) & --25.0,--19.0 &  -- (45) &     --     &   --     &   --     &   --    &    --    &   --   \\
 63  & 286.020 &  +0.040 &  0.92(23) & --25.0,--19.0 & 0.80(28) & --22.48(25)& 1.1 (5)  & 0.9 (5)  &  34(19) &   $<$30  & 126(5) \\
 64  &     N   &     N   &   -- (24) & --25.0,--19.0 &  -- (29) &     --     &   --     &--\vspace{2mm}& --  &    --    &   --   \\
 66  & 286.090 & --0.084 &  0.95(21) & --25.0,--19.0 & 0.56(26) & --21.4 (6) & 1.73(15) & 1.63(15) &  64(14) &   46(16) & 108(5) \\
 67  & 286.077 & --0.020 &  0.74(24) & --25.0,--19.0 & 0.83(29) &     S      &   S      &   S      &  51(15) &   $<$30  &  68(5) \\
 68  &     N   &     N   &   -- (22) & --25.0,--19.0 &  -- (27) &     --     &   --     &   --     &   --    &    --    &   --   \\
 69  & 286.120 & --0.120 &  1.16(24) & --25.0,--19.0 & 0.80(29) & --22.45(36)& 0.96(25) & 0.76(25) &  95(13) &   46(16) & 126(3) \\
 70a & 286.147 & --0.144 &  0.96(23) & --25.0,--19.0 & 0.88(29) & --22.76(39)& 0.87(24) & 0.64(24)\vspace{2mm}&139(12)&$<$30&138(3)\\
 70b & 286.163 & --0.190 &  1.31(19) & --25.0,--19.0 & 0.76(24) & --21.53(22)& 1.44(34) & 1.32(34) &  46(16) &   $<$47  & 101(5) \\
 71  &     N   &     N   &   -- (28) & --25.0,--19.0 &  -- (34) &     --     &   --     &   --     &   --    &    --    &   --   \\
 72  & 286.173 & --0.007 &  0.91(22) & --25.0,--19.0 & 0.87(27) & --21.83(28)& 1.20(22) & 1.05(22) & 112(13) &   45(16) & 135(5) \\
 73  & 286.213 &  +0.166 &  4.87(20) & --25.0,--19.0 & 1.79(24) & --20.69(17)& 0.93(14) & 0.72(14) &  61(14) &   45(16) &  99(3) \\
 76  & 286.327 & --0.360 &  0.83(24) & --25.0,--19.0 & 0.90(29) & --21.6 (9) & 2.1 (9)  & 2.0 (9)\vspace{2mm}&36(18)&$<$38&27(5) \\
 77a & 286.360 & --0.280 &  4.54(20) & --25.0,--19.0 & 2.18(25) & --22.18(18)& 1.07(10) & 0.90(10) &  45(16) &   $<$38  &  37(3) \\
 77b & 286.353 & --0.267 &  3.10(20) & --25.0,--19.0 & 1.63(25) & --22.18(15)& 0.98(15) & 0.79(15) & 101(13) &   $<$45  &  96(3) \\
 77c & 286.380 & --0.237 &  1.53(20) & --25.0,--19.0 & 0.88(24) & --22.80(32)& 1.85 (7) & 1.76 (7) & 101(13) &   $<$45  &  84(3) \\
 77d & 286.330 & --0.292 &  0.97(23) & --25.0,--19.0 & 0.90(28) & --21.30(31)& 1.98(39) & 1.89(39) &  $<$47  &   $<$30  & 166(5) \\
 78a & 286.400 & --0.357 &  0.82(24) & --25.0,--19.0 & 0.88(29) & --20.8 (9) & 1.0 (5)  & 0.8 (5)\vspace{2mm}&$<$30&$<$30&  0(5) \\
 78b & 286.433 & --0.390 &  0.81(25) & --25.0,--19.0 & 0.76(30) & --21.5 (9) & 1.6 (5)  & 1.5 (5)  &  $<$49  &   $<$30  & 135(5) \\
 78c & 286.437 & --0.437 &  0.98(24) & --25.0,--19.0 & 0.94(29) & --22.03(26)& 1.80(11) & 1.70(11) &  $<$39  &   $<$30  & 198(5) \\
 79a &     N   &     N   &   -- (24) & --25.0,--19.0 &  -- (29) &     --     &   --     &   --     &   --    &    --    &   --   \\
 79b &     N   &     N   &   -- (15) & --25.0,--19.0 &  -- (19) &     --     &   --     &   --     &   --    &    --    &   --   \\
 79c & 286.513 & --0.090 &  0.86(24) & --25.0,--19.0 & 0.85(29) & --22.11(33)& 1.82(23) & 1.72(23)\vspace{2mm}&75(14)&$<$30&98(5)\\
 83  & 286.959 & --0.710 &  1.64(38) & --23.5,--12.9 & 0.99(33) & --17.6 (9) & 1.17(22) & 1.01(22)  &  $<$30  &   $<$30  &   0(5) \\
 85a &     N   &     N   &   -- (28) & --23.5,--12.9 &  -- (26) &     --     &   --     &   --     &   --    &    --    &   --   \\
 85b & 286.995 & --0.083 &  1.46(35) & --23.5,--12.9 & 1.11(32) & --18.7(17) & 1.3 (5)  & 1.1 (5)  &  75(14) &   $<$30  & 172(3) \\
 85c &     N   &     N   &   -- (32) & --12.9,--3.3  &  -- (31) &     --     &   --     &   --     &   --    &    --    &   --   \\
117d &     N   &     N   &   -- (22) & --19.5,--15.5 &  -- (33) &     --     &   --     &--\vspace{2mm}& --  &    --    &   --   \\
117e & 288.190 & --1.117 &  0.83(22) & --19.5,--15.5 & 0.79(33) & --17.90(20)& 1.17 (8) & 1.01 (8) & 100(13) &   $<$30  &   0(5) \\
118a & 288.253 & --1.140 &  2.72(22) & --19.5,--15.5 & 1.60(33) & --17.46(13)& 0.83(10) & 0.59(10) &  55(15) &   $<$49  &  45(3) \\
118b &     N   &     N   &   -- (19) & --19.5,--15.5 &  -- (29) &     --     &   --     &   --     &   --    &    --    &   --   \\
118c &     N   &     N   &   -- (22) & --19.5,--15.5 &  -- (33) &     --     &   --     &   --     &   --    &    --    &   --   \\

%% file: table2b.tex
126a & 291.267 & --0.723 &  7.04(26) & --27.5,--21.0 & 2.99(30) & --23.72 (9)& 1.19(14) & 1.04(14) &  $>$55  &   44(16) &  45(5) \\
126b &     N   &     N   &   -- (37) & --27.5,--21.0 &  -- (44) &     --     &   --     &   --     &   --    &    --    &     -- \\
126c & 291.257 & --0.767 &  6.40(28) & --27.5,--21.0 & 2.58(32) & --23.82(22)& 1.19 (9) & 1.04 (9) &  $>$86  &   $>$64  & 120(5) \\
126d &     N   &     N   &   -- (47) & --27.5,--21.0 &  -- (55) &     --     &   --     &   --     &   --    &    --    &     -- \\
126e &     N   &     N   &   -- (43) & --27.5,--21.0 &  -- (50) &     --     &   --     &--\vspace{2mm}& --  &    --    &     -- \\
128a & 291.313 & --0.678 &  8.37(45) & --27.5,--21.0 & 2.77(53) & --24.61(19)& 1.54(15) & 1.43(15) &  45(16) &   $<$45  &   0(3) \\
128b & 291.303 & --0.702 &  2.61(48) & --27.5,--21.0 & 1.72(57) & --24.2 (6) & 1.6 (4)  & 1.5 (4)  & $>$107  &   36(18) &  18(5) \\
208a & 299.543 & --0.341 &  0.94(20) & --41.5,--37.5 & 0.79(30) & --38.46(31)& 0.74(31) & 0.45(31) & 153(12) &   $<$30  &  81(3) \\
208b & 299.536 & --0.315 &  0.65(20) & --41.5,--37.5 & 1.03(30) & --38.2 (5) &   S      &   S      &  $>$45  &   $<$30  & 143(5) \\

%% file: table3a.tex
  5a &  8701 &    0.58(15)   &    5.0(13)  &   --  &     --       &    --       &   --  &    --        &   --      \\
  5b &   --  &     --        &    --       &   --  &     --       &    --       &   --  &    --        &   --      \\
  5c &   --  &     --        &    --       &   --  &     --       &    --       &   --  &    --        &   --      \\
  5d &   --  &     --        &    --       &   --  &     --       &    --       &   --  &    --        &   --      \\
  7a &   --  &     --        &    --       &   --  &     --       &    --       &   --  &    --        &   --\vspace{2mm}\\
 40a &  8565 &    2.10(13)   &   18.0(11)  &  3196 &   1.066(37)  &    3.41(12) &  9925 &  --0.112(71) &   --      \\
 40b &  2819 &    2.36(13)   &    6.66(37) &  7427 &   4.222(37)  &   31.36(27) &  2486 &    0.656(41) &  1.63(10) \\
 40c & 12987 &    0.63(13)   &    8.1(17)  &  7017 &   1.691(37)  &   11.87(26) &  1339 &    1.740(41) &  2.330(55)\\
 40d &  6093 &    4.82(13)   &   29.38(81) &  1183 &   0.881(37)  &    1.043(44)&  1084 &    0.764(41) &  0.828(45)\\
 40e &   --  &     --        &    --       &   --  &     --       &    --       &   --  &    --        &   --\vspace{2mm}\\
 40f &   --  &     --        &    --       &   --  &     --       &    --       &   --  &    --        &   --      \\
 40g &   --  &     --        &    --       &   --  &     --       &    --       &   --  &    --        &   --      \\
 54a & 39131 &   29.0(15)    & 1134(57)    & 16593 &   2.269(7)   &   37.64(11) & 24695 &  --1.941(8)  &   --      \\
 54b & 25349 &   11.7(15)    &  296(37)    & 19841 &   0.722(7)   &   14.32(14) & 15959 &  --0.735(8)  &   --      \\
 54c &  8017 &    0.7(15)    &    5(12)    &  9973 &   0.368(7)   &    3.666(69)& 14945 &    0.185(8)  &  2.77(12)\vspace{2mm}\\
 54d & 21620 &    5.9(15)    &  128(32)    & 14170 &   0.651(7)   &    9.220(98)& 17297 &  --0.498(8)  &   --      \\
 54e & 25635 &    5.3(15)    &  136(37)    & 16856 &   0.471(7)   &    7.93(12) & 20857 &  --0.802(8)  &   --      \\
 54f &  8933 &    7.0(15)    &   63(13)    &  7778 &   1.162(7)   &    9.035(54)&  9523 &  --0.195(8)  &   --      \\
 54g &   --  &     --        &    --       &   --  &     --       &    --       &   --  &    --        &   --      \\
 54h &   --  &     --        &    --       &   --  &     --       &    --       &   --  &    --        &   --\vspace{2mm}\\
 60a &  1363 &  --0.153(64)  &    --       &  4144 &   22.31(14)  &   92.46(57) &  6670 &  --0.15(98)  &   --      \\
 60b &  7302 &  --0.087(64)  &    --       &  3660 &   23.31(14)  &   85.32(51) &  5530 &  --0.06(98)  &   --      \\
 61a & 13249 &  --0.058(31)  &    --       & 15264 &   0.0249(29) &    0.381(45)& 11150 &    0.016(13) &  1.8(15)  \\
 61b &  9337 &  --0.088(31)  &    --       &  8632 &   0.1616(29) &    1.395(25)&  9457 &  --0.161(13) &   --      \\
 62  & 17039 &    0.35(13)   &    5.9(21)  & 10118 &   0.111(15)  &    1.13(15) & 12805 &  --0.025(16) &   --\vspace{2mm}\\
 63  & 12976 &  --0.022(61)  &    --       &  9869 & --0.154(87)  &    --       &  5759 &  --0.32(55)  &   --      \\
 64  &   --  &     --        &    --       &   --  &     --       &    --       &   --  &    --        &   --      \\
 66  & 14895 &    0.442(86)  &    6.6(13)  &  9745 &   0.213(9)   &    2.073(89)&  9652 &    0.166(37) &  1.60(36) \\ 
 67  &   --  &     --        &    --       &   --  &     --       &    --       &   --  &    --        &   --      \\
 68  & 32414 &    0.191(45)  &    6.2(15)  & 47289 &   1.151(60)  &   54.4(28)  & 38773 &    1.365(68) & 52.9(26)\vspace{2mm}\\
 69  &   --  &     --        &    --       &   --  &     --       &    --       &   --  &     --       &   --      \\
 70a & 38262 &    0.897(29)  &   34.3(11)  & 22038 &   0.40(12)   &    8.8(27)  & 31936 &    0.41(23)  & 12.9(74)  \\
 70b & 33407 &    1.159(29)  &   38.72(96) & 16255 &   0.47(25)   &    7.7(41)  & 24806 &    0.090(80) &  2.2(20)  \\
 71  & 13660 &  --0.107(77)  &    --       & 11013 &   0.1869(18) &    2.059(19)&  9728 &  --0.24(18)  &   --      \\
 72  & 35428 &    0.122(46)  &    4.3(16)  & 31768 &   0.200(11)  &    6.36(36) & 44600 &  --0.0198(25)&   --\vspace{2mm}\\
 73  & 27890 &    0.5033(35) &   14.038(97)& 20101 &   2.896(23)  &   58.22(46) & 25601 &    0.438(43) & 11.2(11)  \\
 76  & 21667 &    0.0199(90) &    0.43(20) & 26169 &   0.589(83)  &   15.4(22)  & 20401 &  --0.20(23)  &   --      \\
 77a &  6874 &    1.97(26)   &   13.5(18)  &  7762 &   1.616(37)  &   12.55(29) &  5869 &    0.488(51) &  2.86(30) \\
 77b & 12460 &    0.01(12)   &    0.1(14)  & 14473 &   1.676(37)  &   24.26(53) & 26230 &    0.620(51) & 16.3(13)  \\
 77c & 11596 &    1.94(12)   &   22.5(13)  & 19028 &   1.586(37)  &   30.18(70) &  7017 &    0.256(51) &  1.80(35)\vspace{2mm}\\
 77d & 10347 &  --0.16(12)   &    --       &  4840 &   0.281(37)  &    1.36(18) &  7758 &    0.188(61) &  1.46(47) \\
 78a &   --  &     --        &    --       &   --  &     --       &    --       &   --  &     --       &   --      \\
 78b &   --  &     --        &    --       &   --  &     --       &    --       &   --  &     --       &   --      \\
 78c &   --  &     --        &    --       &   --  &     --       &    --       &   --  &     --       &   --      \\
 79a &   --  &     --        &    --       &   --  &     --       &    --       &   --  &     --       &   --\vspace{2mm}\\
 79b &   --  &     --        &    --       &   --  &     --       &    --       &   --  &     --       &   --      \\
 79c &   --  &     --        &    --       &   --  &     --       &    --       &   --  &     --       &   --      \\
 83  & 19632 &    1.25(25)   &   24.5(49)  &  8203 &   0.030(59)  &    0.25(48) &  8567 &  --0.39(23)  &   --      \\
 85a & 10647 &     --        &    --       & 11864 &     --       &    --       &  9584 &  --0.381(25) &   --      \\
 85b & 11961 &     --        &    --       & 16439 &   0.590(66)  &    9.7(11)  & 17858 &    0.213(25) &  3.81(45)\vspace{2mm}\\
 85c &  5409 &     --        &    --       & 11796 &   2.991(66)  &   35.28(78) &  6526 &    0.080(25) &  0.52(16) \\
117d &   --  &     --        &    --       &   --  &     --       &    --       &   --  &     --       &   --      \\
117e &   --  &     --        &    --       &   --  &     --       &    --       &   --  &     --       &   --      \\
118a &  1827 &  --0.60(13)   &    --       &  1827 &    0.31(15)  &    0.57(27) &  1827 &    0.45(19)  &  0.81(34) \\
118b & 13681 &  --0.26(13)   &    --       & 29983 &  --0.50(55)  &    --       &  2212 &  --0.71(23)  &   --      \\

%% file: table3b.tex
118c & 25658 &  --0.32(13)   &    --       & 26030 &  --0.36(55)  &    --       & 15925 &  --0.06(23)  &   --      \\
123a & 11504 &    0.148(20)  &    1.71(23) &  8869 &     --       &    --       &  6899 &  --0.01(10)  &   --      \\
123b & 14545 &  --0.063(20)  &    --       &  7427 &     --       &    --       &  8835 &    0.15(10)  &  1.31(89) \\
123c & 12987 &     --        &    --       &  9291 &     --       &    --       &  7017 &    0.18(10)  &  1.26(71) \\
123d &  3645 &     --        &    --       &  7117 &    0.64(19)  &    4.5(13)  &  6706 &    0.29(10)  &  1.92(68)\vspace{2mm}\\
126a & 56157 &   85.17(61)   & 4783(34)    & 29782 &   12.8(16)   &  382(48)    & 72722 &  --6.8(26)   &   --      \\
126b & 29695 &   32.80(27)   &  973.9(79)  & 23051 &    5.4(21)   &  125(48)    & 18052 &    3.2(14)   & 57(25)    \\
126c &  9257 &   12.06(27)   &  111.6(25)  & 20812 &    2.3(21)   &   48(44)    & 11880 &    4.2(14)   & 50(16)    \\
126d & 17795 &    1.83(27)   &   32.6(47)  &  7117 &    3.8(21)   &   27(15)    & 20649 &    2.7(14)   & 55(29)    \\
126e &  3645 &     --        &    --       &  7117 &    3.9(21)   &   28(15)    & 18639 &    3.3(14)   & 62(26)\vspace{2mm}\\
128a & 25351 &   19.4(10)    &  492(26)    &  8869 &   75.9(21)   &  674(19)    & 11945 &    2.4(14)   & 29(16)    \\
128b & 31898 &   30.6(10)    &  977(33)    & 35054 &    8.3(21)   &  289(73)    & 20380 &    0.9(14)   & 19(28)    \\
208a & 32709 &  --0.48(15)   &    --       & 37008 &  --1.15(17)  &    --       & 46299 &  --0.8(14)   &   --      \\
208b & 23836 &  --0.08(15)   &    --       & 21789 &    0.12(17)  &    2.6(38)  & 23053 &     --       &   --      \\